 \documentclass[]{aa1}		  

\usepackage{natbib}
\bibpunct{(}{)}{;}{a}{}{,}

\usepackage{lscape}
\usepackage{graphicx}
\usepackage{longtable}
\usepackage{supertabular}

\usepackage{txfonts}
\begin{document}

   \title{The seeds of supermassive black holes and the role of\\ 
   				local radiation and metal spreading}
   \titlerunning{DCBHs}
   \authorrunning{U. Maio et al.}

   \author{Umberto Maio\inst{1,2},
   	   Stefano Borgani\inst{2,3,4}
           Benedetta Ciardi\inst{5}
          \and
          Margarita Petkova\inst{6,7}
          }

   \offprints{U. Maio,\\ e-mail: umaio@aip.de}

   \institute{
	Leibniz Institute for Astrophysics, An der Sternwarte 16, 14482 Potsdam, Germany
	\and
   	INAF--Osservatorio astronomico di Treiste, via G. Tiepolo 11, 34143 Trieste, Italy
	\and
	Department of Physics, University of Trieste, Piazzale Europa 1, 34128 Trieste, Italy
	\and
	INFN-Sezione di Trieste, via Valerio 2, 34127 Trieste, Italy
      \and		
	Max Planck Institute for Astrophysics, Karl-Schwarzschild-Str. 1, 85748 Garching, Germany
	\and
	Faculty of Physics of the University of Munich, Scheinerstr. 1, 81679 Munich, Germany
	\and
	Excellence Cluster Universe, Boltzmannstr. 2, 85748 Garching bei Muenchen, Germany
             }



\abstract
{}
{We present cosmological hydrodynamical simulations including atomic and molecular non-equilibrium chemistry, multi-frequency radiative transfer (0.7-100~eV sampled over 150 frequency bins) and stellar population evolution to investigate the host candidates of the seeds of supermassive black holes coming from direct collapse of gas in primordial haloes (direct-collapse black holes, DCBHs).}
{We consistently address the role played by atomic and molecular cooling, stellar radiation and metal spreading of C, N, O, Ne, Mg, Si, S, Ca, Fe, etc. from primordial sources, as well as their implications for nearby quiescent proto-galaxies under different assumptions for early source emissivity, initial mass function and metal yields.}
{We find that putative DCBH host candidates need powerful primordial stellar generations, since common solar-like stars and hot OB-type stars are neither able to determine the conditions for direct collapse nor capable of building up a dissociating Lyman-Werner background radiation field.
Thermal and molecular features of the identified DCBH host candidates in the scenario with very massive primordial stars seem favourable, with illuminating Lyman-Werner intensities featuring values of $1$-$50\,J_{21}$.
Nevertheless, additional non-linear processes, such as merger events, substructure formation, rotational motions and photo-evaporation, should inhibit pure DCBH formation in 2/3 of the cases.
Local turbulence may delay gas direct collapse almost irrespectively from other environmental conditions.
The impact of large Lyman-Werner fluxes at distances smaller than $\sim 5\,\rm kpc$ is severely limited by metal pollution.
}
{}


\maketitle



\section{Introduction}\label{Sect:introduction}


The appearance of massive black holes is one of the most remarkable events in the first billion years. While standard stellar evolution models predict the formation of black holes with masses comparable to the Sun up to hundreds solar masses, there is currently no general consensus about supermassive black holes.
Observational programmes have led to detection of supermassive black holes with billions of solar masses up to $z \simeq 7.5$  \citep{Fan2001, Fan2006, Mortlock2011, Wu2015, Banados2017}.
Theoretical analyses based on gas accretion on primordial stars have failed in reproducing such large masses in such a short lapse of time: even large metal-free stars of $10$--$1000$ solar masses are not able to leave remnant black-hole seeds having more than 400 solar masses \citep{Hirano2014}.
Black-hole seeds of masses around $10^4$--$10^6\,\rm M_\odot$ have been conjectured \citep{Rees1984, LoebRasio1994} as possible seeds of supermassive black holes. Despite never being observed nor yet fully understood, they could be good candidates to grow up to the desired masses in less than a billion years.
\\
An attracting scenario for the formation of massive black holes is the so-called ``direct-collapse'', a rapid collapse of the gas residing in primordial haloes \cite[][etc.]{BrommLoeb2003, Begelman2006, Mayer2010, Latif2013}.
Albeit direct-collapse black holes (DCBHs) could explain the quasar population at $z>6$, the conditions to form them are very peculiar.
Indeed, they are expected to assemble in atomic cooling haloes hosting inflowing pristine material (i.e. with no heavy elements nor dust), in the presence of a strong H$_2$ dissociating UV field in the Lyman-Werner (LW) band, [11.2, 13.6]~eV.
Under these hypotheses, cooling below $10^4\,\rm K$, normally driven by H$_2$ molecules or metals \citep{Maio2007, Maio2013}, is not possible, hence fragmentation is completely inhibited and the halo gas content could collapse into a DCBH by gravitational instability.
The currently established critical level of dissociating LW radiation to be effective ranges between intensities (in units of $10^{-21} \,\rm erg/s/cm^2/Hz/sr$, hereafter $J_{21}$)
$J_{\rm LW} \sim J_{21}$ and $ \sim 1000 J_{21}$ 
\citep{Shang2010, WG2011, Sugimura2014, JD2017}.
Lately, \cite{Habouzit2016} have shown that fiducial values around $ 30-300 J_{21} $ are required to obtain a number of DCBHs compatible with observations of supermassive black holes.
The amount of photons emitted in the LW band depends strongly on the adopted spectral properties of primordial sources \citep{Sugimura2014}, as well as on H$_2$ photodissociation and  H$^-$ photodetachment rates  \citep{WG2017}.
\\
While values of the order of $J_{21}$ should be sufficient to prevent primordial metal-free runaway cooling \cite[][]{Yoshida2003, Valiante2017}, critical intensities for DCBH formation are still controversial.
In practice, the establishment of  a LW background has the net effect of delaying gas collapse and the epoch of DCBH formation \citep{M2001, Oshea2008}.
However, the role of additional local radiation has not been fully assessed, yet, despite an intense burst of LW photons seems required to completely suppress primordial star formation in nearby galaxies.
\\
The implications of molecular self-shielding are still under debate \cite[see][for recent updates]{GnedinDraine2014, GnedinDraine2016Erratum, Hartwig2015}, because it is not clear if gas self-shielding can significantly preserve molecules formed in pristine environments and limit DCBH formation.
\\
The presence of an early population of cosmic rays could provide enough free electrons to promote the formation of molecular hydrogen \citep{Jasche2007, Leite2017} and hence inhibit the birth of DCBHs.
\\
Similarly, the emission of UV and X-ray photons should enhance H$_2$ formation and increase the amount of LW radiation required to form DCBHs \citep{Inayoshi2015, Inayoshi2016, Latif2015}.
\\
The statistical occurrence of DCBHs is still obscure: different studies have proposed number densities varying from 1 DCBH per 10 Mpc$^3$ comoving volume to 1 DCBH per Gpc$^3$ volume \citep{Habouzit2016}.
Since DCBHs are accreting objects, it is likely that they will have a hot corona, but there is no solid information about its emission at high (X-rays) energies.
\\
Observationally, no DCBHs have been identified so far and initial speculations based on HST, VLT and Keck data are now ruled out by [CII] detections consistent with normal star forming galaxies \citep{Matthee2017}.
\\
Our understanding of DCBHs is still very limited.
As an example, little is known about:
the exact mass distribution and growth mechanism;
the physical properties of the hosting haloes; 
the final fate of a DCBH, such as its possible ejection from the hosting structure or its inclusion into larger massive black holes; 
the inhibiting role of local gas fragmentation as consequence of star formation and metal pollution from heavy elements, acquired either in situ or via minor mergers;
the effects of merger-induced vigorous turbulence halting collapse of pristine material.
Furthermore, the effects of radiation from primordial population~III (popIII) stars and the following population~II-I (popII-I) regime might vary strongly depending on the assumptions about spectral energy distribution (SED) and initial mass function (IMF).
\\
Given the lack of definitive answers to these open questions, throughout this work we will investigate some of the above topics by employing cosmological N-body hydrodynamical simulations including non-equilibrium chemistry calculations and full radiative transfer from PopIII and PopII-I stellar sources.
We will explore chemical and thermal implications of the different populations on DCBH host halo candidates.
\\
The paper is organized as follows. Details on the numerical implementation and the data analysis are given in Sect.~\ref{Sect:method}; results are presented in Sect.~\ref{Sect:results} and discussed in Sect.~\ref{Sect:discussion}; conclusions are summarised in Sect.~\ref{Sect:conclusions}.


\section{Method}\label{Sect:method}


In the following subsections, we briefly describe the most important features of the cosmological calculations we have performed (Sect.~\ref{Sect:simulations}), as well as the selection criteria for our analyses (Sect.~\ref{Sect:selection}).

\subsection{Simulations}\label{Sect:simulations}

The numerical calculations performed in this work are based on radiative hydrodynamical calculations carried out via the parallel numerical code P-Gadget3, an updated version of P-Gadget2 \cite[][]{Springel2005}.
The code implementation combines several physical processes and, in particular, besides gravity and smoothed-particle hydrodynamics (SPH), contains a self-consistent treatment of non-equilibrium chemistry, metal cooling, low-temperature cooling by molecules and fine-structure lines \cite[][]{Maio2007, Maio2010} as well as a multi-frequency implementation of photon propagation based on the Eddington tensor scheme \cite[][]{PS2009, PS2011, PM2012,Maio2016} which takes into account radiative transfer from 150 frequency bins in the energy range [0.7, 100]~eV --  including 
H, He, D, H$_2$, HD, HeH$^+$ transitions as well as 
LW band ([11.2, 13.6]~eV), 
near-IR (at energies $\lesssim\rm 1.7~eV$) and 
UV ($\sim\rm [3, 100]~eV$) radiation.\footnote{
Visible photons with wavelengths between 
$\sim 400\,\rm nm$ and $700\,\rm nm$  have energies in the range 
$1.7$--$3$~eV.
}
\\
Stellar evolution is followed for a range of stellar masses and initial metallicities.
Stars with masses above $8\, \rm M_\odot$ explode as SNe~II and inject a kinetic energy of $10^{51}\,\rm erg$ in the surrounding medium.
Lower-mass stars evolve through AGB or SNe~Ia phase \cite[][]{Tornatore2007} with consequent mass loss.
Explosion energies of massive ($\rm >100\,M_\odot$) PopIII stars range between $10^{51}$ and $10^{53}\,\rm erg$, depending on the mass.
Metal yields for stars with different masses and initial metallicities are traced for a number of species (He, C, N, O, Ne, Mg, Si, S, Ca, Fe, etc.) according to the input tables listed e.g. in the final paragraph of Sect.~2.1 of \cite{Maio2016}.
\\
Star formation takes place stochastically in particles with densities above a threshold of $1 \, \rm cm^{-3}$ and gas and heavy elements are ejected by star forming regions via winds (at $500\,\rm km/s$) \cite[][]{SpringelHernquist2003, Maio2009}.
\\
Metal diffusion in the surrounding medium is mimicked by smoothing individual metallicities over the neighbouring particles in the SPH kernel.
\\
We use the full radiative-transfer simulations performed by \cite{Maio2016} in boxes of 0.5~Mpc$/h$ (comoving) a side.
They sample gas and dark-matter fields with $128^3$ particles for each species, which results in gas and dark-matter resolutions of $6.6 \times 10^2\,\rm M_\odot/{\it h}$ and 
$4.3 \times 10^3\,\rm M_\odot/{\it h}$, respectively, and comoving softening length of $0.2\,\rm kpc/{\it h}$ (i.e. $20\,\rm pc/{\it h}$ at $z=9$).
We note that for a precise picture radiative hydrodynamical simulations should resolve the small structures collapsing at early times below kpc-scales. In terms of space resolution, the softening length used in this work is enough to provide a realistic description of collapsing material at high $z$.
\\
Sources of radiative transfer are distinguished into popIII and popII-I stars, according to the underlying gas metallicity, $Z$.
Stars forming in pristine environments or in regions with $Z<Z_{\rm crit}=10^{-4}\,Z_\odot$ are assumed to be popIII, otherwise they are assumed to be popII-I.
The initial mass function (IMF) adopted for these latter is always a Salpeter IMF over the range ${\rm [0.1, 100]~M_\odot}$, while their input spectral energy distribution (SED) is a black body with effective temperature $T_{\rm eff} = 10^4\,\rm K$, well suited to describe low-mass stars.
\\
Due to our ignorance on the properties of primordial stars, we consider three cases, as also summarised in Table~\ref{tab:assumptions}:
\begin{table*}  
\centering
\caption{
Model assumptions for the three radiative scenarios.
}
\label{tab:assumptions} 
\begin{tabular}{l | cccccc} 
\hline
\hline
Model 		& PopIII IMF					&  PopIII IMF & PopIII BB 			& PopII IMF					&  PopII IMF	& PopII BB \\ 
	 		& range $\rm [M_\odot]$		&  slope 		& $T_{\rm eff}$~[K] 	& range $\rm [M_\odot]$ 	&  slope 		& $T_{\rm eff}$~[K] \\ 
\hline
TH.1e5	& [100, 500]	& $-2.35$ & $10^5$				& [0.1, 100]	& $-2.35$	& $10^4$ \\ 
SL.4e4	& [0.1, 100] 	& $-2.35$ & $4\times 10^4$ 	& [0.1, 100]	& $-2.35$	& $10^4$ \\ 
SL.1e4	& [0.1, 100] 	& $-2.35$ & $10^4$ 				& [0.1, 100] 	& $-2.35$	& $10^4$ \\ 
\hline
\end{tabular}
\end{table*}
i)  very massive stars with top-heavy popIII IMF (with slope $-2.35$ and range ${\rm [100, 500]~M_\odot}$) that emit as a black body with $T_{\rm eff} = 10^5\,\rm K$ (TH.1e5);
ii) massive hot stars\footnote{
The hottest early-type O star in the Milky Way is HD~93129A with an effective temperature of $5.2 \times 10^4\,\rm K$, while the closest O star to Earth is $\theta^1$ Orionis C with an effective temperature of $4.5 \times 10^4 \, \rm K$ \cite[see ][and references therein]{Maio2016}.
}
with Salpeter popIII IMF (with slope $-2.35$ and range ${\rm [0.1, 100]~M_\odot}$) that emit as a black body with $T_{\rm eff} = 4\times 10^4\,\rm K$ (SL.4e4);
iii) regular stars with Salpeter popIII IMF (with slope $-2.35$ and range ${\rm [0.1, 100]~M_\odot}$) that emit as a black body with $T_{\rm eff} = 10^4\,\rm K$ (SL.1e4).
Estimating the exact amounts of ionising photons produced by a star is not a trivial issue, since it requires detailed stellar modelling to quantify total luminosity and spectral properties as a function of the stellar lifetime.
Thus, in the radiative transfer calculations, we assume an ionising luminosity for top-heavy popIII stars of $10^{51}$ photons per second, while for popII or lower-mass popIII sources we assume $10^{49}$ photons per second.
Since each star particle represents a simple stellar population, the emissivity for each star particle is normalized by weighting over the corresponding IMF.
Density-dependent gas self-shielding, which alters rates of the non-equilibrium chemical network, is evaluated following the seminal work by \cite{DraineBertoldi1996}, as mentioned in \cite{Maio2016}.
Self-shielding is effective in a small range of physical conditions (at large gas densities and gas temperatures around or below a few thousands Kelvin). Hence, additional dependences on metallicity can be neglected \cite[as already shown by e.g.][]{Sugimura2014}.
\\
The radiative rates implied by the different SEDs in the different cases are computed consistently with the assumptions and employed, jointly with the relevant collisional rates, to get the correct non-equilibrium abundances from the differential equations describing the evolution of the species number densities and of the photon number density in each frequency bin.
For sake of convergence, the integration of the chemical equations is performed on a timescale which is 1/10th the actual timestep \cite[]{Anninos1997}.
\\
At each snapshot, cosmic structures are identified by means of a friends-of-friends algorithm with a linking length of $20$ per cent the mean inter-particle separation. Substructures are identified by the Subfind algorithm \cite[see][and references therein]{Dolag2009} and are post-processed to trace:
masses, positions, radii, velocities, star formation rates, mass-weighted temperatures, abundances of e$^-$, H, H$^+$, H$^-$, He, He$^+$, He$^{++}$, H$_2$, H$^+_2$, D, D$^+$, HD, HeH$^+$, C, N, O, Ne, Mg, Si, S, Ca, Fe, etc., angular momentum, substructures and all the relevant physical properties of each object.
\\
We adopt a $\Lambda$CDM background cosmological model with present-day expansion parameter normalised to 100~$\rm km/s/Mpc$ of $h=0.7$.
Baryon, matter and cosmological-constant parameters are assumed to be $\Omega_{0,b}=0.04$, $\Omega_{0,m}=0.3$, $\Omega_{0,\Lambda}=0.7$, respectively.
Adopted spectral parameters are $\sigma_8=0.8$ for the $z=0$ mass variance within 8~$\rm Mpc/{\it h}$ radius and $n=1$ for the slope of the primordial power spectrum.
\\ 
We note that our choices for initial conditions, box size and resolution are determined by the necessary trade-off between the required accuracy of the physical descriptions implemented in the code and the numerical feasibility of the runs. The set-up adopted here satisfies such constraints.
We refer the interested reader to \cite{PM2012} and \cite{Maio2016} for more details.
\\
The simulation data considered in this work have redshifts 
$z = 14.5, 11.5, 9.5,  9.0,  8.5$, corresponding to cosmic times of about 
$    0.28, 0.38, 0.5, 0.54, 0.58 \, \rm Gyr$.


\subsection{Selection criteria for DCBH host candidates } \label{Sect:selection}

The formation of a direct-collapse black hole is a difficult event.
Popular analytical models require a number of hypotheses to allow the gas to collapse without fragmenting \cite[see e.g.][for details and reviews]{
BrommLoeb2003, 
Begelman2006, 
Mayer2010,
Choi2015}.
\\
Host haloes should have null star formation rate to assure that there is no ongoing gas fragmentation nor metal enrichment.\\
Then, they should have pristine chemical composition to rule out cooling by heavy elements.\\
The host structure should also be lighted up by a strong radiation field in the LW band to prevent molecule (mainly H$_2$) formation and consequent cooling.
In pristine media without molecular content, H and He collisions would be able to bring gas temperatures down to only $\sim 8000 \,\rm K$ \cite[][]{Oh2002, Smith2017}.
This implies a temperature floor below which the gas cannot cool due to the lack of metallic and molecular coolants.\\
As it will be clear by the basic properties of early haloes (see also next section), common solar-like stars are able to dissociate H$_2$ molecules down to mean fractional values of the order of $10^{-13}$.\\
At the same time, though, the hosting dark-matter mass should be at least $\sim 2\times 10^6\,\rm M_\odot$, since gas in lower-mass haloes does not collapse and can be susceptible to photo-evaporation or nearby stellar feedback \cite[e.g.][etc.]{Whalen2004, Maio2011, Maio2016, Jeon2012, Wise2012, Wise2012prad, Wise2014, Kannan2014, deSouza2014, deSouza2015, DF2018}.
Furthermore, smaller haloes are not suitable to form massive black hole seeds, because of the deficiency of available gas.
\\
Due to such basic constraints, we adopt the following commonly used criteria for the identification of DCBH host candidates in our simulations, choosing haloes with:
\begin{itemize}
\item
null star formation rate ($\rm SFR=0$);
\item
pristine gas ($Z=0$);
\item
mass-weighted gas temperatures higher than $ 8 \times 10^3\,\rm K$;
\item
mean H$_2$ content $ x_{mol} < 10^{-13}$, as proxy for H$_2$ destruction by external radiation;
\item
minimum dark-matter mass of $2\times 10^6\,\rm M_\odot$.
\end{itemize}
\noindent
The first two conditions are easily fulfilled in early epochs, because at those times only relatively few haloes experience star formation and metal enrichment, while, besides very unusual cases, dust production is probably in its earliest phases \cite[][]{Mancini2015}.
On the contrary, the third and fourth conditions are strongly related to the presence of a background or local radiation field that heats the gas and dissociates H$_2$ molecules.
\\
Studies in the literature have clearly shown that the main effect of a uniform dissociating LW background is a shift in the masses and timescales before collapse \cite[][]{Wise2005, Ahn2009, Visbal2014}.
The effects on DCBH formation are small or modest for intensity values around a few up to hundreds times $J_{21}$.
They are dramatic for much bigger values, preventing DCBH formation by the end of the first Gyr when the radiation field exceeds $ \sim 1000 J_{21}$ \cite[][]{Shang2010, Regan2017}.
\\
More complicated is the role of local radiation, for which there is no simple foreseeable trend, that, in fact, strongly depends on the properties and environment of the local radiative sources (SED, emitting power, lifetime, isolated location or clustered regions).
\\
In this paper we account self-consistently for the LW radiation originating by the radiative emission of formed stars and for the consequent build-up of a LW background. Given the small boxes, though, we do not account for the effect of sources located further away. In fact, their contribution is expected to be lower than the local contribution \cite[][]{Ciardi2000}.
Here, we focus on the intriguing implications of local LW radiation from different types of radiative sources.
\\
We stress that the conventional limits outlined above come from popular analytical arguments for DCBH formation and they should be simply considered as necessary conditions.
It is not clear whether they are also sufficient conditions, since additional non-linear processes, such as mergers, substructure formation, rotational motions and/or turbulence might halt direct collapse or even enhance star formation.
These additional phenomena will be addressed throughout this paper.


\section{Results}\label{Sect:results}


In this section we present the main results from our analysis and illustrate the evolutionary pathways of cosmic gaseous systems which could directly collapse into a massive black hole.


\subsection{Basic halo properties} \label{Sect:basic}

To understand the basic properties of the halos that could be able to host DCBH events, we start our investigation by looking at their typical dark-matter masses, chemical content, thermal conditions and star formation rate.

The halo samples at different redshifts contain objects with dark-matter masses between $\sim 10^5 \,\rm M_\odot$ and $10^8 \,\rm M_\odot$.
Their baryonic properties are affected by the assumed features of the primordial stellar populations, as well as of the emitting spectrum.

\begin{figure}
\includegraphics[width=0.5\textwidth]{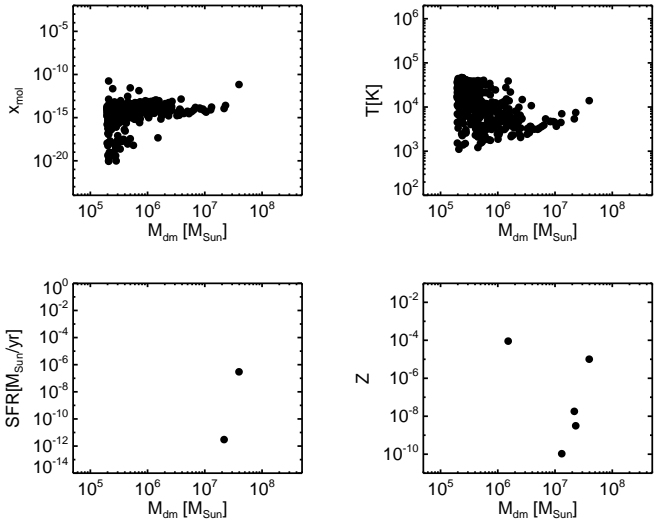}\\
\caption[]{\small Mean H$_2$ content, $x_{\rm mol}$, temperature, $T$, star formation rate, SFR, and metallicity, $Z$, as a function of the dark-matter mass, $M_{\rm dm}$, of haloes at $z=9$ for the run with a $T_{\rm eff}=10^5\, \rm K$ black body as popIII SED. }
\label{fig:basic017.sub}
\end{figure}

In Fig.~\ref{fig:basic017.sub}, halo properties at $z=9$ are shown for the run with top-heavy popIII IMF and black body spectrum with $T_{\rm eff} = 10^5\, \rm K$ (TH.1e5 model).
At this epoch there are two star forming sites with masses of about $2 \times 10^7\,\rm M_\odot$ and $4 \times 10^7\,\rm M_\odot$, respectively.
The same stellar populations that provide UV photons are also responsible for enriching nearby regions up to metallicities of $\sim 10^{-2} Z_\odot$ once they explode as supernovae.
The radiative sources are responsible for enriching nearby regions up to metallicities of $\sim 10^{-2} Z_\odot$. As a result, metal spreading involves both the halo hosting star formation and four additional halos in which star formation is not taking place.
The powerful radiation of the emitted photons is responsible for destroying most of the molecular content in the simulated volume. The residual H$_2$ fraction ($x_{\rm mol}$) usually lies below $10^{-10}$ and reaches values as low as $10^{-20}$ in the smaller unshielded haloes.
The gas mass-weighted temperatures vary correspondingly between $10^3\,\rm K$ and a few times $10^4\,\rm K$ due to radiative heating from primordial stars.
The haloes mostly affected are the ones with small masses -- below a few $10^6\,\rm M_\odot$ -- that are not dense enough to cool against photo-heating, are not able to efficiently self-shield and suffer strong evaporation effects \cite[][]{Maio2016}.
Most of the objects have been heated to temperatures around $10^4\,\rm K$, corresponding to average molecular fractions of $x_{\rm mol} \sim 10^{-13}$--$10^{-15}$.
Haloes with temperatures as low as $\sim 10^{3}\,\rm K$ still retain certain amounts of H$_2$, resulting in $x_{\rm mol} \sim 10^{-10}$--$10^{-12}$.
The biggest halo undergoes star formation, has a mass of $\sim 4\times 10^7\,\rm M_\odot$, an average fraction $x_{\rm mol}\simeq 10^{-11}$ and a mass-weighted gas temperature $T\simeq 10^4\,\rm K$.

\begin{figure}
\includegraphics[width=0.5\textwidth]{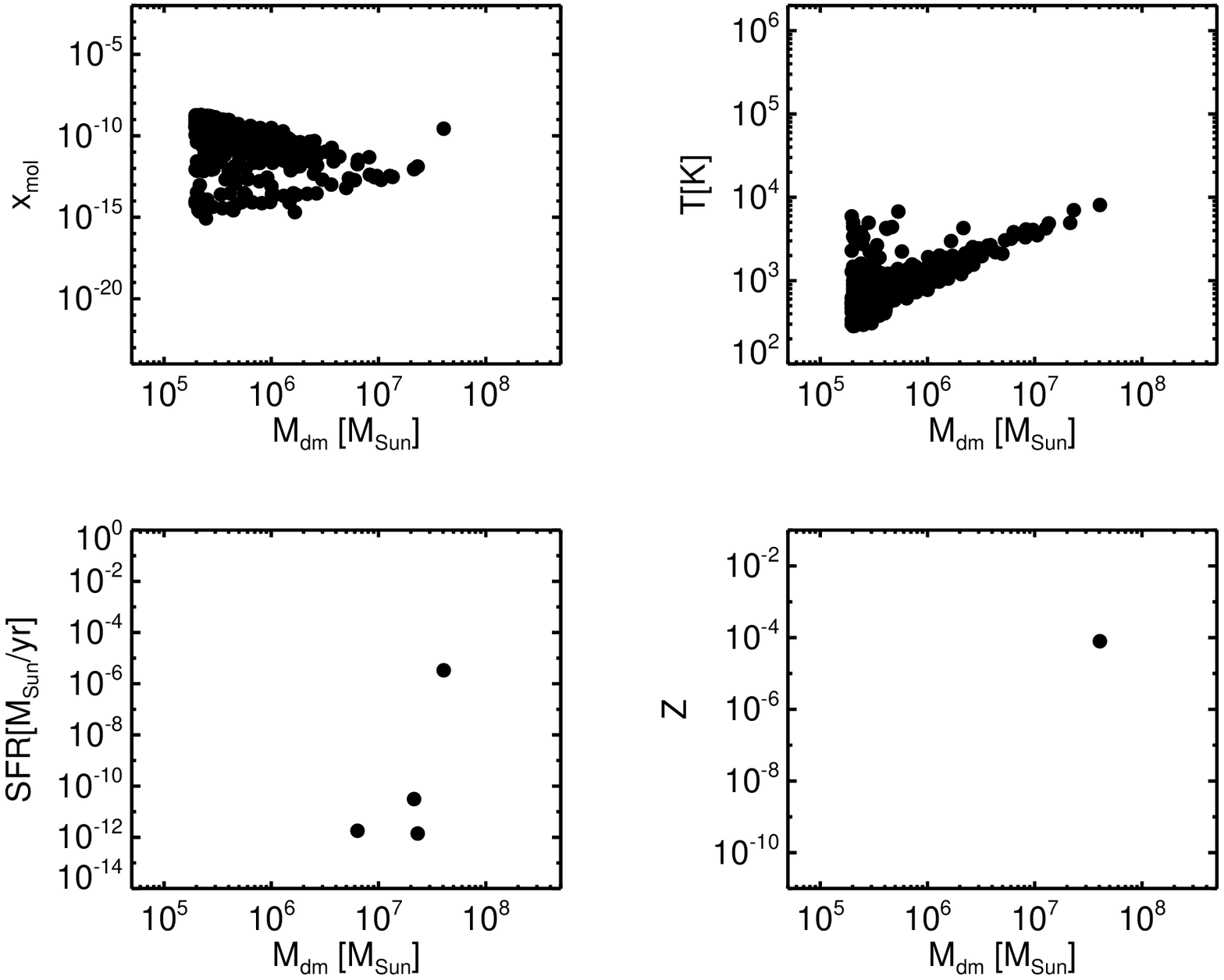}\\
\caption[]{\small As Fig.~\ref{fig:basic017.sub} for the run with a $T_{\rm eff}=4\times10^4\,\rm K$ black body as popIII SED.}
\label{fig:basic017.sub.lowmass.OB}
\end{figure}

Different assumptions for the popIII IMF and SED have clear implications on the basic halo properties.
In Fig. ~\ref{fig:basic017.sub.lowmass.OB} results corresponding to the run with Salpeter-like popIII IMF and $T_{\rm eff}=4\times10^4\,\rm K$ black body (SL.4e4 model) are displayed.
In this case, there are four star forming haloes, i.e. twice as much as the TH.1e5 case, although only the most massive halo with mass of $\sim 4 \times 10^7\,\rm M_\odot$ is found to be enriched at $Z \simeq 10^{-2} Z_\odot$.
The other three haloes (with masses between $\sim 7\times 10^6 \,\rm M_\odot$ and $\sim 2\times 10^7 \,\rm M_\odot$) do not feature signatures of metal enrichment, yet.
This is not surprising, because, once compared to the previous TH.1e5 model, the SEDs of the SL.4e4 model are less powerful (up to 2 dex) and, despite their longer stellar lifetimes (up to 1 dex), radiative feedback is not able to rapidly shut off star formation in distant haloes and metal spreading is less efficient in enriching nearby haloes.
The trend for the average molecular content in each halo, $x_{\rm mol}$, shows that radiative feedback in the SL.4e4 scenario reduces the molecular fraction down to $10^{-8}$--$10^{-15}$, but $x_{\rm mol}$ never reaches values of the order of $10^{-20}$, as in the TH.1e5 case.
This means that the gas is not heated up to very high values and stays confined below $10^4\,\rm K$, as shown by the trend for $T$ as function of mass.\footnote{
Diffuse gas is below $10^4\,\rm K$ as a result of early cosmic expansion.
}
In particular, photon propagation seems to play a little role for the thermal behaviour of the haloes.
There is a well-defined trend of increasing temperature for increasing mass, whose corresponding molecular content shows typical fractions $x_{\rm mol} \sim 10^{-8}$--$10^{-12}$.
The few haloes that are affected by radiation deviate from the displayed increasing trend (low masses and gas temperatures of $2\times 10^3$--$10^4\,\rm K$) and suffer molecule destruction with 
$x_{\rm mol}$ values going down to $\sim 10^{-13}$--$10^{-15}$.

\begin{figure}
\includegraphics[width=0.5\textwidth]{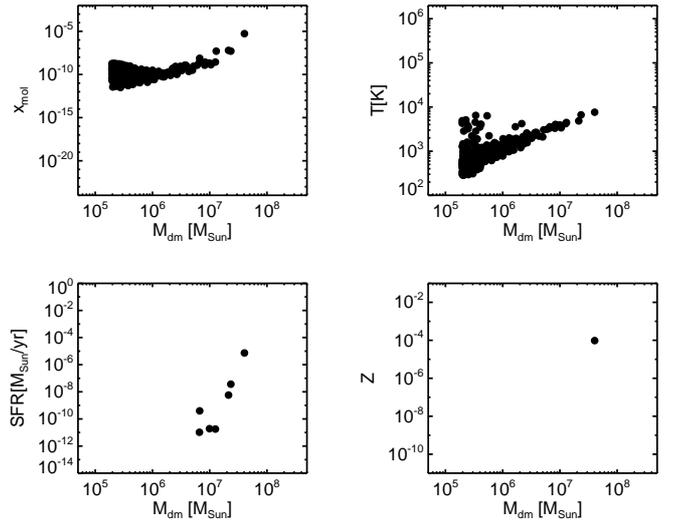}\\
\caption[]{\small As Fig.~\ref{fig:basic017.sub} for the run with a $T_{\rm eff}=10^4\,\rm K$ black body as popIII SED. }
\label{fig:basic017.sub.lowmass}
\end{figure}

The most conservative scenario with a Salpeter-like popIII IMF and a  $T_{\rm eff}=10^4\,\rm K$ black body as popIII SED (SL.1e4) is shown in Fig.~\ref{fig:basic017.sub.lowmass}.
The SL.1e4 model is the least powerful in terms of radiation emitted.
As a consequence, it predicts more star forming haloes (7) than in the TH.1e5 and SL.4e4 models, as well as localised metal enrichment in one single halo with $Z\simeq 10^{-2}\,\rm Z_\odot$, consistently with the previous considerations.
In this scenario, radiative effects are negligible, as clearly visible from the trend of both $x_{\rm mol}$ and $T$.
Molecules are not significantly dissociated and thus average molecular fractions never decline below $10^{-12}$ at $z=9$ and barely reach $10^{-13}$ at later times.
The trend for mass-weighted gas temperatures is little affected by emitted photons, too.
In this case, the chemical and thermal evolution are mainly led by cosmological growth and mechanical feedback, rather than radiative feedback.

We mention that basic host properties at higher redshift do not show evident differences among the models considered, due to the limited structure evolution at early times.
In all the cases, there is no or little star formation and metal enrichment; there is no relevant effect from radiative feedback; average molecular fractions are close to initial-condition values, i.e. $x_{\rm mol} \sim 10^{-4}$--$10^{-6}$; and the haloes are smaller, with mass-weighted gas temperatures between a few hundreds and a few thousands K.
\\


\begin{table*}  
\centering 
\caption{ 
Mean halo properties at $z \ge 9$ for the three different models adopted.
From left to right, different columns indicate:
name of the model considered, 
mean stellar mass in solar units (1), 
mean fraction of haloes hosting metal-enriched star formation (2), 
mean star formation efficiency (3), 
mean gas fraction (4), 
mean absolute UV magnitude in the AB system at $1500\AA$ (5), 
mean bolometric luminosity in solar units (6), 
mean number of ionising photons per second (7).
}
\label{tab:properties}
\begin{tabular}{l | ccccccc}
\hline
\hline
Model 	&	Log($M_\star/M_\odot$)	 &	$ f_{\rm host} $ 	& 	Log($M_\star/M_{\rm gas}$)	&	$f_{\rm gas}$	&	$\rm M_{\rm UV}$	&	Log($L_{\rm bol}/L_\odot$) &  Log($\dot{N}_{\rm ph,ion} / {\rm ph\,\,s^{-1}}$) \\
			 &		(1) 			&         	(2) 			&  			(3) 		& 			(4) 		& 		(5)   	&		(6)		&		(7) \\
\hline
TH.1e5	&	$2.40$	& 	1 	 	&   $-4.04$	&	$0.0831$	&	$-8.47$	&	$6.68$	&	$ 51.3 $ \\ 
SL.4e4	&	$2.73$ 	&	0.18	&	$-3.84$	&	$0.0884$ 	&	$-7.14$	&	$5.48$	&	$ 50.5 $ \\ 
SL.1e4	&	$2.78$	&	0.12	&   $-3.81$	&	$0.0885$	& 	$-7.14$	&	$5.43 $ 	&	$ 50.1 $ \\
\hline
\end{tabular}
\end{table*}
We summarise mean halo properties for star forming haloes in the different models in Table~\ref{tab:properties}.
Statistics are shown for three samples corresponding to the three different scenarios considered here, TH.1e5, SL.4e4 and SL.1e4.
Each sample consists of all the haloes found at redshift $z \ge 9$.
The different columns refer to the mean values of stellar mass ($M_\star$) in solar units, fraction of haloes hosting metal-enriched star formation ($f_{\rm host}$), star formation efficiency ($M_\star/M_{\rm gas}$), gas fraction ($f_{\rm gas}$), absolute UV magnitude in the AB system at $1500\AA$ ($\rm M_{UV}$), bolometric luminosity ($L_{\rm bol}$) in solar units and number of ionising photons per second ($\dot N_{\rm ph,ion}$).
AB magnitudes and luminosities are computed by employing the spectral templates for the emission at $1500\AA$, as a function of stellar lifetimes and metallicities, from the stellar population synthesis code GALAXEV \cite[][]{BC2003}.
We use the star particle properties as inputs, adopt the instantaneous burst model, assume a Salpeter IMF and do not consider any nebular emission.
Since GALAXEV is calibrated for enriched stellar populations with conventional low-mass IMFs (either Salpeter or Chabrier) the resulting magnitudes and luminosities lack the contribution of harder radiation from powerful popIII sources, that for a $10^5\,\rm K$ black body accounts for roughly one dex increased emission at $1500\,\AA$, and that we consider for massive popIII stars.
In the TH.1e5 scenario, where star formation is more inhibited by early powerful popIII stars, the resulting mean stellar mass is slightly smaller, although, due to the higher emitting power, the mean magnitude and bolometric-luminosity estimates are brighter than in SL.4e4 or SL.1e4.
In these latter cases, star formation suppression due to radiative feedback is milder, hence mean stellar masses are $\sim 0.3$~dex larger, mean absolute AB magnitudes more than one unit fainter and mean bolometric luminosities about one dex dimmer.
The effects of radiative feedback in the different cases can be better revealed from star formation efficiencies and gas fractions.
The TH.1e5 scenario has a lower mean star formation efficiency, $M_\star/M_{\rm gas}$, due to stronger gas heating and cooling suppression linked to powerful popIII sources.
In the SL.4e4 and SL.1e4 scenarios mean star formation efficiencies, $M_\star/M_{\rm gas}$, are a factor $\sim 1.6$ larger, because of the more limited impact of early solar-like stars on the surrounding gas.
This is also reflected by their $\sim 6$ per cent larger mean gas fractions, that, in average, are less subject to photo-heating and/or photo-evaporation than in the TH.1e5 case.
The fraction of haloes hosting metal-enriched star formation shows an opposite trend, highlighting the effects of metal spreading: the more powerful the source the higher its ability to enrich and contaminate the medium out to larger distances.
For this reason, the mean $f_{host}$ varies from unity in the TH.1e5 case down to $0.18$ and $0.12$ in the SL.4e4 and SL.1e4 cases, respectively.
The average rate of ionising photons is estimated by
$\dot N_{\rm ph,ion} = f_{\rm esc} Q_i M_\star$,
with
$f_{esc}$ escape fraction, 
$Q_i$ ionization parameter giving the number of ionising photons per second per unit mass of simple stellar population, and 
$M_\star$ stellar mass.
An escape fraction of $f_{\rm esc} = 0.5$ is adopted, consistently with expectations for star forming haloes in the mass range around $10^7\,\rm M_\odot$ \cite[][]{Wise2014}.\footnote{
We note that \cite{Wise2014} suggest values of 
$f_{\rm esc} \simeq 0.5$ for haloes in the mass tange $10^{6.25}$-$10^{7.25}\,\rm M_\odot$,
$f_{\rm esc} \simeq 0.3$ for haloes with mass $\sim 10^{7.5}\,\rm M_\odot$,
$f_{\rm esc} \simeq 0.1$-$0.25$ for haloes with mass $\sim 10^{8}\,\rm M_\odot$ and
$f_{\rm esc} \simeq 0.05$ for haloes with mass $ \sim 10^{8.5}\,\rm M_\odot$.
}
$Q_i$ parameters are taken from the tabulated values of the evolutionary synthesis model by \cite{Schaerer2003} for starbursts with given IMF and metallicity.
For our SL.1e4, SL.4e4 and TH.1e5 models we use \cite{Schaerer2003}'s case A, case B and case C, respectively.
While the two scenarios for regular and massive stars SL.1e4 and SL.4e4 adopt a Salpeter IMF similar to \cite{Schaerer2003}'s case A and case B, our extreme case TH.1e5 has no close equivalent in \cite{Schaerer2003} and we have to rely on case C therein.
TH.1e5 model results to be more powerful than SL.1e4 and SL.4e4 models by $\sim 1\,\rm dex$.
We have to stress, though, that $f_{\rm esc}$  is a poorly known parameter in the literature and its value might span over a wide range.
While \cite{Wise2014} suggest average values of the order of 50~per cent, other authors \cite[such as][]{Yoshida2007} suggest values closer to unity, i.e. a factor of 2 larger, and dependent both on the particular stellar mass range considered and on the features of the environment.
Because of these uncertainties, the expected $\dot N_{\rm ph,ion}$ values could vary sensibly.


\subsection{DCBH host candidates} \label{Sect:candidates}

\begin{figure}
\includegraphics[width=0.46\textwidth]{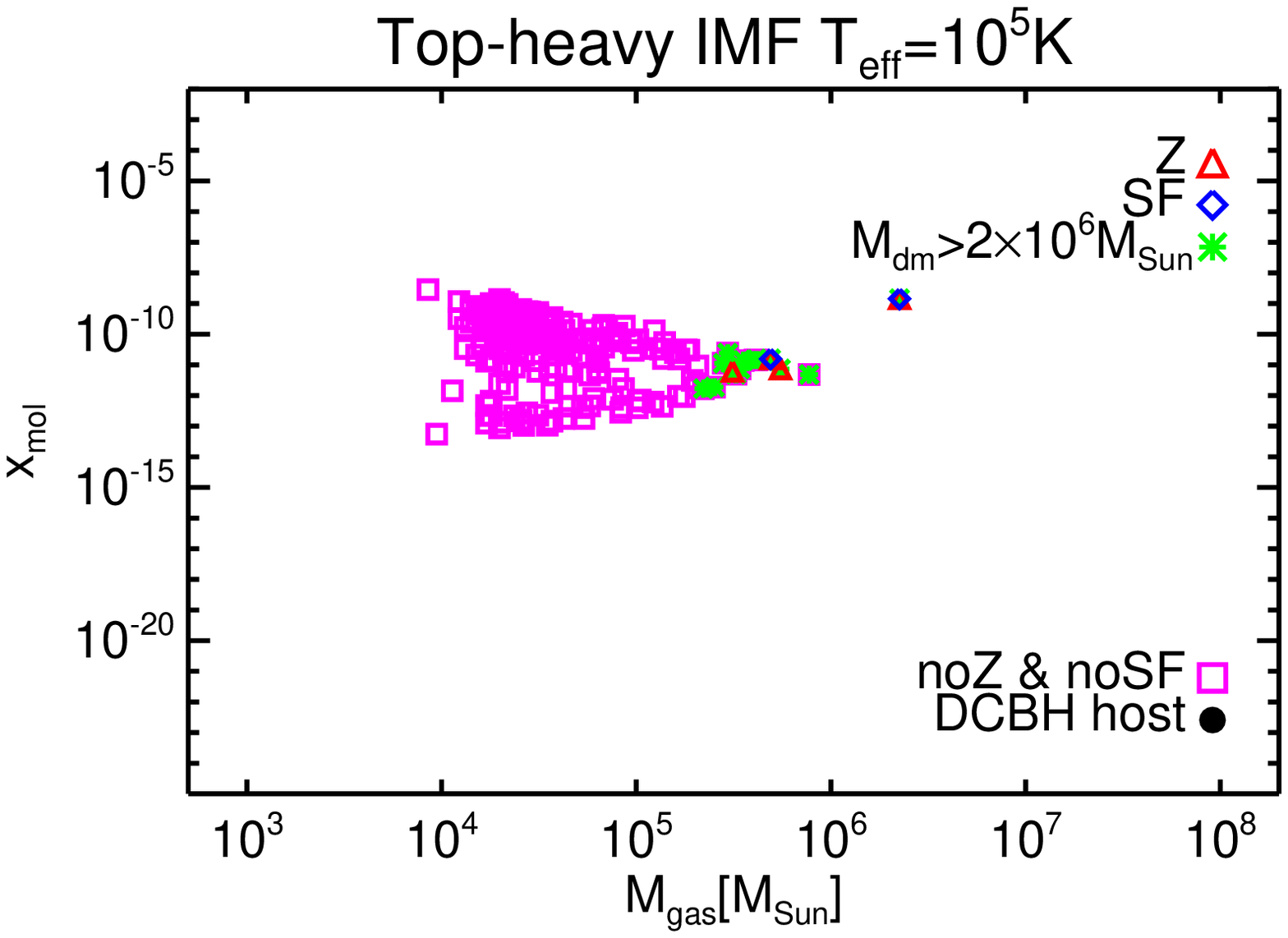}\\
\includegraphics[width=0.46\textwidth]{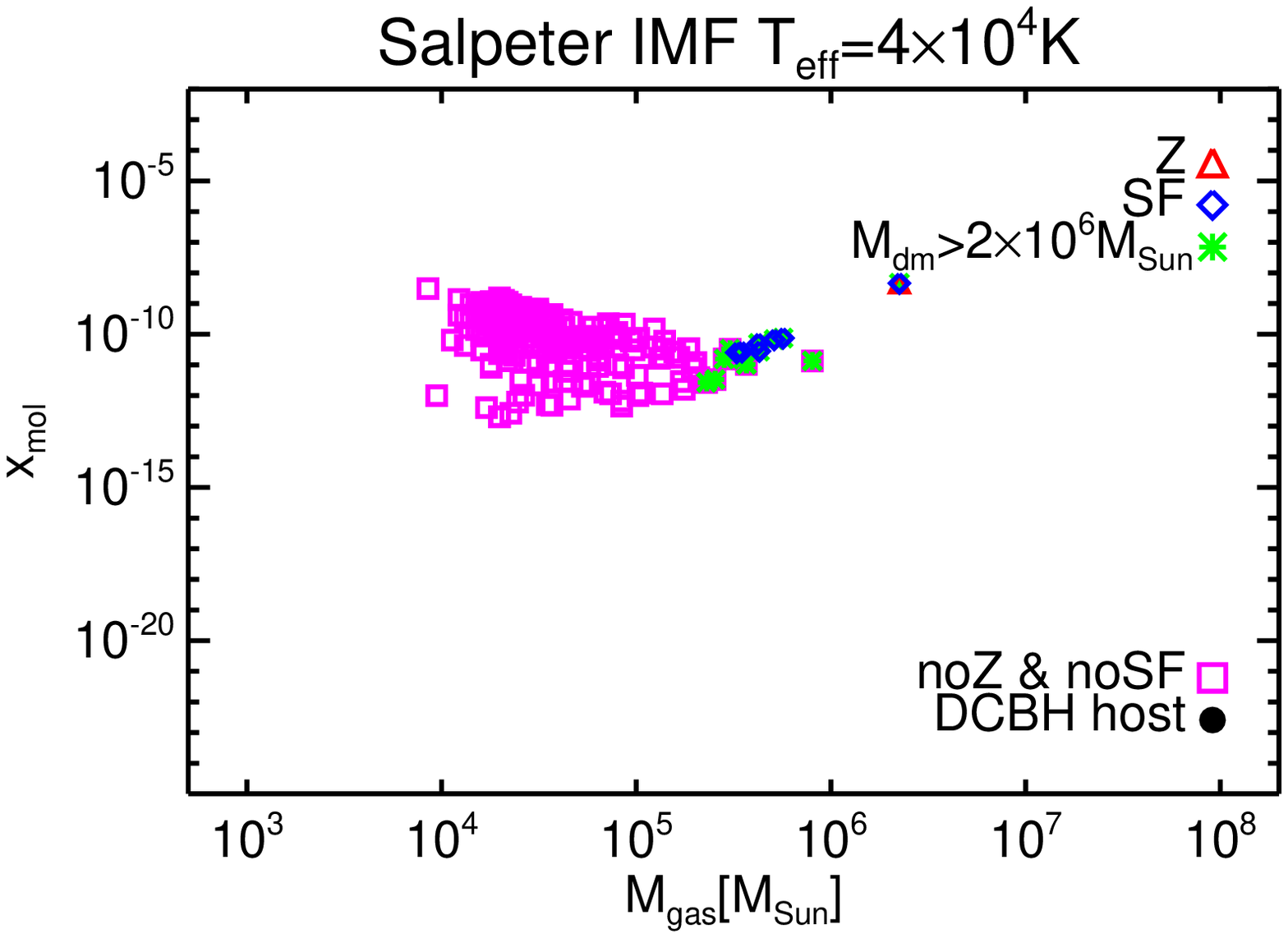}\\
\includegraphics[width=0.46\textwidth]{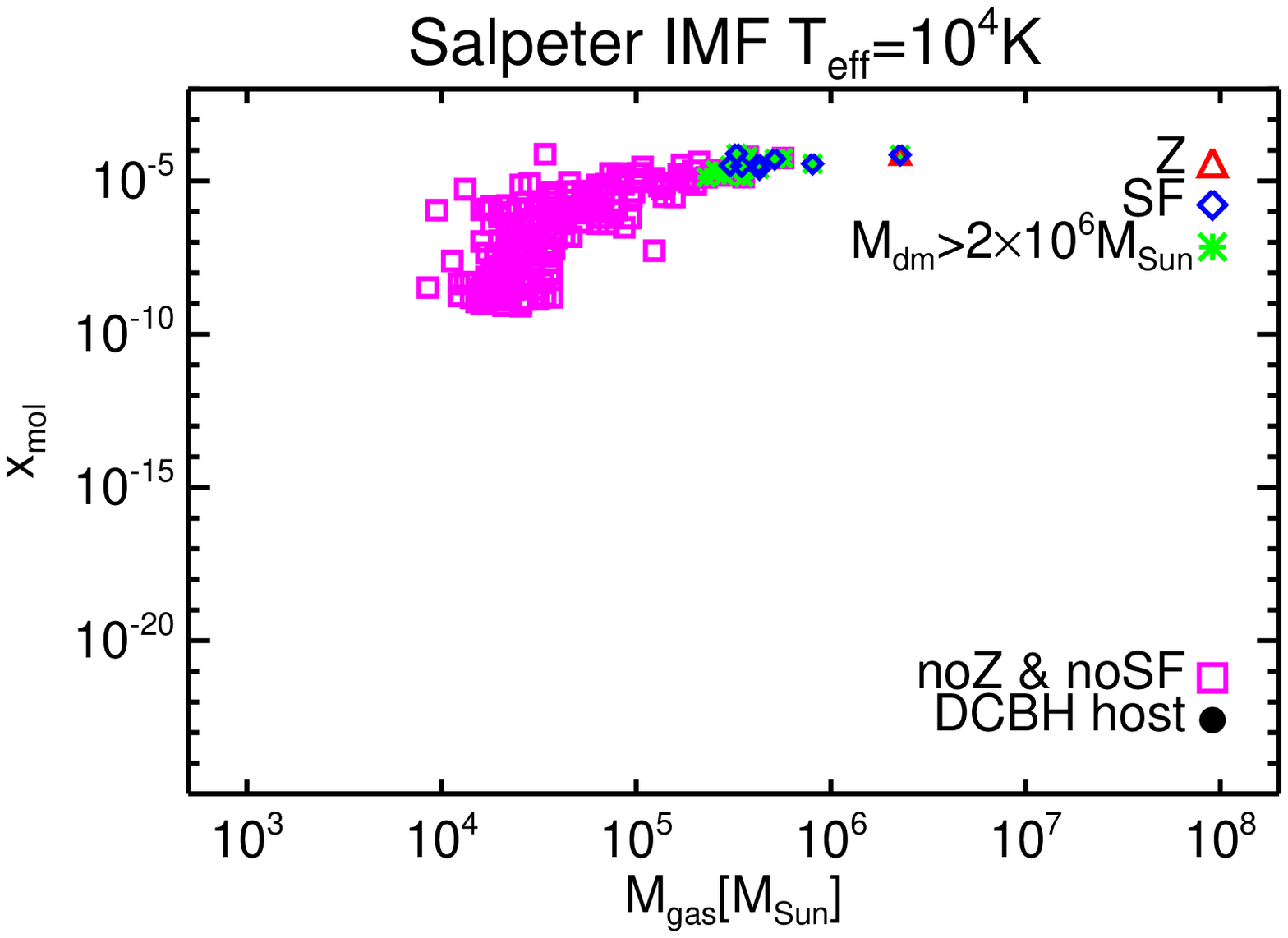}\\
\caption[]{\small
Gas mass versus molecular fraction of the simulated haloes at $z=11.5$ for runs with different popIII SEDs: TH.1e5 (top panel), SL.4e4 (middle panel) and SL.1e4 (bottom panel).
Different symbols refer to different types of haloes: metal enriched haloes (red triangles), star forming haloes (blue diamonds), haloes with dark-matter mass larger than $2 \times 10^6\,\rm M_\odot$ (green asterisks), pristine non-star-forming haloes (magenta squares) and DCBH host candidates (black bullets), that have no metals, no star formation, dark-matter mass larger than $2 \times 10^6\,\rm M_\odot$, gas temperature larger than $8\times 10^3\,\rm K$ and molecular fraction lower than $10^{-13}$.
No DCBH host candidates are present at this epoch in any of the radiative models.
}
\label{fig:DCBH011}
\end{figure}

\begin{figure}
\includegraphics[width=0.46\textwidth]{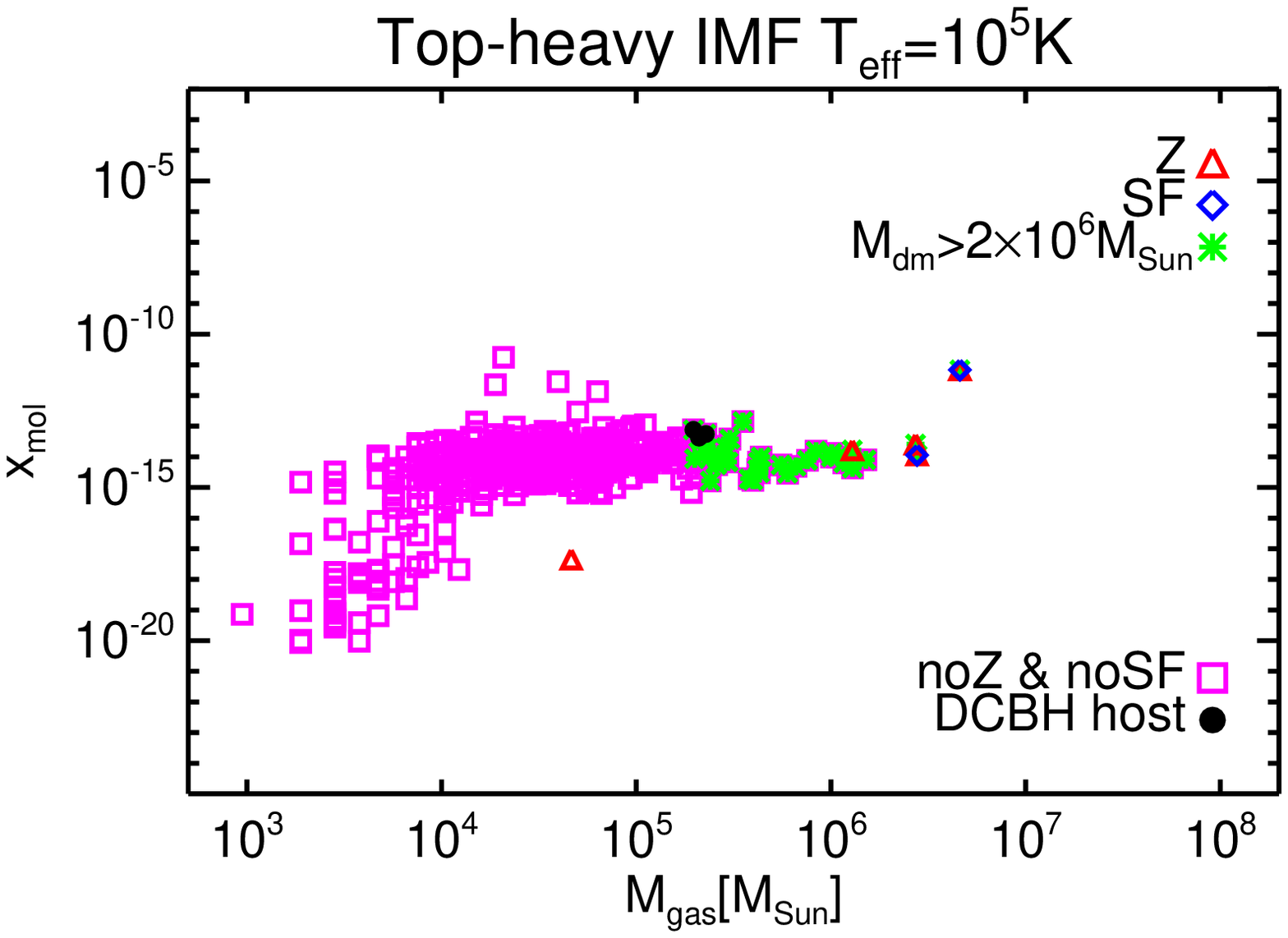}\\
\includegraphics[width=0.46\textwidth]{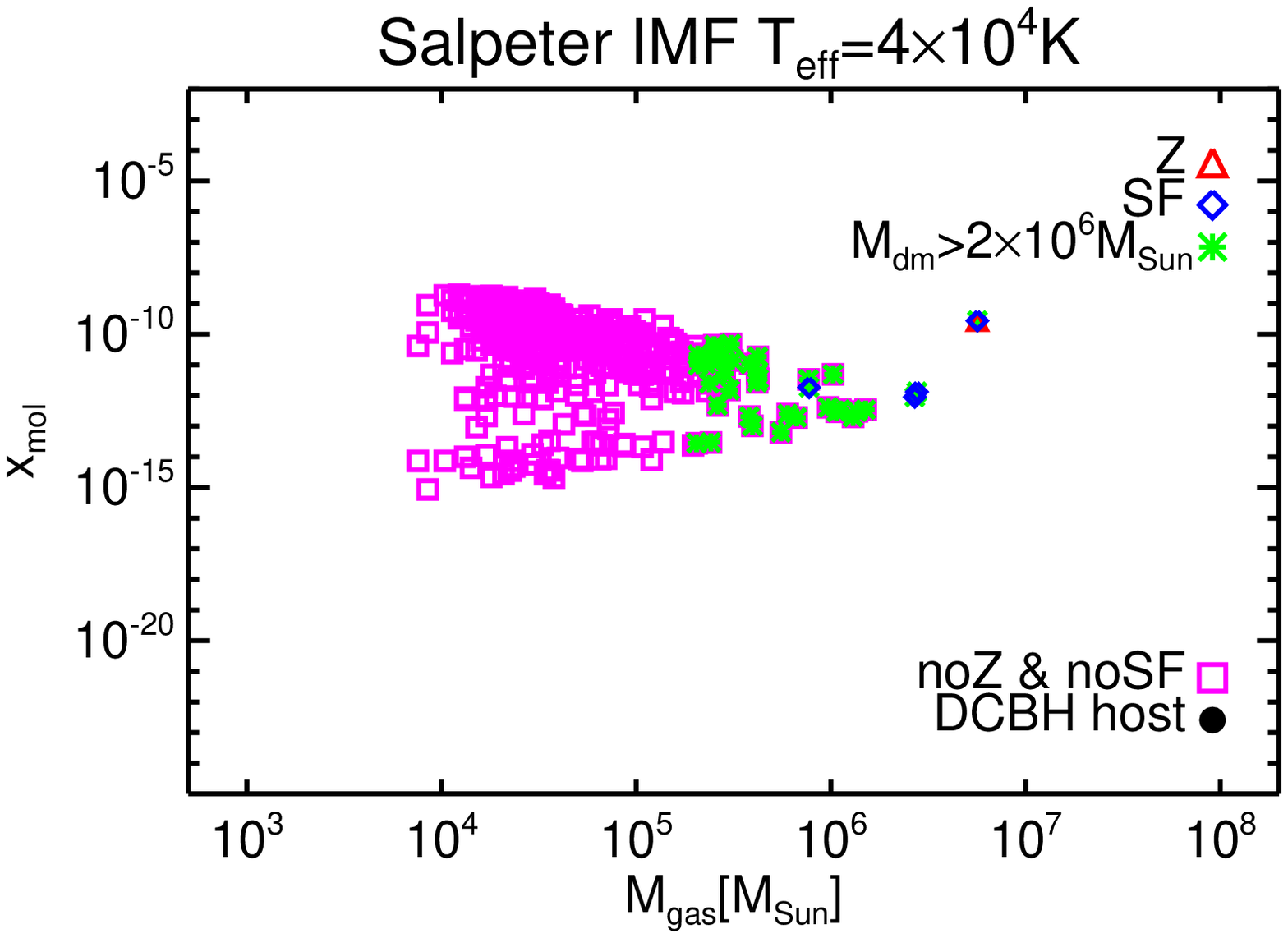}\\
\includegraphics[width=0.46\textwidth]{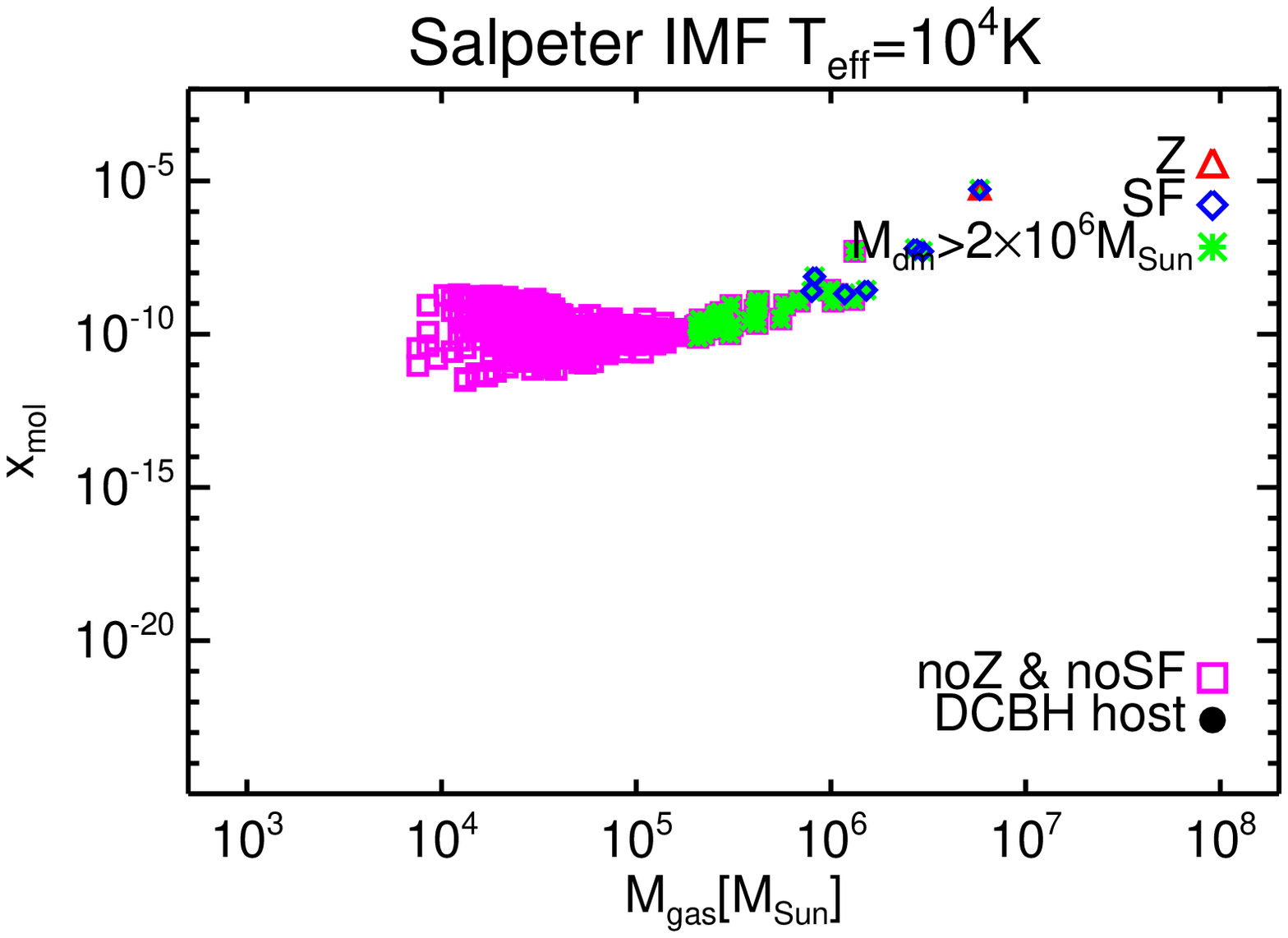}\\
\caption[]{\small
Same as Fig.~\ref{fig:DCBH011}, but at $z=9$. There are 3 possible DCBH host candidates in the top panel.
}
\label{fig:DCBH017}
\end{figure}

To understand the properties of the haloes hosting DCBH events in the three models considered in this work we select the simulated candidates by referring to the conditions listed and discussed in Sect.~\ref{Sect:selection}.
The resulting trends and requirements for DCBH formation are shown in Fig.~\ref{fig:DCBH011} and Fig.~\ref{fig:DCBH017} for $z=11.5$ and $z=9$, respectively.
They refer to the baryon properties that need to be checked to investigate the possibility of a direct collapse of the gas.
In each figure, the host gas mass is plotted against the molecular fraction for
the TH.1e5 model (top), 
the SL.4e4 model (middle) and
the SL.1e4 model (bottom).
Since only gas (and not dark matter) would collapse directly into a black hole, the gas masses quoted in the figures are comparable to the expected mass of the DCBH which would be born from such process.
Metal enriched haloes are denoted by red triangles, star forming haloes by blue diamonds and haloes more massive than $2 \times 10^6\,\rm M_\odot$ by green asterisks.
Magenta squares refer to pristine non-star-forming haloes, while the resulting DCBH host candidates are highlighted by bullet points.
The halo population at very early times (first hundreds of million yrs after the Big Bang) is dominated by pristine non-star-forming small haloes that retain their molecular content irrespectively from the radiative model considered for primordial stars.
Larger amounts of molecules start to form only in halos with dark-matter mass higher than $2 \times 10^6\,\rm M_\odot$ and gas content of about $3 \times 10^5\,\rm M_\odot$.
Because of the local metal pollution, ongoing star formation and large molecular fraction, these primordial halos are not suitable candidates for hosting a DCBH, since local gas is going to cool below 8000~K and fragment further
\cite[][]{McCourt2016}.
Most of the haloes are still H$_2$ rich and too small to induce a direct gas collapse and no DCBH candidates are found at these epochs.

Results at $z=11.5$ are shown in Fig.~\ref{fig:DCBH011}, about $380$ million yrs after the Big Bang.
The first stars have formed at  $z\simeq 14.5$ and have evolved for about $\sim 100$ million yrs, while new stars are born in the meantime.
\\
In the top panel (TH.1e5), we see that photon propagation dissociates molecules up to levels as low as $x_{\rm mol} \sim 10^{-8} $--$10^{-14}$ and that, due to the powerful stellar feedback, there are only 2 star forming haloes and 4 enriched haloes, 2 of which are quiescent non-star-forming objects polluted by nearby spreading events.\\
The SL.4e4 model (centre) produces similarly strong variations (with $x_{\rm mol} \sim 10^{-8} $--$10^{-13}$), although mechanical and radiative feedback is not so extreme to push metals into other haloes (they rather remain confined in their birth place) and to shut off star formation in the larger ones.
Consequently, there are several massive objects (green asterisks) with typical dark masses above $2 \times 10^6\,\rm M_\odot$ and typical gas masses between a few times $10^5$ and $10^6\,\rm M_\odot$, but none of them is a good DCBH host candidate.
In fact, these massive-enough haloes feature values of $x_{\rm mol}$ that are always higher than $10^{-13}$ and range around $x_{\rm mol}\simeq 10^{-12}$--$10^{-10}$.
Considering that lower-mass objects reach smaller $x_{\rm mol}$ values, it is clear that shielding effects in the denser regions of the bigger haloes play a crucial role to prevent DCBH formation (see also sect.~\ref{Sect:simulations} and \ref{Sect:basic}).\\
The trends in the bottom panel follow from the weaker sources assumed for this case (SL.1e4).
Thus, most of the haloes with dark-matter mass $ > 2 \times 10^6\,\rm M_\odot$ undergo star formation and/or get enriched with metals, still keeping average molecular fractions of the order of $x_{\rm mol} \sim 10^{-6}$--$10^{-4}$.
Smaller haloes suffer minimal radiative effects and $x_{\rm mol}$ always stays above $10^{-10}$.
The bulk of the haloes in this scenario is either ``large" and star forming or small and quiescent, therefore they are unlikely to host DCBHs.
\\
Environmental effects (mergers, feedback and ongoing cosmological growth)  contribute to spreading the trends in the plots.

We show the results at $z=9$ in Fig.~\ref{fig:DCBH017}.
This redshift correspond to a cosmic time of half Gyr, i.e. slightly more than $100$ million yrs after $z=11.5$.
At such epoch, the differences among the three models are striking.
\\
In the TH.1e5 case (top panel), powerful radiative emissions from primordial stars have heavily dissociated H$_2$, so that most of the haloes have $x_{\rm mol} \sim 10^{-15}$ and the smaller ones get down to $10^{-20}$.
The two largest haloes (blue diamonds) with gas masses of about $2$--$4\times 10^6\,\rm M_\odot$ are forming stars and are metal enriched, as well as two other smaller nearby object (red triangles).
The quiescent pristine haloes with dark-matter mass larger than $2\times 10^6\,\rm M_\odot$, despite their low molecular fraction, are mostly cold, with gas temperatures below $8\times 10^3\,\rm K$ (see sect.~\ref{Sect:basic}).
Thus, they cannot host DCBH formation.
Only three objects with dark mass of roughly $2\times 10^6\,\rm M_\odot$ and gas mass of about $2$--$3 \times 10^5\,\rm M_\odot$ are hotter and poor of molecules, hence they are good DCBH host candidates.
We note that the minimum temperature threshold of $8\times 10^3\,\rm K$ is mainly led by H collisional cooling \cite[e.g.][]{Oh2002, Smith2017}, active during cosmic structure formation, when diffuse cosmic gas falls into the growing potential wells of dark-matter haloes and gets shock-heated to typical temperatures of the order of $10^4$-$10^5\,\rm K$. In absence of additional coolants, it is not possible to bring temperatures below $ \sim 8\times 10^3\,\rm K $.
In the quiescent haloes mentioned above, gas is still at a diffuse stage, as demonstrated by their small masses. Thus, the hosted gas simply follows the thermal cosmic expansion, hence it is by no means able to collapse nor to survive nearby feedback events.
Only haloes with dark-matter masses larger than $\sim 2 \times 10^6\,\rm M_\odot \times $ are going to host collapsing events and survive nearby star formation or photo-evaporation feedback 
\cite[][]{Whalen2004, Jeon2012, Wise2012, Wise2012prad, Wise2014, Maio2016}.
\\
In the SL.4e4 case (middle panel), the behaviour of the gas is bimodal, with many pristine quiescent haloes having still an average molecular fraction $x_{\rm mol}>10^{-13}$ and a few others lying at $x_{\rm mol}<10^{-13}$. These latter are the ones that surround star forming haloes (blue diamonds) and result affected by the nearby radiative feedback, although not by chemical feedback.
These objects are too small to host DCBHs, though.
Some of the largest haloes are undergoing star formation and/or metal enrichment, however, the remaining ones (green asterisks) are still cold or feature $x_{\rm mol}>10^{-13}$.
Thus, no eligible DCBH hosts are found for this model.
\\
In the SL.1e4 case (bottom panel) the scenario is much less dramatic for the environment of star forming halos.
Indeed, radiation from weaker solar-like sources affects very marginally the thermal and chemical properties of the local gas and has little implications for external haloes.
For this reason, star formation is not severely inhibited by radiation and all the objects with gas masses higher than $\sim 10^6\,\rm M_\odot$ form stars and are locally enriched by metals.
The entire population of primordial haloes has an average molecular content that is always above the threshold of $10^{-13}$ and therefore it cannot host DCBH formation.
In this case, gas molecular evolution is mainly led by mergers and feedback effects, while photon propagation act as a minor character.
\\
In general, cosmological evolution in the first half Gyr is responsible for spreading the values displayed in the plots, mostly in the low-mass end, that is very susceptible to environmental processes.


\subsection{DCBH host candidate location} \label{Sect:locations}

\begin{figure}
\includegraphics[width=0.45\textwidth]{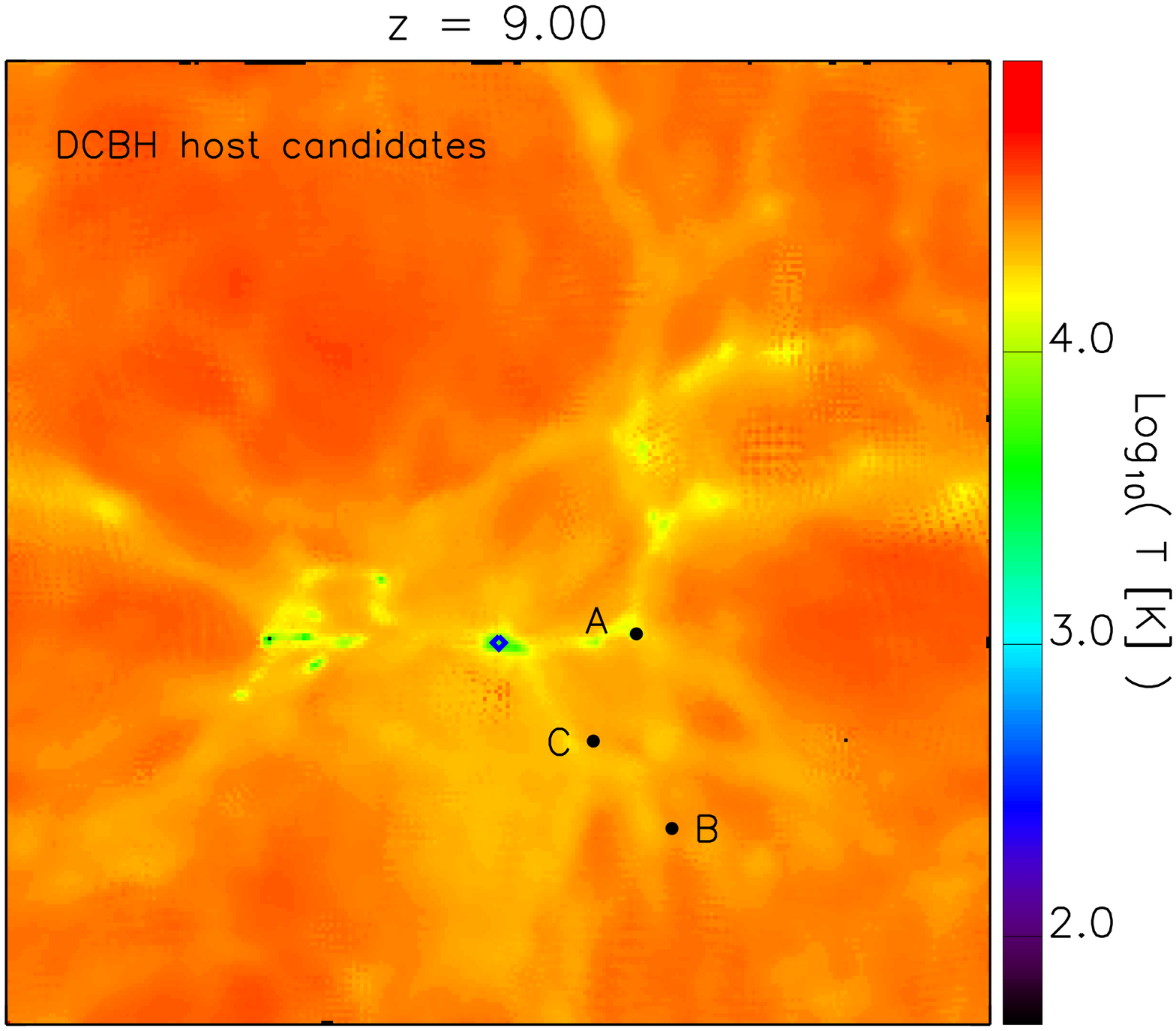}
\vspace{-0.35cm}\\
\includegraphics[width=0.45\textwidth]{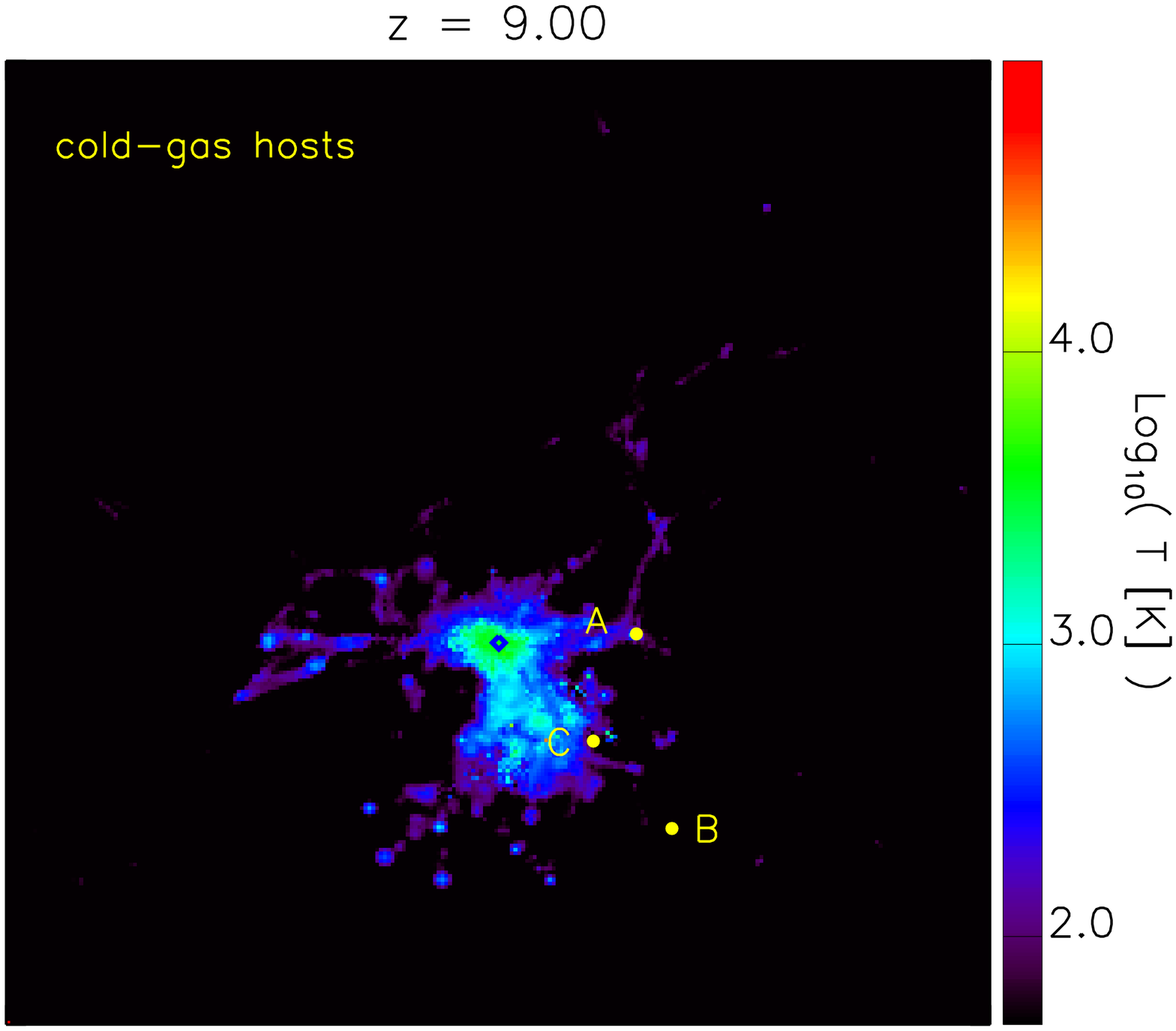}
\vspace{-0.35cm}\\
\includegraphics[width=0.45\textwidth]{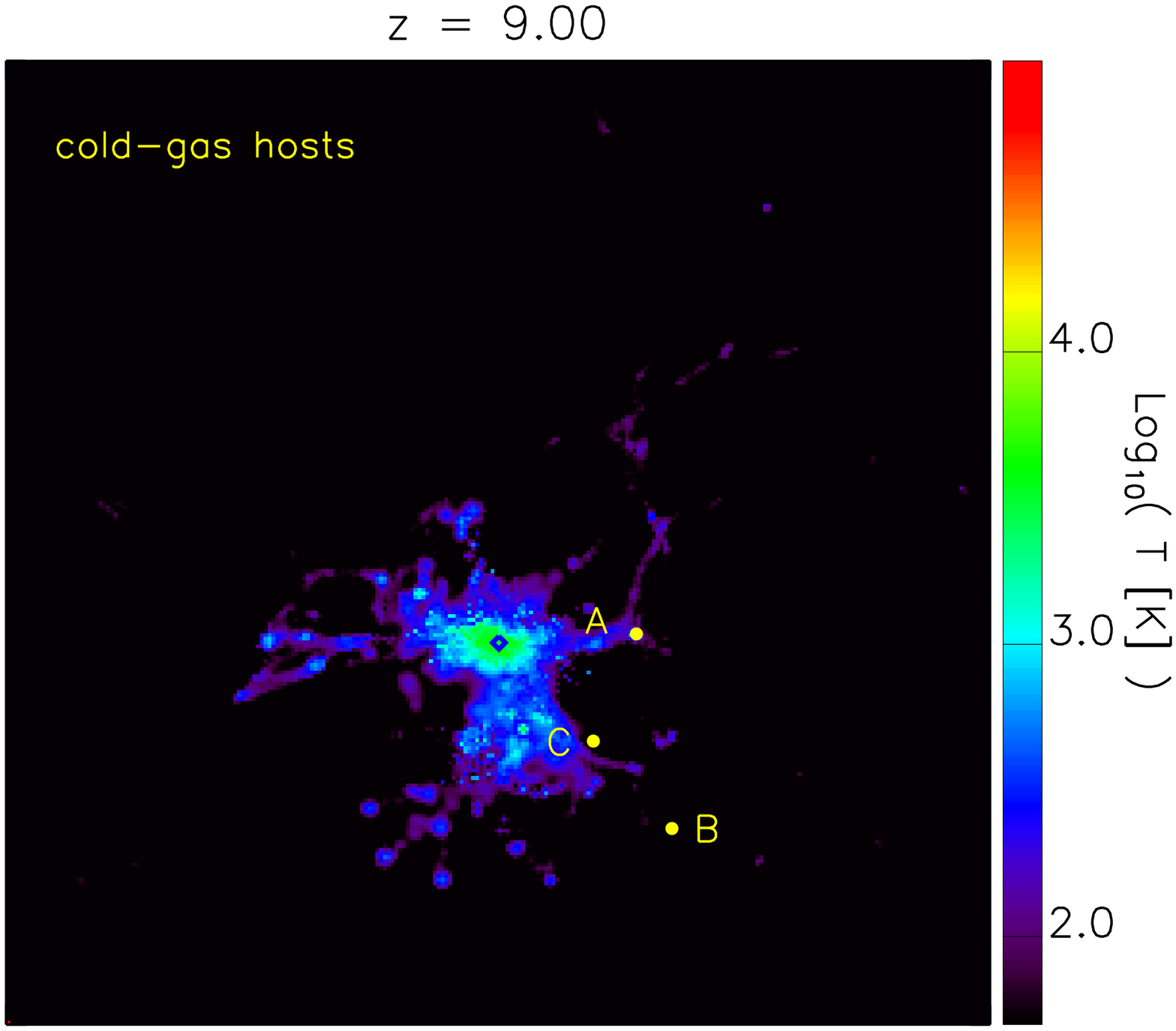}\\
\caption[]{\small
{\it Top.}
Mass-weighted temperature map where we have marked the position of the DCBH host candidates at $z=9$ for the run with top-heavy popIII SED (TH.1e5).
The map is a projection of the simulated structures on the $xy$ plane centred on the middle of the $z$ axis and within a slice of width equal to 1/20th the box length.
The DCBH host candidates are denoted by black bullets and the letters A, B and C.
{\it Centre.}
Same as top panel, but for the run with OB-like popIII SED (SL.4e4). The positions of the haloes marked by the yellow bullets and the letters A, B and C are the same as in the top, but in this case they are not DCBH host candidates.
{\it Bottom.}
Same as top panel, but for the run with standard solar-like popIII SED (SL.1e4). Also in this case, the haloes marked by the yellow bullets and the letters A, B and C are not DCBH host candidates.
}
\label{fig:groups017}
\end{figure}
In Fig.~\ref{fig:groups017}, we display the position where DCBH host candidates are found (bullet points) with respect to the dissociating radiative sources (blue diamond) at redshift $z=9$ in the TH.1e5 model (top).
These hosts have similar total masses slightly larger than $2\times 10^6\,\rm M_\odot$.
The three candidates are found at comoving (physical) distances of about
72  (7.2) kpc/{\it h},
290  (29) kpc/{\it h} and 
80  (8) kpc/{\it h}
from the central radiative source.
For sake of clarity, they have been marked by letters A, B and C and we will refer to them in the following as candidates A, B and C, respectively.
Since temperature is the main driver of DCBH formation, the temperature map allows us to get hints about the thermal properties of cosmic gas in different environments.\footnote{
The cosmic structure is well visible also from the various maps in \cite{Maio2016}.
}
Interestingly, DCBH candidates are all located near cosmic filaments that preserve mass-weighted temperatures of $\sim 10^4\,\rm K$.
Isolated haloes are found not to be suitable DCBH host candidates.
This is not surprising, because early isolated objects are usually smaller (hence they do not satisfy the mass requirement for DCBH formation) and thinner (hence they can severely suffer radiative heating and photo-evaporation effects, instead of gas collapse).
In clustered environments, such as filaments or filament intersections, there is a wider variety of halo masses, thermal conditions and molecular content, so there are higher chances that DCBH formation requirements are fulfilled.

As a comparison, in the figure we also check the thermal conditions of the corresponding haloes in the SL.4e4 (centre) and SL.1e4 (bottom) models.
In both cases the medium is typically colder because of more efficient molecular cooling and weaker radiative fluxes.
In fact, haloes denoted by A, B and C (yellow bullets) host colder gas denote haloes with mass-weighted temperatures below $10^4 \,\rm K$ and ranging between roughly $10^2$ and $10^3\,\rm K$.
In the SL.4e4 scenario, haloes A and C have mass-weighted temperatures of $\sim 10^3\,\rm K$, while halo B features values $\lesssim 10^2\, \rm K$.
In the weaker SL.1e4 scenario, instead, only halo A has a mass-weighted temperature of $\sim 10^3\,\rm K$, while haloes B and C are both at $\lesssim 10^2\,\rm K$.
Given these typical thermal conditions, the gas in these haloes results too cold to experience direct collapse, irrespectively from the hosting dark-matter mass, chemical composition and local star formation.
The main reason why the temperatures of these haloes are lower than in the previous case lie in the adopted features of stellar sources. In these two cases, central sources are not powerful enough to reach haloes at large distances and to dissociate their molecular content. Therefore, the hosts A, B and C evolve almost unaffected by them, can retain cold gas and will probably fragment in the next epochs. With such thermal conditions they are unlikely to turn into DCBHs.
These structures could be, instead, small diffuse cold objects that are just assembling and represent the theoretical counterparts of currently debated early damped Lyman-alpha systems or dwarf galaxies forming at the end of reionisation \cite[for further discussions on these topics we refer the interested reader to available works in the literature, such as the ones by][]{Simcoe2012, MaioDLA2013, Keating2014, Bosman2015,Bosman2017, Garcia2017}.


\subsection{DCBH host candidate radiative properties} \label{Sect:radiation}

\begin{figure}
\centering
\includegraphics[width=0.46\textwidth]{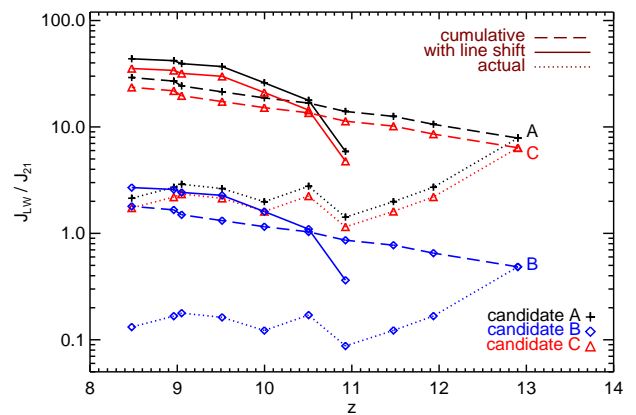} \\
\caption[]{\small
Spectral intensity in the LW band, $J_{\rm LW}$, in units of $J_{21}$ at the locations of the three DCBH host candidates (A, B and C) identified at $z=9$ as function of redshift, $z$.
For each candidate, as indicated by the legends, the figure shows the actual amount of LW radiation produced at each snapshot (dotted lines), the cumulative LW radiation resulting from the sum of all the LW radiation produced until any given redshift $z$ (dashed lines) and the radiation entering the LW band estimated by including redshifted photons from earlier times (solid lines).
}
\label{fig:J}
\end{figure}

The DCBH host candidates are exposed to external local radiation, therefore, it is interesting to estimate the average spectral intensity, $J_\nu$, they experience at different times and frequencies, $\nu$.
As we are interested in the LW band, if not otherwise specified we will refer to this frequency range for the next calculations of $J_\nu$.
To this aim, we focus on the locations of the three hosts identified at $z=9$ at physical distances from the radiative sources of 7.2~kpc/{\it h} (A), 29~kpc/{\it h} (B) and 8~kpc/{\it h} (C), respectively.
Due to the cosine law for isotropic radiation, they observe a spectral intensity $J_\nu = F_\nu / \pi$, where $F_\nu$ is the average monochromatic flux and $\pi$ is the value of the solid angle under which each of the three host candidates `sees' the radiation.\footnote{
The value of $\pi$ comes from the integration of the cosine law over the solid angle [0, $\pi/2$] $\times$ [0, $2\pi$].
}
The average monochromatic flux emitted by radiative sources is given by luminosity divided by surface area and frequency bin.
Under spherical approximation this leads to:
\begin{equation}
\label{eq:J}
J_\nu = \frac{F_\nu}{\pi} = \frac{ \dot{N}_{\rm ph} h_p \nu } {4 \pi^2 r^2 \Delta \nu}
\end{equation}
with
$\dot{N}_{\rm ph}$ number of ionising photons per second, 
$h_p$ Planck constant,
$\nu$ central frequency of the considered frequency range,
$r$ distance and
$\Delta \nu$ frequency bin.
To have an estimate of the expected order of magnitude for
$\nu$ LW central frequency (corresponding to 12.4~eV) and 
$\Delta \nu$ LW band (corresponding to [11.2, 13.6]~eV),
it is convenient to rewrite the previous expression as
\begin{equation}
\label{eq:Jnumbers}
J_{\rm LW} \simeq
9.1 \!\times\! 10^{-21}
\left(  \frac{ \dot{N}_{\rm ph} }{10^{50}\, \rm s^{-1}}  \right)
\left(  \frac{\rm kpc }{r}  \right)^2
~ \rm erg/s/cm^2\!/Hz/sr.
\end{equation}
The redshift evolution of the spectral intensity, in units of $J_{21} = 10^{-21}\,\rm erg/s/cm^2\!/Hz/sr$, to which the three halos, A, B and C, are exposed to, is displayed in Fig.~\ref{fig:J}.
We note that such commonly adopted reference unit, $J_{21} $, represents a quite large spectral intensity, being equal to 100 Jansky per steradiant.

The actual production of LW  photons (due to radiation emitted by sources at that specific redshift), considering the processes of stellar mass growth and loss, is described by the dotted lines, while their cumulative distribution is given by the dashed lines.
In order to quantify the exact amounts of LW radiation influencing the DCBH candidates at $z=9$, the solid lines include the integrated effect of line shift into the LW band of LW photons produced at earlier times.
For the sake of simplicity, we treat the LW central line at 12.4~eV as representative of the whole LW band. This is not a crude approximation, because the range of $\Delta\nu$ around the central frequency is rather small.
Thus, we impose a redshift constrain for $z_{\rm ref} = 9$ of $\left| \Delta z / (1+z_{\rm ref}) \right| < \Delta\nu/\nu \simeq 0.2$, or $z \lesssim 11$.
This means that, roughly speaking, LW photons produced until $z\sim11$ result redshifted in the LW band itself at $z=9$, hence increasing the total amount of radiative intensity experienced by the three involved structures.
We additionally take into account harder UV energies that can be shifted into the LW band.
We limit our calculations only to a few representative lines\footnote{
The lines considered in the text are of interest, because they correspond to ionization energies of D, H$_2$, He,  HeH$^+$ and He$^+$ species, respectively.
We have verified that, given the steep decrease of the spectral shape at high frequencies, these specific details are not crucial.
}
at about 15, 15.4, 24.6, 44.5, 54.4~eV.
They increase the resulting LW intensity according to their relative contribution to the adopted black-body shape, $\propto \nu^3 \left[ 1- \exp( h_p \nu / k T) \right]^{-1}$, rescaled by redshift as $ \left[ (1+z)/(1+z_{\rm ref}) \right]^{-4} $.
These lines are produced at earlier times, when they have higher frequency than the LW central frequency.
During cosmological evolution they get redshifted into the LW range and become dimmer.

The values experienced by the three DCBH host candidates are quite heterogeneous.
Candidates A and C feature actual values (dotted lines) that are always above $J_{21}$, with an initial burst of $J_{\rm LW} \sim 10 J_{21}$ and $z=9$ values of $2$-$3 J_{21}$.
Candidate B instead is located further away and is exposed to radiation of spectral intensity around only $\sim 0.1$-$1 J_{21}$.
In all the three cases the cumulative amount of LW photons emitted (dashed lines) is always higher than $2 J_{21}$ at $z=9$ and sums up to $20$-$30 J_{21}$ for candidates A and C.
When including line shift (solid lines) these latter ones result exposed to values of $J_{\rm LW} \simeq 50 J_{21}$ and $J_{\rm LW} \simeq 40 J_{21}$, respectively.
In this case, the radiation entering the LW band from previous epochs is accounted for.
In general, line shift into the LW band from earlier sources appears to have a certain relevance (a factor of $\sim 2$).
Despite the large uncertainties on the exact critical level of LW radiation, we definitely find the three candidates in conditions where resolved molecular cooling is severely inhibited (i.e. when molecular gas is exposed to $J_{\rm LW} > J_{21}$) and that favour DCBH formation.
We highlight that the presence of metal spreading inhibits DCBH formation where $J_{\rm LW}$ is larger, i.e. closer to stellar sources, around or below kpc distance (see eq.~\ref{eq:Jnumbers}).
Differently from more idealised setups, these finding strengthens the role of metal pollution for assessing DCBH formation within numerical three-dimensional studies \cite[see, however,][]{Valiante2016}.
When checking similar radiative properties for the other two runs, we find $J_{\rm LW}$ values that are always smaller than $J_{21}$ due to the 100-times lower adopted $\dot{N}_{\rm ph}$.
This explains why the two runs with solar-like and OB-type sources fail in producing viable DCBH host candidates and is consistent with the thermodynamical trends presented previously.

Finally, as a warning, we point out that literature works are very uncertain about $J_{\rm LW}$ critical values and different authors give different LW thresholds varying in the large range between 1 and $10^5 J_{21}$. Thus, there is no general consensus about the exact critical flux.
For example, \cite{Shang2010} run three-dimensional simulations of pristine gas and manage to identify different candidates in presence of LW fluxes of $1$-$1000 J_{21}$, for standard stars, and of $1$-$10^5 J_{21}$, for primordial stars.
For the production of the observed population of $z \sim 6$ black holes, large values of the order of $\sim 10^4$-$10^5J_{21}$ are considerably too high and it is by no means clear whether such fluxes are in fact required for the formation of massive objects.
In addition, the stabilising impact of viscous heating alleviates the need for a strong UV background to keep the gas atomic and objects more massive than $\rm 10^4\,M_\odot$ can form even with moderate values of $J_{\rm LW} \sim 100 J_{21}$ 
\cite[][]{LatifSchleicher2015}.
Analyses of simulations by e.g. \cite{Latif2014b} conclude that the typical mean flux may vary from halo to halo, although values larger than $\sim 1500 J_{21}$ are rare, while massive objects still form for radiative fluxes of $10$-$500 J_{21}$ \cite[][]{Latif2014}.
\cite{Habouzit2016} claim that, depending on the feedback scheme and the metal spreading implementation, the critical flux must be at least one or two orders of magnitude lower than predicted by pure-chemistry models.
This work practically sets an upper limit of $J_{\rm LW} \ll 1000 J_{21}$ \cite[in clear contrast with e.g.][]{Sugimura2014} and favours smaller values around $30$-$300J_{21}$.
\cite{Yoshida2003} even suggest values of the order of $J_{21}$ to prevent primordial gas runaway cooling
\cite[for a deeper discussion we refer the interested reader to the review by][]{Valiante2017}.
Furthermore, all the studies finding very large $J_{\rm LW}$ thresholds in the vicinity of star forming regions typically neglect stellar evolution and metal pollution from nearby sources. This can seriously affect the results about critical $J_{\rm LW}$ values. Given the large uncertainties, we have no reason to favour the results of some authors with respect to other authors and, in practice, the only general agreement among the various literature studies is that required critical $J_{\rm LW}$ values should be between $J_{21}$ and $10^3 J_{21}$.


\subsection{DCBH host candidate structure}
\begin{figure*}
{\Large Candidate A}
\vspace{-0.3cm}\\
\includegraphics[width=0.4\textwidth]{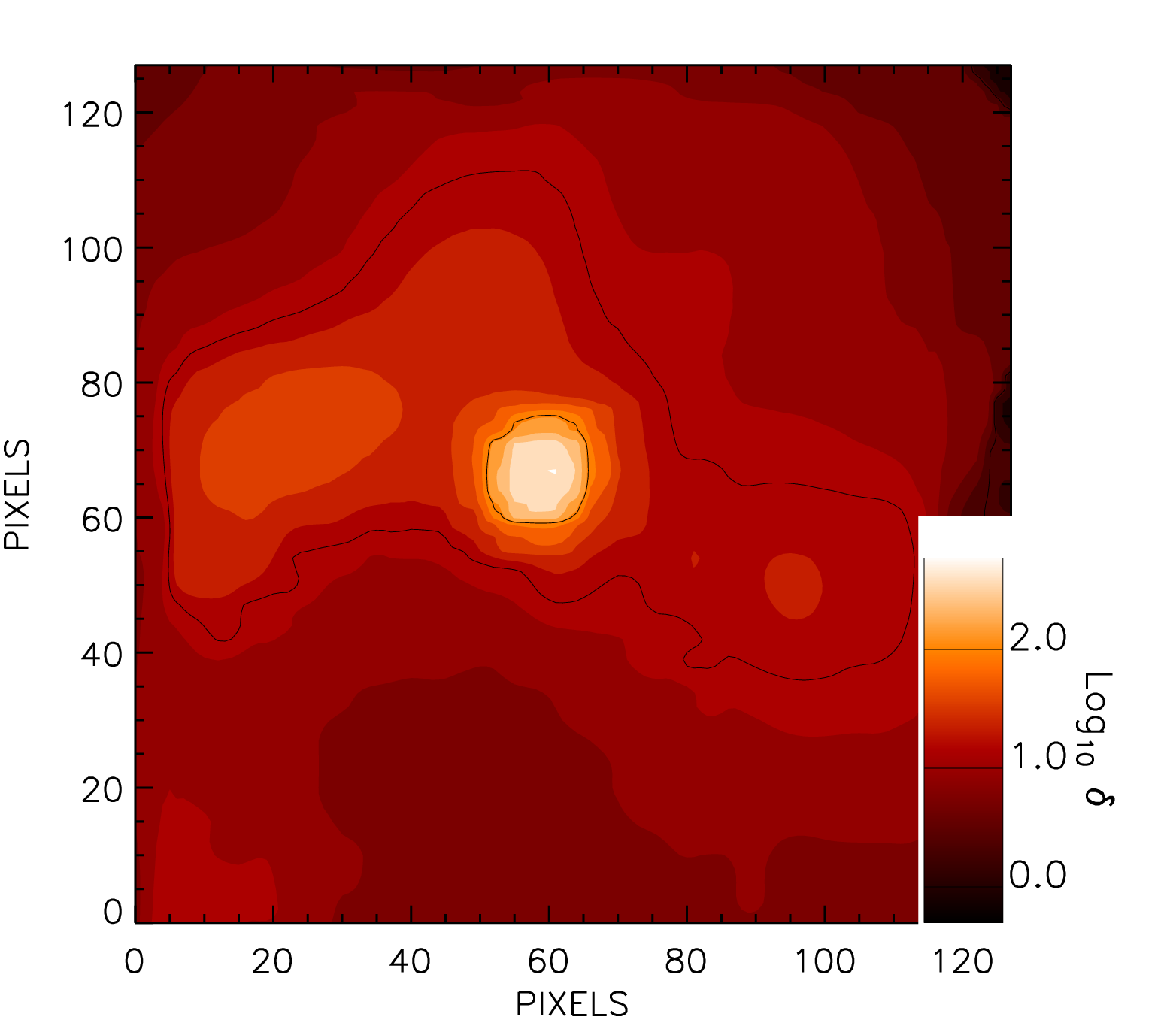}  	
\includegraphics[width=0.5\textwidth]{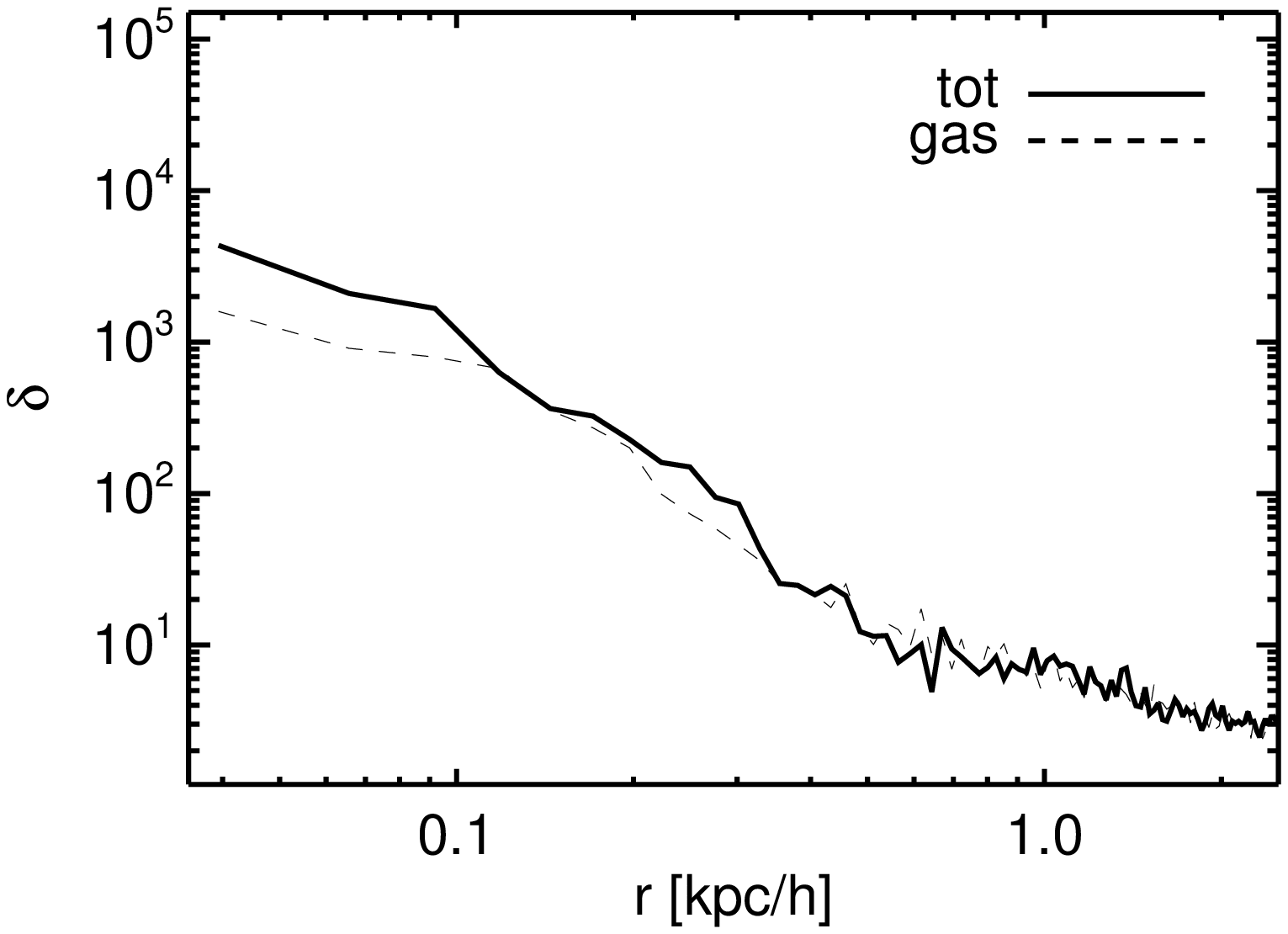}\\  	
{\Large Candidate B}
\vspace{-0.3cm}\\
\includegraphics[width=0.4\textwidth]{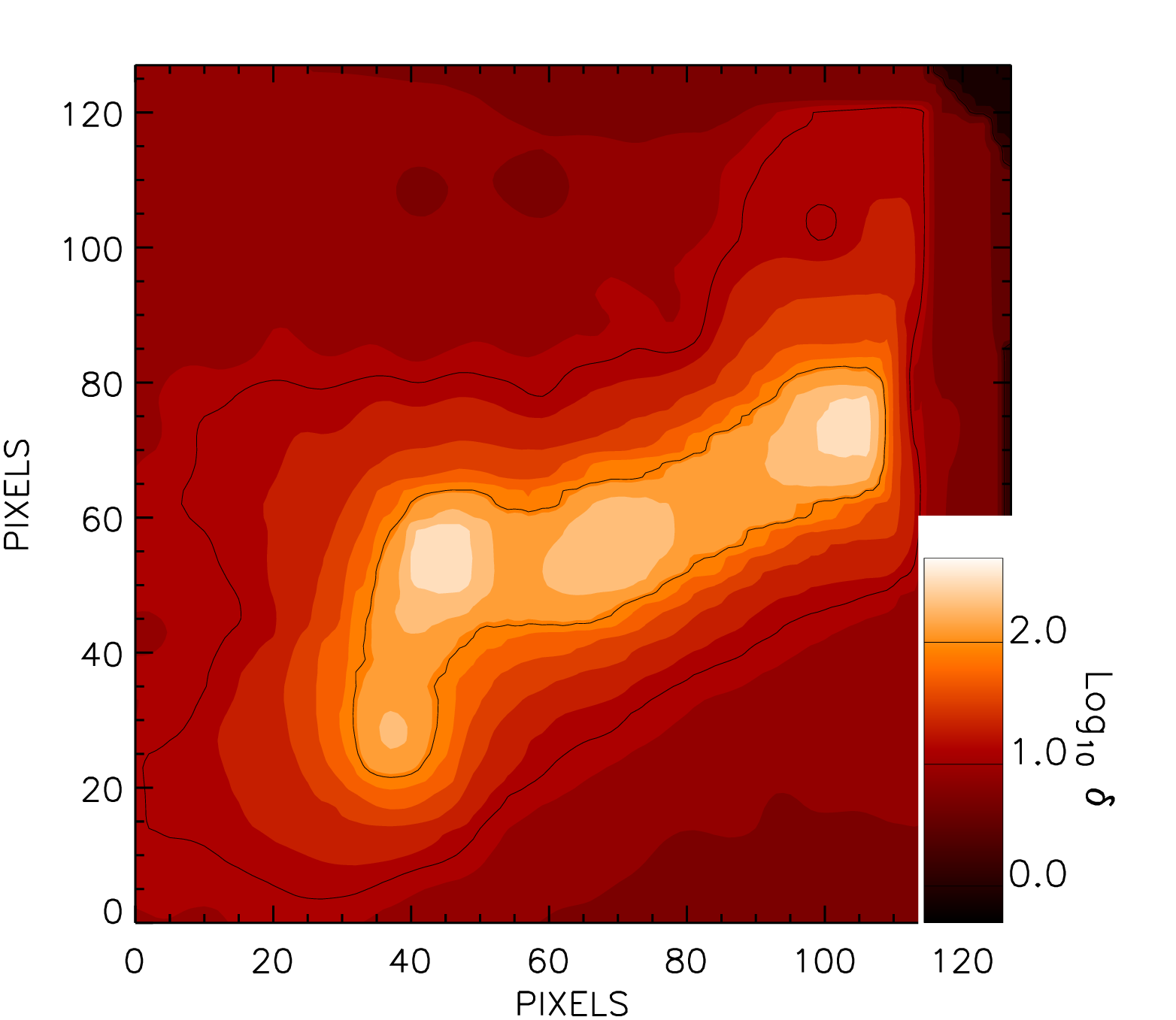}  	
\includegraphics[width=0.5\textwidth]{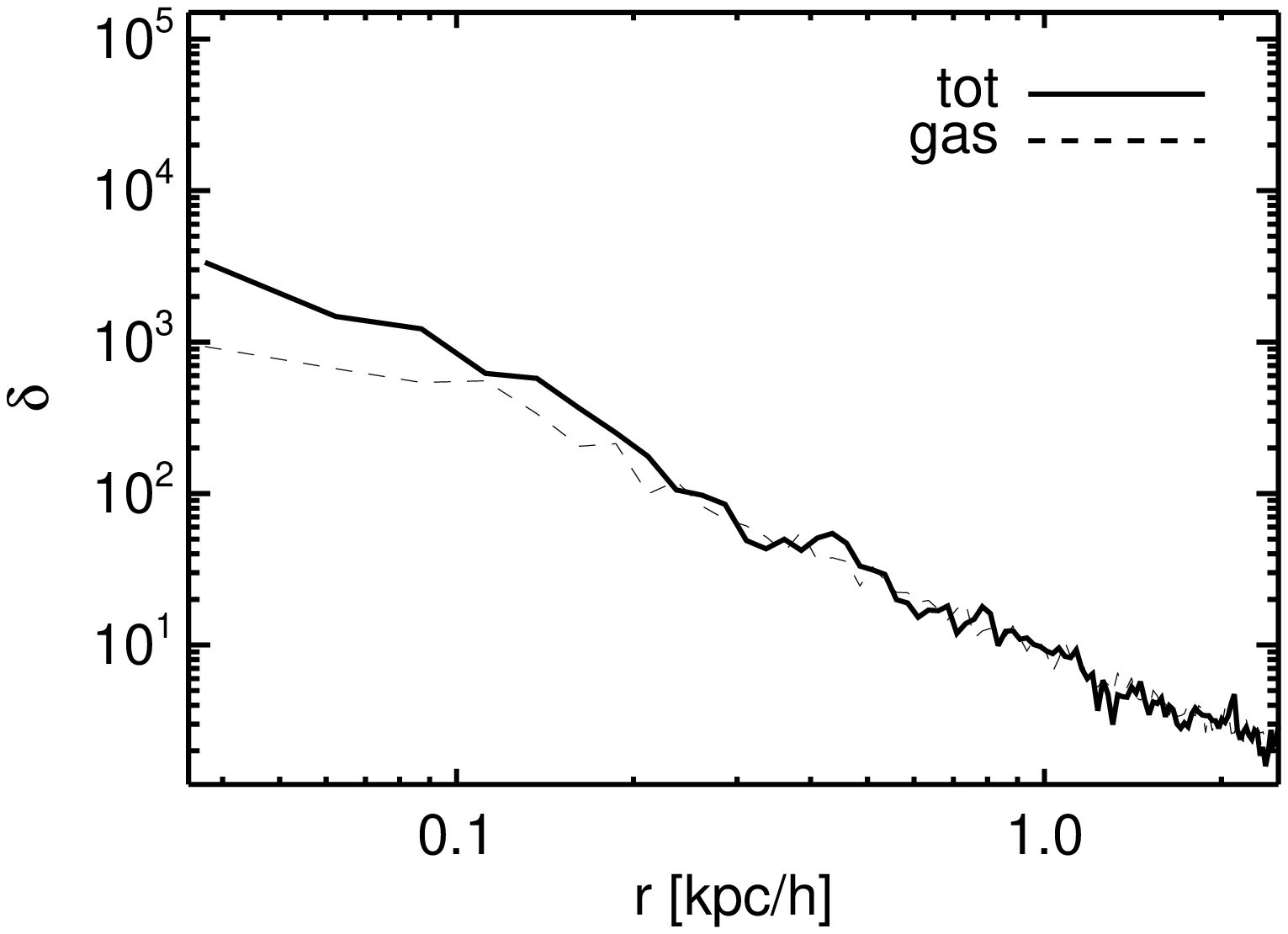}\\  	
{\Large Candidate C}
\vspace{-0.3cm}\\
\includegraphics[width=0.4\textwidth]{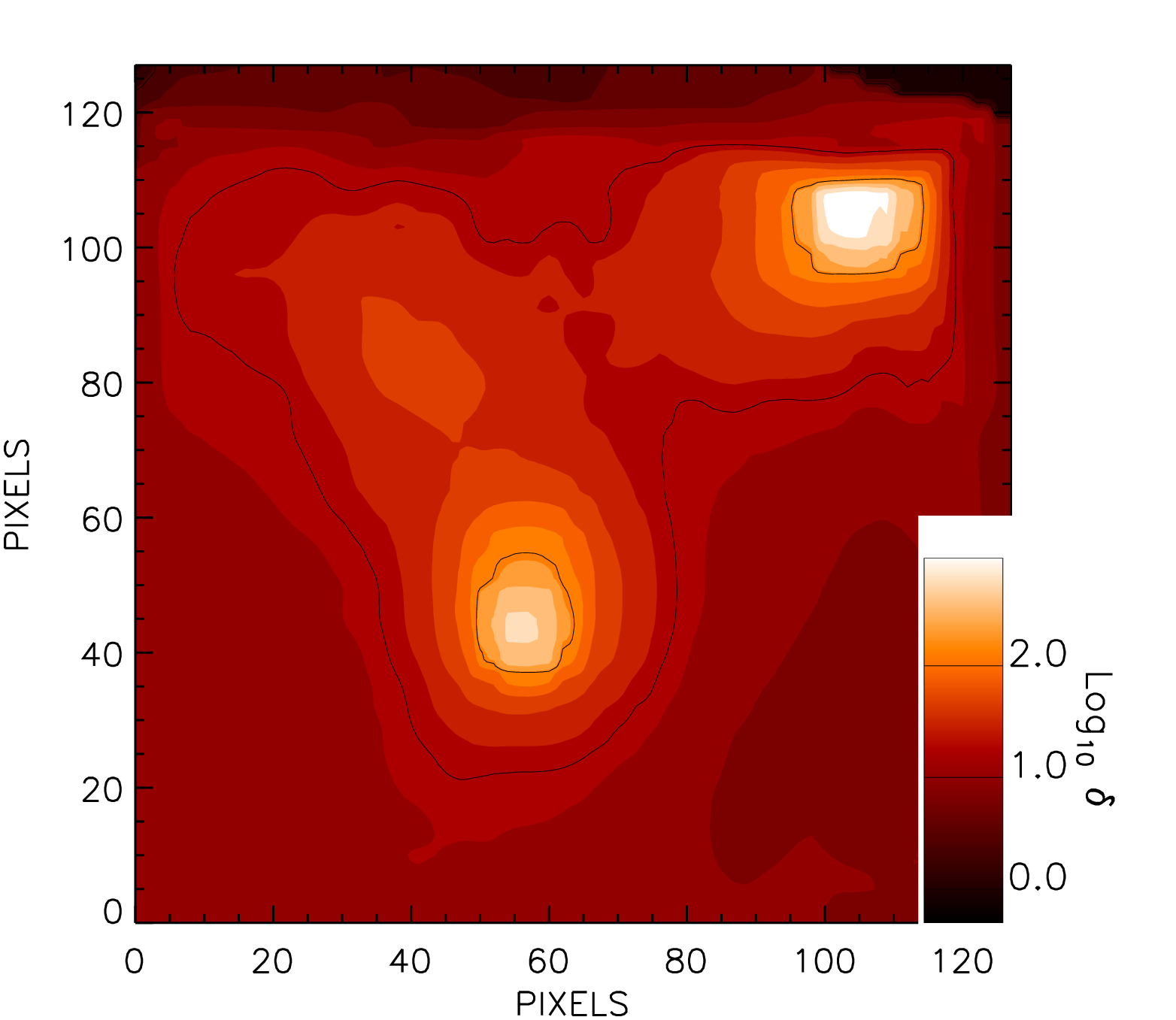}   	
\includegraphics[width=0.5\textwidth]{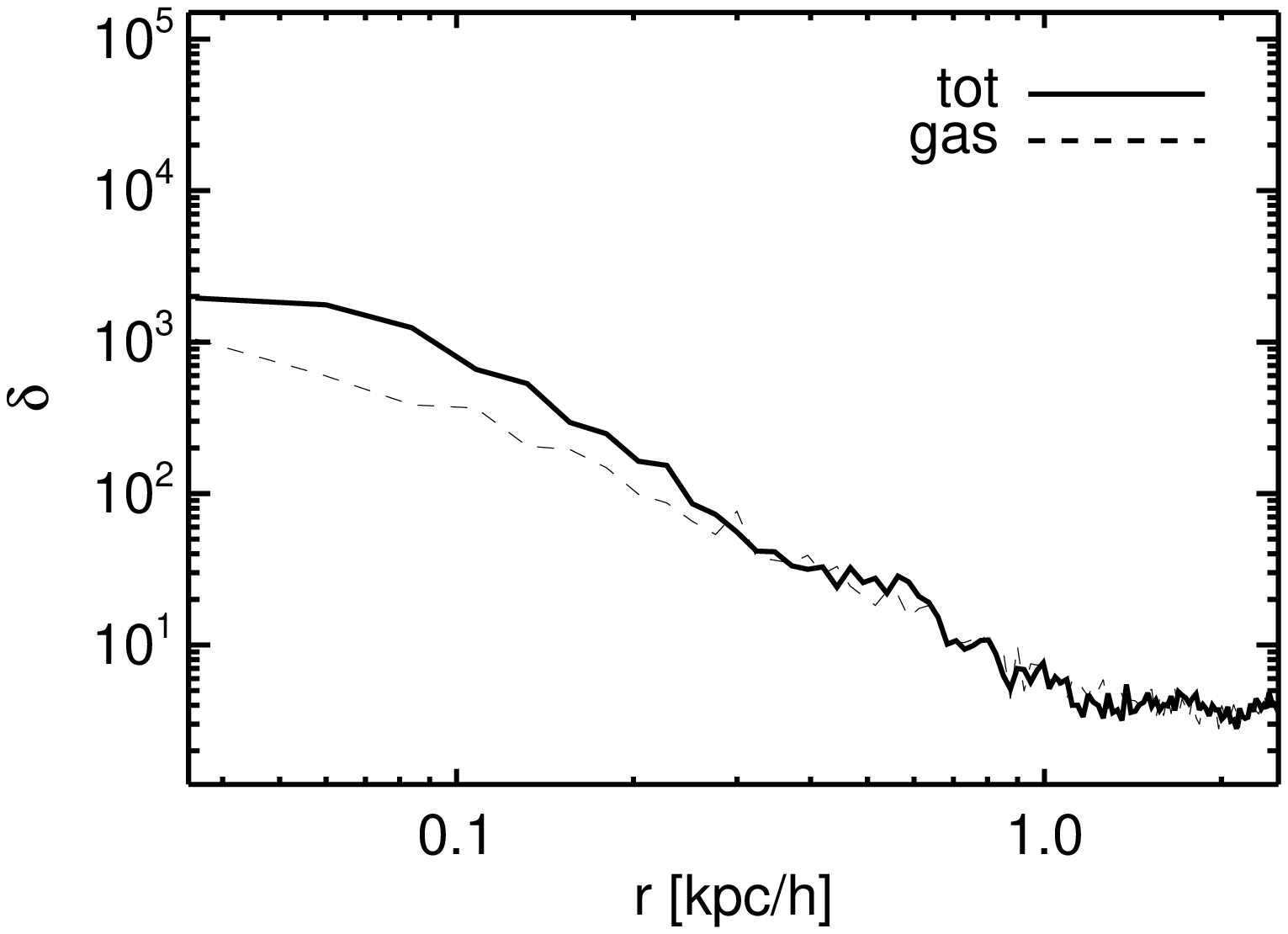}\\  	
\caption[]{\small
{\it Left:}
Maps of the gas over-density with respect to the mean, $\delta$, of three DCBH host candidates at $z=9$ for the run with a $T_{\rm eff}=10^5$~K black body as popIII SED. The maps are obtained via projection of each candidate on its $xy$ plane containing the vertical coordinate of the centre of mass and smoothed over a grid of 128 pixels a side. Three equally spaced isocontour levels are overplotted in solid black lines. The colour scale is the $\rm Log_{10}~\delta$.
{\it Right:}
Over-density profiles, $\delta$, of gas (dashed lines) and total-matter (solid lines) as function of the physical radial distance, $r$, for candidate A (top), B (centre) and C (bottom).
}
\label{fig:maps}
\end{figure*}
To discuss the possibility that a halo candidate selected with the basic necessary requirements mentioned above turns in fact into a DCBH host, it is important to understand whether the local halo structural properties could have an impact on the actual gas direct collapse.
Indeed, mergers and/or substructure formation alter the gas thermal state and lead to fragmentation, inhibiting a ``pure" direct collapse.
For this reason we check the presence of substructures in the three halo candidates.
\\
The biggest halo at $z=9$ (candidate A) has no sub-haloes, while the other two candidates (candidate B and C) are composed by one major halo and one smaller sub-halo.\footnote{In these cases, the positions refer to those of the main halos.}
The shape of the three candidates can be retrieved from the left panels of Fig.~\ref{fig:maps}, where gas over-density maps, $\delta$, of the DCBH host candidates at $z=9$ are displayed.
For each candidate, we show the projection on the $xy$ plane containing the vertical coordinate of the centre of mass.
From the $\delta$ distributions, three equally spaced contour levels are derived and overplotted in black solid lines.
Candidate A is a quiescent halo featuring a fairly spherical structure, mostly on the $xy$ plane.
On the contrary, candidates B and C show more irregular shapes because of the presence of a main halo and a smaller sub-halo.
The presence of a smaller halo (on the right) constituting candidate B is well visible from the central stream bridging it with the main halo (on the left).
Candidate C, instead, is clearly constituted by two distinct bound objects.
The projection highlights that active interactions between the two components (merging event) are taking place and are responsible for the asymmetric compression of the material surrounding them.
\\
Despite their local molecular content being rather low and the typical gas temperature being around $10^4\, \rm K$, the multiple structure of candidates B and C suggests that the occurrence of a pure direct collapse is unlikely for them and that they are not favoured DCBH hosts.

On the contrary, candidate A is composed by one single quiescent gaseous structure with mean temperature of about $10^4\,\rm K$ and is not subject to evident fragmentation ($x_{mol} \ll 1$) nor merger activity.
\\
The presence of satellites in 2 out of 3 cases implies that the formation of a pure DCBH is a rare event, even in those haloes where the basic criteria mentioned in previous Sect.~\ref{Sect:selection} are met.

The right panels of Fig.~\ref{fig:maps} show corresponding over-density profiles of the candidates A (top), B (centre) and C (bottom) as function of the radius normalised to the virial one.
Gas profiles and total-matter profiles are displayed in the three panels by dashed and solid lines, respectively.
The three profiles are computed over radial shells around the centre of mass of each halo, therefore, the differences in the shapes of the objects are smeared out.
Over-densities reach values $\gtrsim 10^3$ (consistently with the maps on the left panels) and in all cases the declining trend for both gas and total content is well visible.
In particular, gas behaviour is very regular due to the lack of ongoing star formation and feedback effects.
These results are consistent with quiescent haloes, since active star forming structures are expected to have larger central over-densities, $\delta$.
For example, molecular cooling ignites catastrophic runaway and collapse at typical number densities of $\sim 1 \,\rm cm^{-3} $, i.e. at values roughly $10^5$ times the critical density (with a modest $z$ dependence).\footnote{
This is why many works have modelled star forming haloes via over-density criteria of the order of $10^5$ \cite[as e.g.][]{Wise2012}.
}

The profiles of the total-matter content is led by dark-matter, while the gas profiles show some divergence towards the centre of about a factor of $2$ for candidates A and C, and of a factor of $3$ for candidate B.
Smaller deviations from the total trend can be observed even at larger distances, where $\delta \sim 10$-$10^2$.

Since the candidates B and C contain substructures, it is unlikely that they will collapse directly and form a DCBH, because, in order to do so, their substructures should merge during collapse. However, mergers are commonly accompanied by gas compression and shocks, that cause H$_2$ (re)formation capable of stimulating violent thin-shell instabilities even in the presence of intense LW flux and for a significant range of shock velocities, as well as a boost of HD abundances  determined by the increased H$_2$ fractions \cite[][]{Shapiro1987, Ahn2007, Whalen2008, PM2012}.
More quantitatively, the cooling timescale, $t_{\rm cool}$, for H-dominated monoatomic gas around $10^4\,\rm K$ can be written as
\begin{equation}
t_{\rm cool} = 
\frac{3}{2} \frac{k_{\rm B} T}{ \Lambda n_{\rm H}}
\simeq
10^3 \frac{ \left(T/10^4{\rm K} \right) }{ \left( \Lambda/10^{-22} {\rm erg\,s^{-1}\,cm^3} \right) n_{\rm H} }  ~ {\rm yr},
\end{equation}
where $k_{\rm B}$ is the Boltzmann constant, $T$ the gas temperature, $\Lambda$ the cooling function and $n_{\rm H}$ the gas H number density \cite[see e.g. Fig.~4 in][]{Maio2007}.
During merger events, gas compression, molecule reformation and tidal effects are likely to make the gas fragment into several clumps, instead of leading the formation of a massive black hole seed
\cite[mediated by the collapse of a single supermassive star;][]{Hosokawa2012, Hosokawa2013, Hosokawa2016}.


\subsection{DCBH host thermal and chemical characteristics}

\begin{figure*}
{\Large Candidate A}
\vspace{-0.3cm}\\
\includegraphics[width=0.45\textwidth]{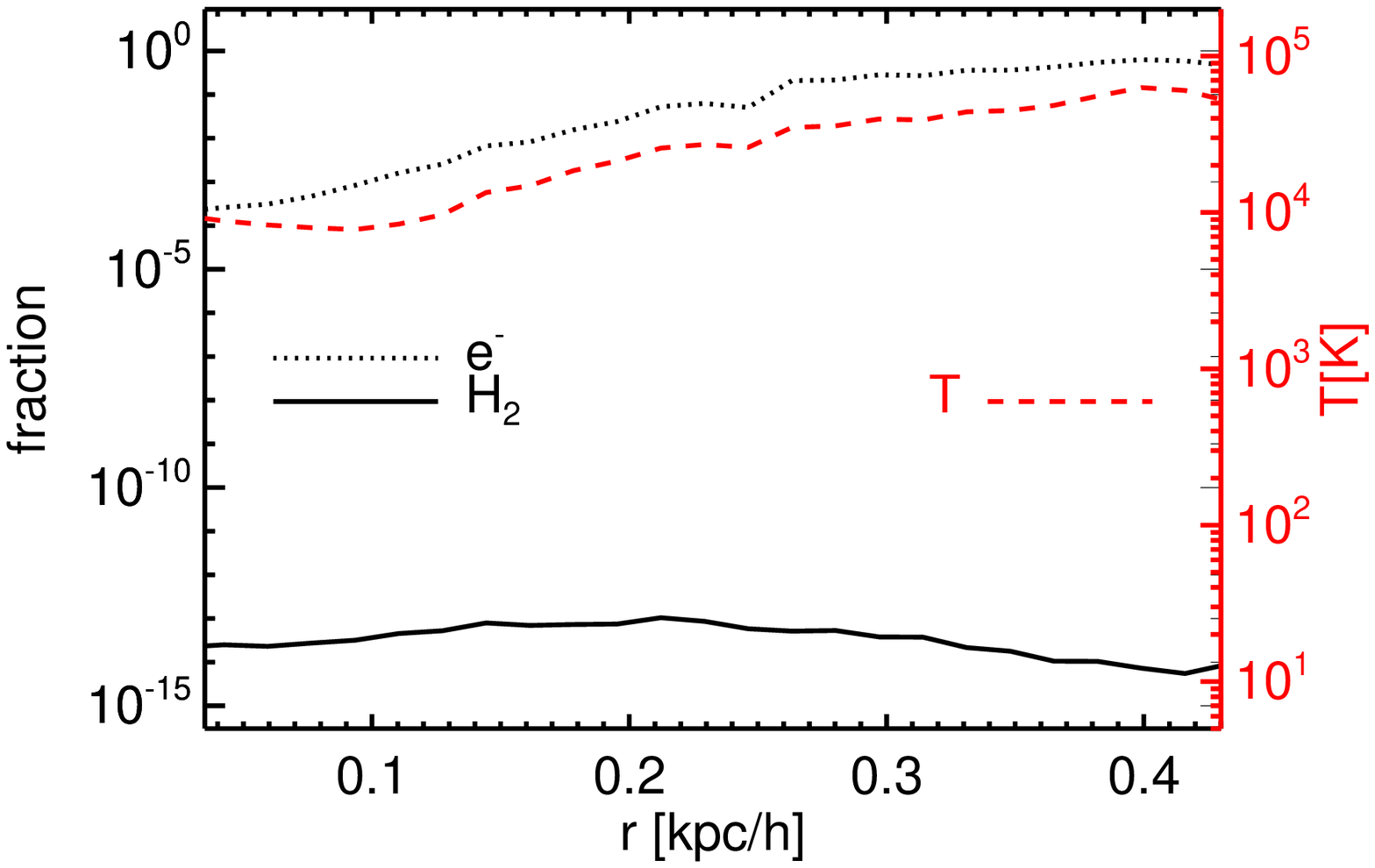}
\includegraphics[width=0.45\textwidth]{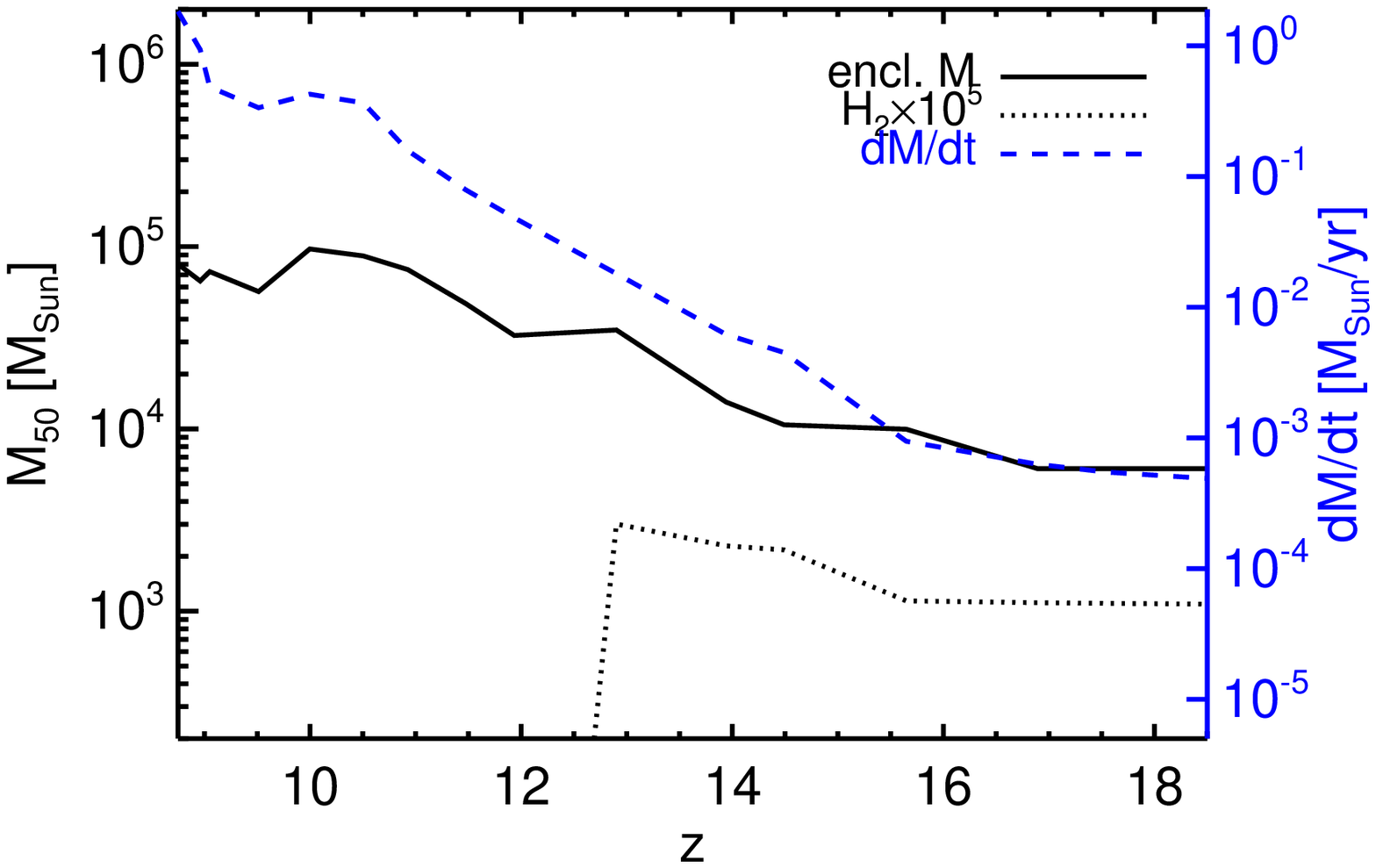}\\
{\Large Candidate B}
\vspace{-0.3cm}\\
\includegraphics[width=0.45\textwidth]{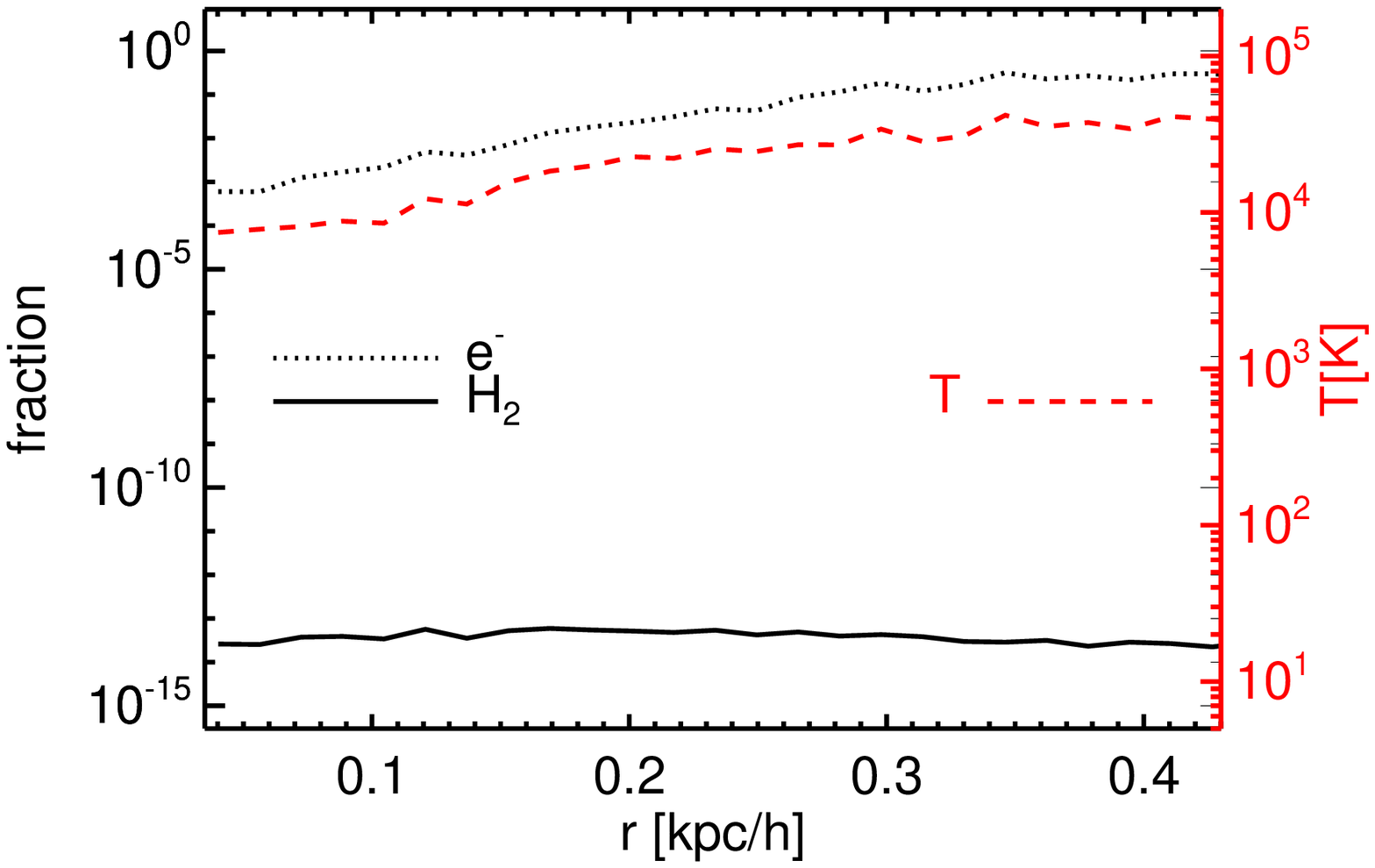}
\includegraphics[width=0.45\textwidth]{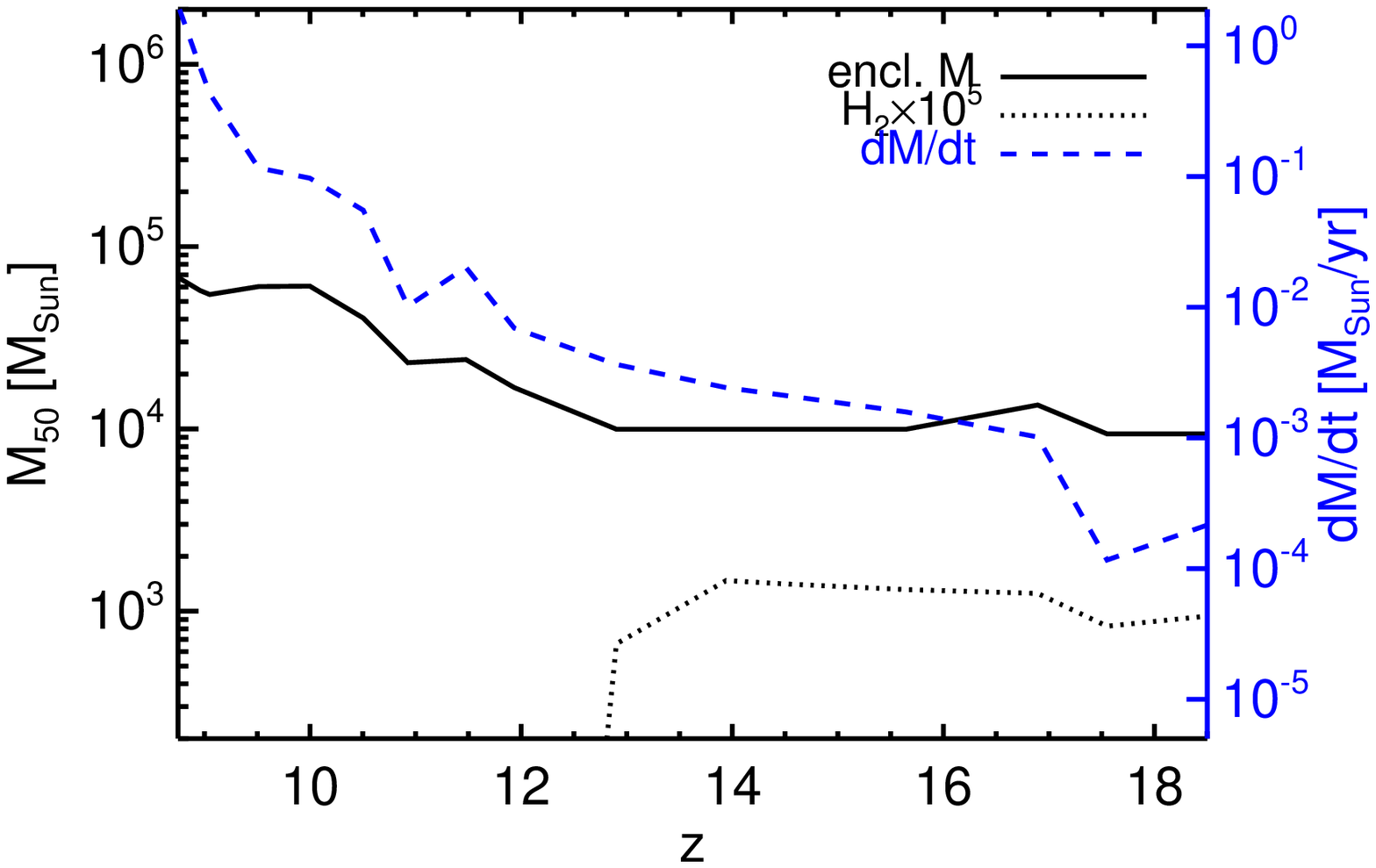}\\
{\Large Candidate C}
\vspace{-0.3cm}\\
\includegraphics[width=0.45\textwidth]{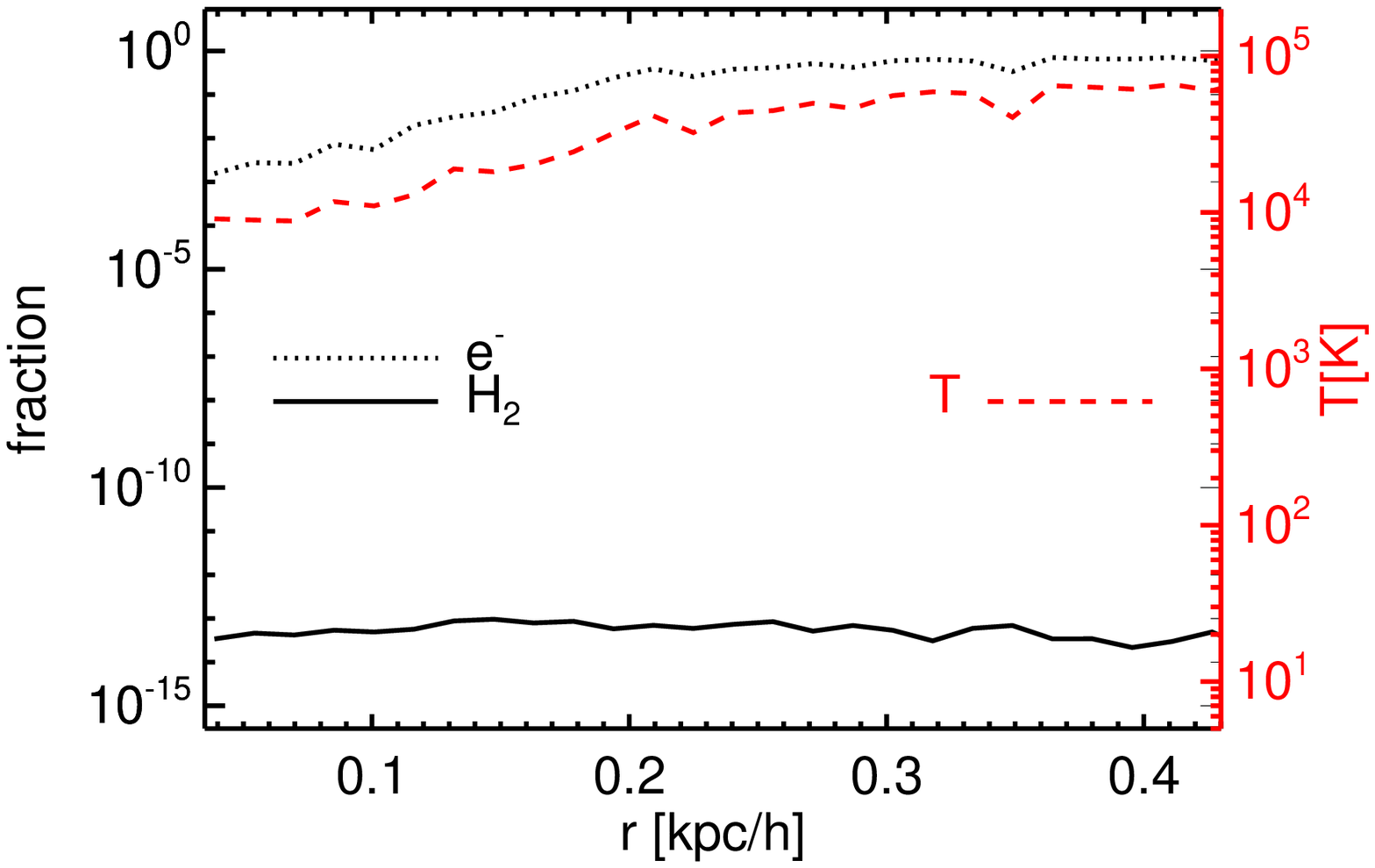}
\includegraphics[width=0.45\textwidth]{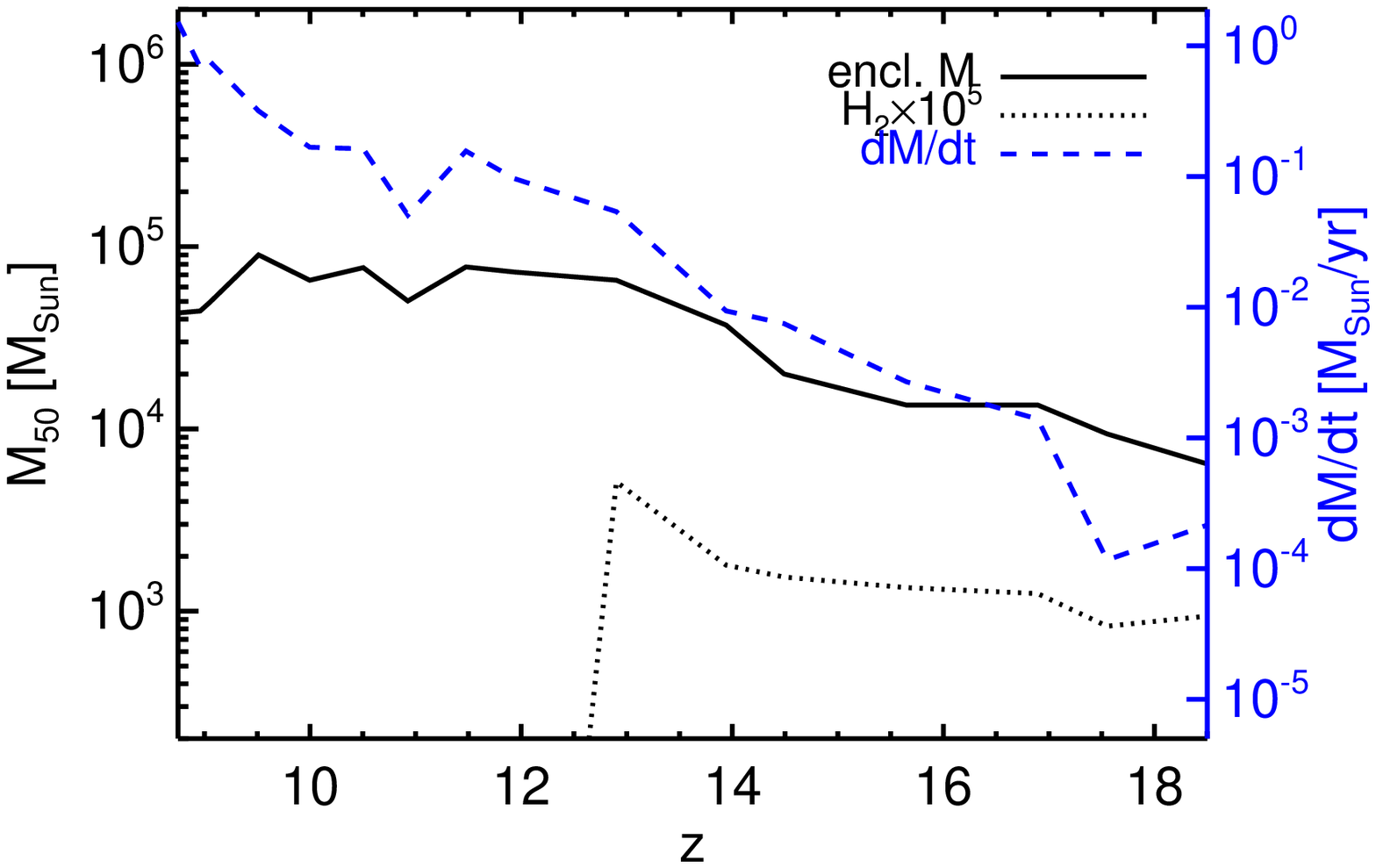}\\
\caption[]{\small
{\it Left:}
Radial profile of electron fraction (dotted lines), H$_2$ molecular fraction (solid lines) and mass-weighted temperature (dashed lines) for candidate A (top), B (middle) and C (bottom). The right scale (in red) refers to temperature values in Kelvin.
{\it Right:}
Evolution as a function of redshift, $z$, of the enclosed mass within the innermost 50~pc/{\it h} (solid lines), of the H$_2$ molecular mass times $10^5$ (dotted lines) and of the expected inflow rate (dashed lines). The right scale (in blue) refers to the values of the inflow rate in solar masses per year.
}
\label{fig:ct-character}
\end{figure*} 
In Fig.~\ref{fig:ct-character} we show the main chemical and thermal characteristics of the three candidates.\\
On the left column we show radial profiles of electron fraction, H$_2$ molecular fraction and mass-weighted temperature for the candidate A (top raw), B (middle raw) and C (bottom raw).
On the right column, the redshift evolution of the mass within the innermost 50~physical pc/{\it h}, $M_{50}$, is displayed for the enclosed material, as well as for the corresponding H$_2$ molecular mass and inflow rates.\\
The radial profiles are computed under spherical approximation and by assuming that the center of the halo corresponds to the position of its most bound particle.\footnote{We have verified that the center of mass of each halo is typically very close to the most bound particle.}
The resulting shapes are very similar for all three candidates.
Due to dissociating radiation, residual H$_2$ fractions are always tiny, with values about $10^{-14}$ (left column in Fig.~\ref{fig:ct-character}), while electron abundances vary between $\sim 10^{-4}$ (in the neutral central regions with $T \lesssim 10^4\,\rm K$) and fractions of unity (around or beyond the virial radii, $\gtrsim 0.3 \,\rm kpc/{\it h}$, where $T \gtrsim 10^4\,\rm K$ and H atoms start getting ionized).
These trends are consistent with the profiles of gas temperature (dashed line, right scale in the panels) that give typical values of about or slightly lower than $10^4\,\rm K$ in the innermost denser regions and slightly larger in the diffuse outskirts.
This is in line with the corresponding gas density profiles of these haloes (dashed lines in previous Fig.~\ref{fig:maps}) and consistent with previous studies by e.g. \cite{Kitayama2004}, \cite{Whalen2004} and \cite{Abel2007}, who showed that ionising shocks from massive popIII sources can leave behind a warm $\sim 3 \times 10^4\,\rm K$ diffuse medium.
Ionisation and recombination processes around $10^4\,\rm K$ are very fast and give origin to the patterns observed in the three profiles of the electron fraction.
Given the negligible amount of available H$_2$ molecules, the gas of these pristine haloes cannot ignite metal-free cooling nor fragmentation below $\sim 8\times 10^3\,\rm K$.
\\
On the right column of Fig.~\ref{fig:ct-character} the redshift evolution of the three candidates, A, B and C, is computed by tracing in time their progenitors and by evaluating the mass in a sphere of $50\,\rm pc/{\it h}$ radius around the most bound particle, $M_{50}$, at each snapshot, for both the enclosed halo mass and the H$_2$ content.
We also compute the corresponding mass inflow rate, $\dot M_{in} \sim c_s^3 / G $, with $c_s$ sound speed and $G$ gravitation constant \cite[][]{Oshea2007, Oshea2008}.
Mass evolution appears to be quite smooth.
The progenitors of the considered haloes feature $M_{50}$ values increasing by one order of magnitude, from roughly $10^4\,\rm M_\odot$ at $z\simeq 20$ up to almost $10^5\,\rm M_\odot$ at $z\simeq 9$.
In particular, 
candidate A reaches $M_{50} \simeq 8 \times 10^4 \,\rm M_\odot $, 
candidate B $M_{50} \simeq 7 \times 10^4 \,\rm M_\odot $ and 
candidate C $M_{50} \simeq 4 \times 10^4 \,\rm M_\odot $.
The evolution of candidate B is the most regular one, instead candidates A and C, that are closer to the star forming regions, suffer more photo-heating at lower redshifts, as visible from the small decline of $M_{50}$ at $z\sim 10 $ and its subsequent stabilisation.
Since the total mass of these candidates is around $10^6\,\rm M_\odot$, the amount of material enclosed in the innermost $50\,\rm pc/{\it h}$ constitutes almost 1/10th the halo mass.
The most striking event is definitely the dramatic decrease of H$_2$ molecules, that, during the star formation episodes at $z \gtrsim 10 $, are destroyed down to fractions $ < 10^{-13}$ by the dissociating radiation coming from the newly established star forming regions.
As a comparison, mass profiles do not show such sudden variations.
The expected inflow rates (right scale in the plots) evolve accordingly to the mass growth and are sensitive to the thermal state of the gas.
The final value achieved by candidates A and B is almost  $2\,\rm M_\odot/yr$, while candidates C features $ \dot M_{in} \simeq 1\,\rm M_\odot/yr$.
These values are reached only in the final part of their evolution, though, since before $z\sim 10$ typical expectations are $\dot M_{in}  \ll 10^{-1} \,\rm M_\odot/yr$.
We warn the reader that exact estimates for the inflow rates bear dependences on assumptions and environment.
Standard commonly used rates of $0.975~c_s^3/G$ refer to spherical hydrostatic isothermal clouds accreting at constant rate \cite[][]{Shu1977, Hunter1977}.
Violent inflow rates of $46.84~c_s^3/G$ are expected for \cite{Larson1969}-\cite{Penston1969} self-similar solutions after a discontinuous jump from an initial value of $29~c_s^3/G$ \cite[][]{WS1985}.
Time-dependent accretion rates with initial central values of $47~c_s^3/G$ were suggested by \cite{FosterChevalier1993} who also found a rapid decline at later times to $ \lesssim10~c_s^3/G$.
Modest inflow rates in more realistic axisymmetric MHD contracting cloud undergoing runaway collapse may vary up to maximum rates of $40~c_s^3/G$, with late-phase lower limit $\sim 2.5~c_s^3/G$ and time-averaged value of $\sim 4~c_s^3/G$ \cite[][]{GS1993a, GS1993b,Tomisaka1996, Safier1997}.
These values are not dramatically different from more recent studies of magnetic and non-magnetic clouds \cite[][]{MellonLi2008,MachidaDoi2013,Susa2015}.
With respect to these considerations, the values showed in the right plots of Fig.~\ref{fig:ct-character} might represent a lower limit and actual rates can episodically reach values of a few to tens times higher.
\\ 
These trends clearly suggest that the three candidates A, B and C can remain metal free during their entire lifetimes, because they are sufficiently far from star forming haloes and metal-free star formation is prevented by the fact that destroyed H$_2$ molecules in the halo progenitors are not able to form again by $z\sim 9$.


\subsection{DCBH host candidate rotational patterns}

Independently of the structure of the haloes, rotational patterns play an important role for the occurrence of gas collapse.
In fact, gas rotational motions will halt direct collapse and prevent DCBH formation.
For this reason we investigate the angular momentum per unit mass of the material hosted by the DCBH host candidates, $\mathbf j = \mathbf r \times \mathbf v$, with $\mathbf r$ and $\mathbf v$ particle position and velocity vectors \cite[][]{BinneyTremaine2008}.
For each particle, we compute the circularity $\varepsilon = j_{\rm z}/j_{\rm circ}$ distributions, being $j_{\rm z}$ the component of the angular momentum perpendicular to the plane of rotation and $j_{\rm circ}$ the expected angular momentum of a circular orbit in the gravitational potential determined by the enclosed mass, $M$, at the same radial distance, $r$, i.e.:
$j_{\rm circ} = r~v_{\rm circ}$, with 
$v_{\rm circ} = \left[ G M(r)/r \right]^{1/2} $.
Obviously, $\varepsilon $ is sensitive to the rotational patterns of the constituting halo particles and can give us hints about the dynamics of the material inside the hosting halo.
Values of $\varepsilon \simeq 0$ represent test particles that have negligible angular momentum and that can eventually collapse.
The closer is $\varepsilon $ to unity, the closer is the motion of a test particle to a stable (i.e. {\it non-collapsing}) circular orbit.
Larger $\varepsilon $ values refer to unbound particles, that move along escaping orbits.

\begin{figure}
\centering
\includegraphics[width=0.35\textwidth]{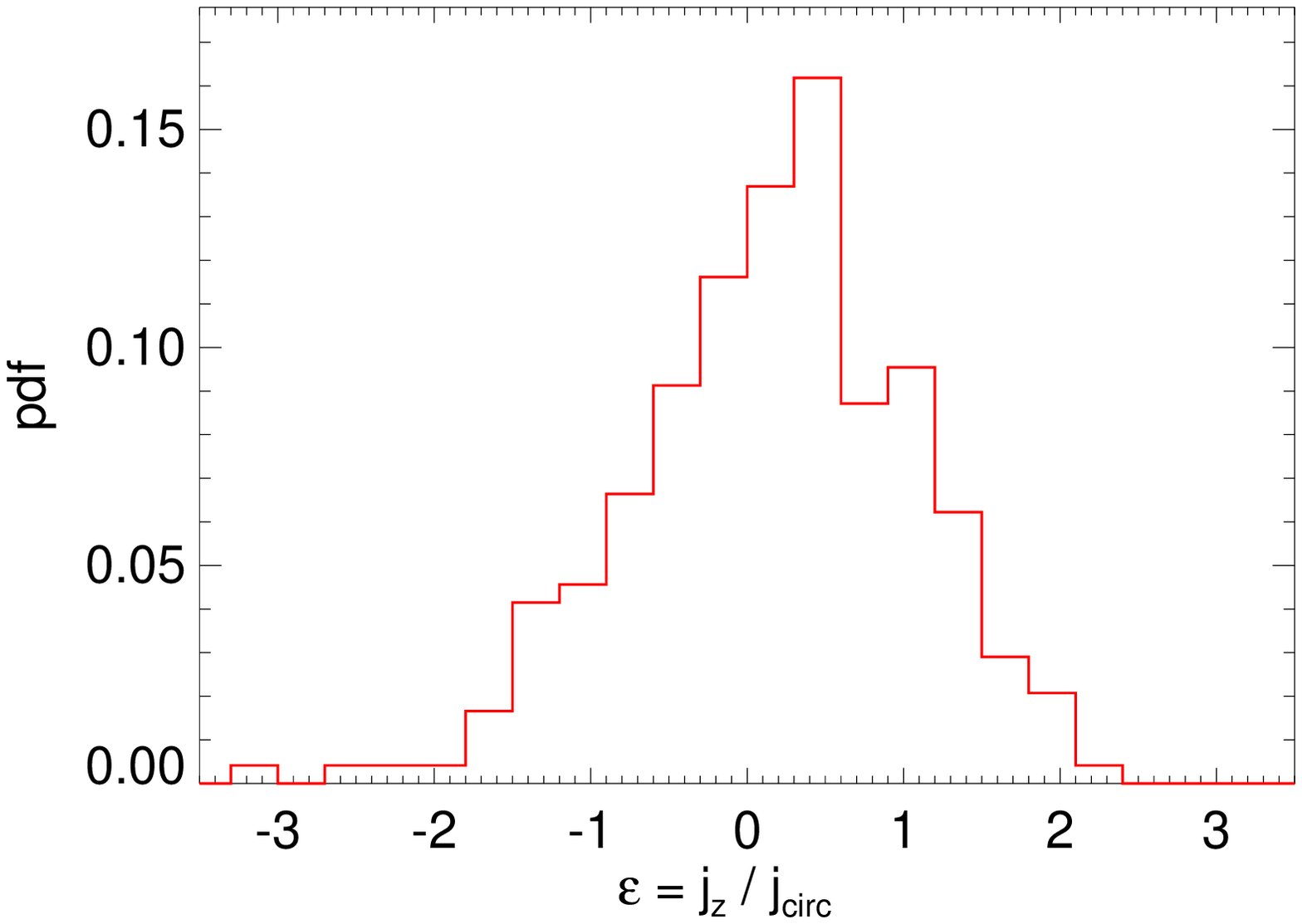}
\vspace{-0.4cm}\\
\includegraphics[width=0.35\textwidth]{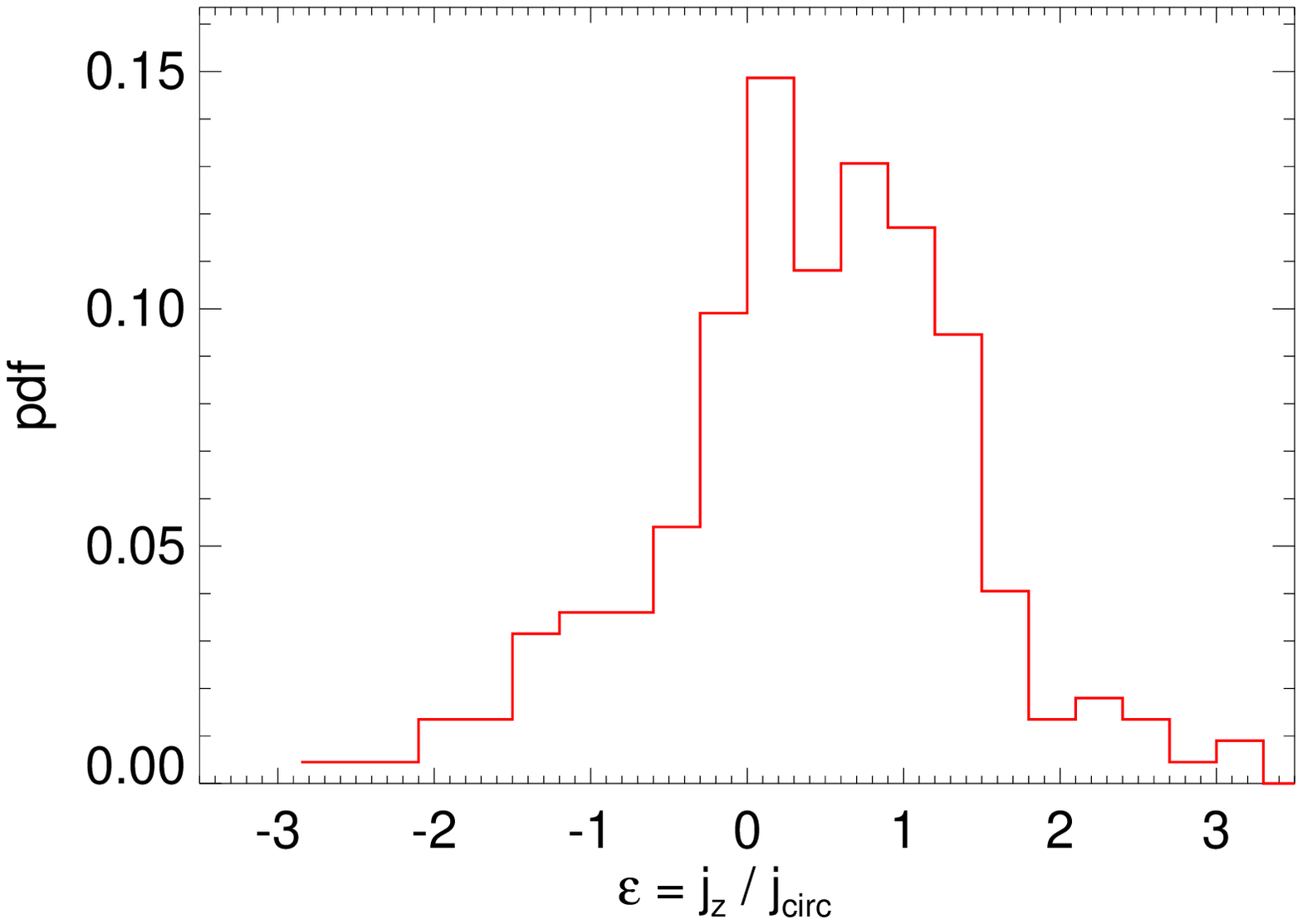}
\vspace{-0.4cm}\\
\includegraphics[width=0.35\textwidth]{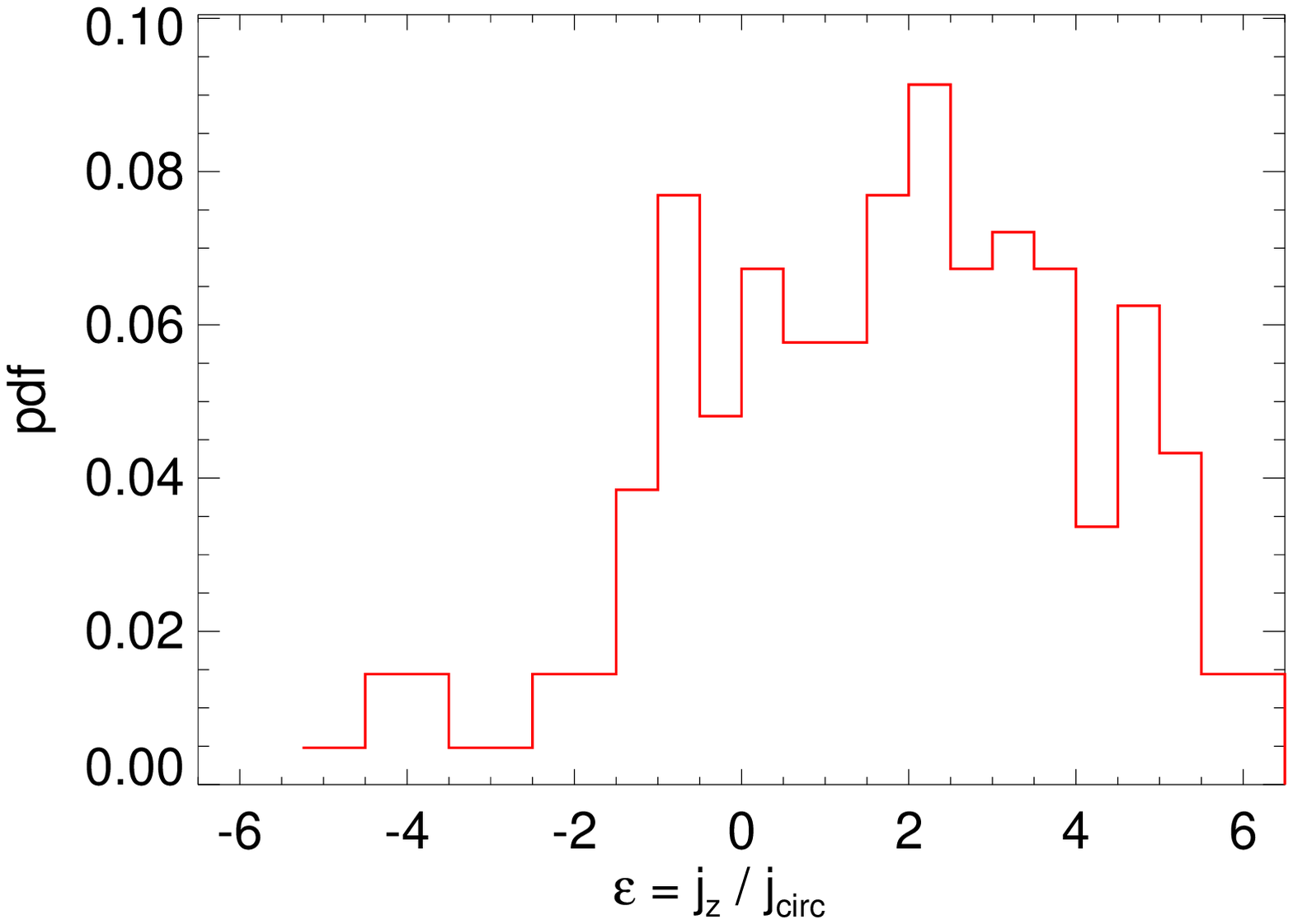}\\
\caption[]{\small
Distributions of the gas circularity, $\varepsilon$, for the three DCBH host candidates A (top), B (centre) and C (bottom) at $z=9$ for the run with a $T_{\rm eff}=10^5$~K black body as popIII SED.
}
\label{fig:eps}
\end{figure}

\begin{figure}
\centering
\includegraphics[width=0.35\textwidth]{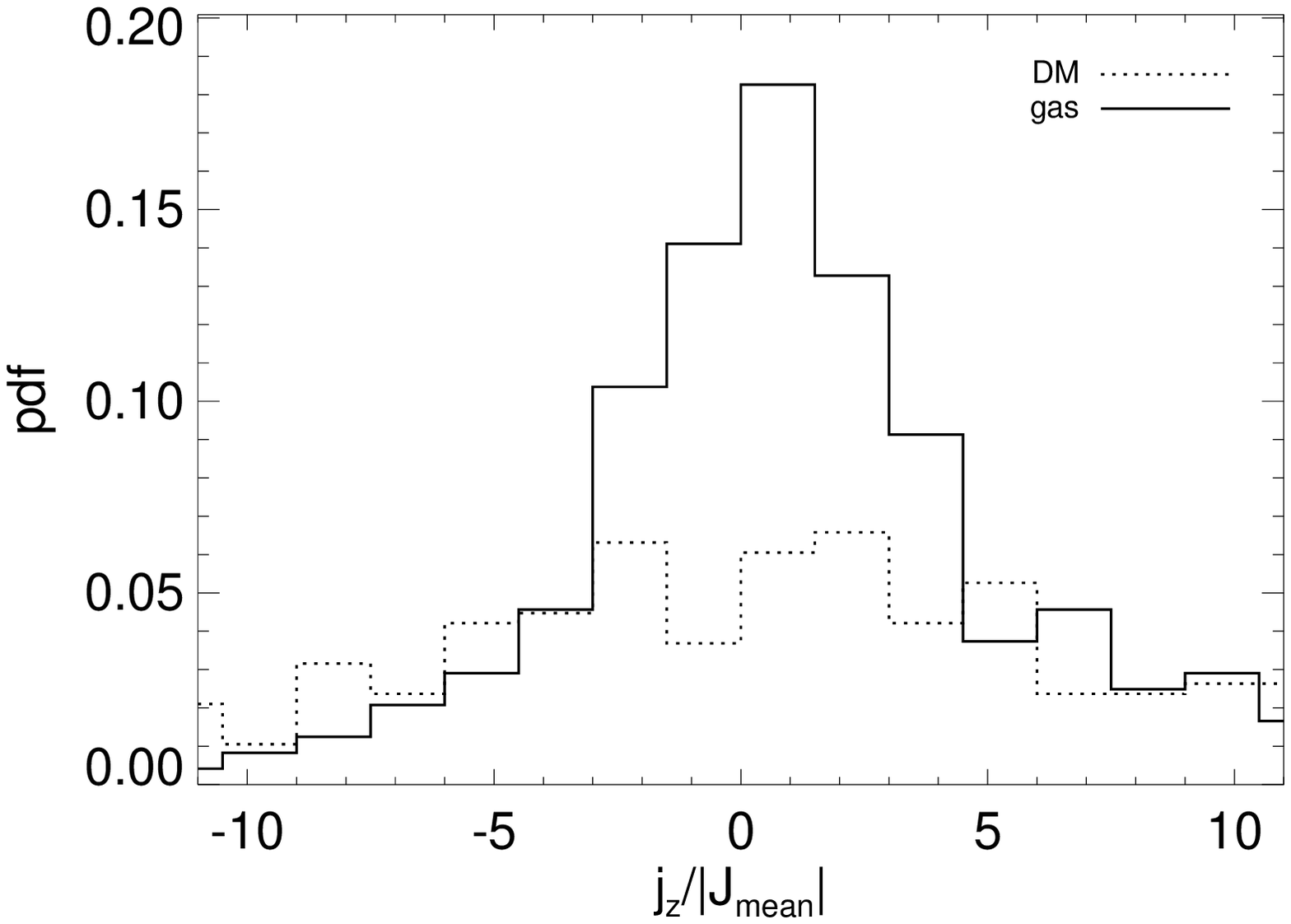}
\vspace{-0.4cm}\\
\includegraphics[width=0.35\textwidth]{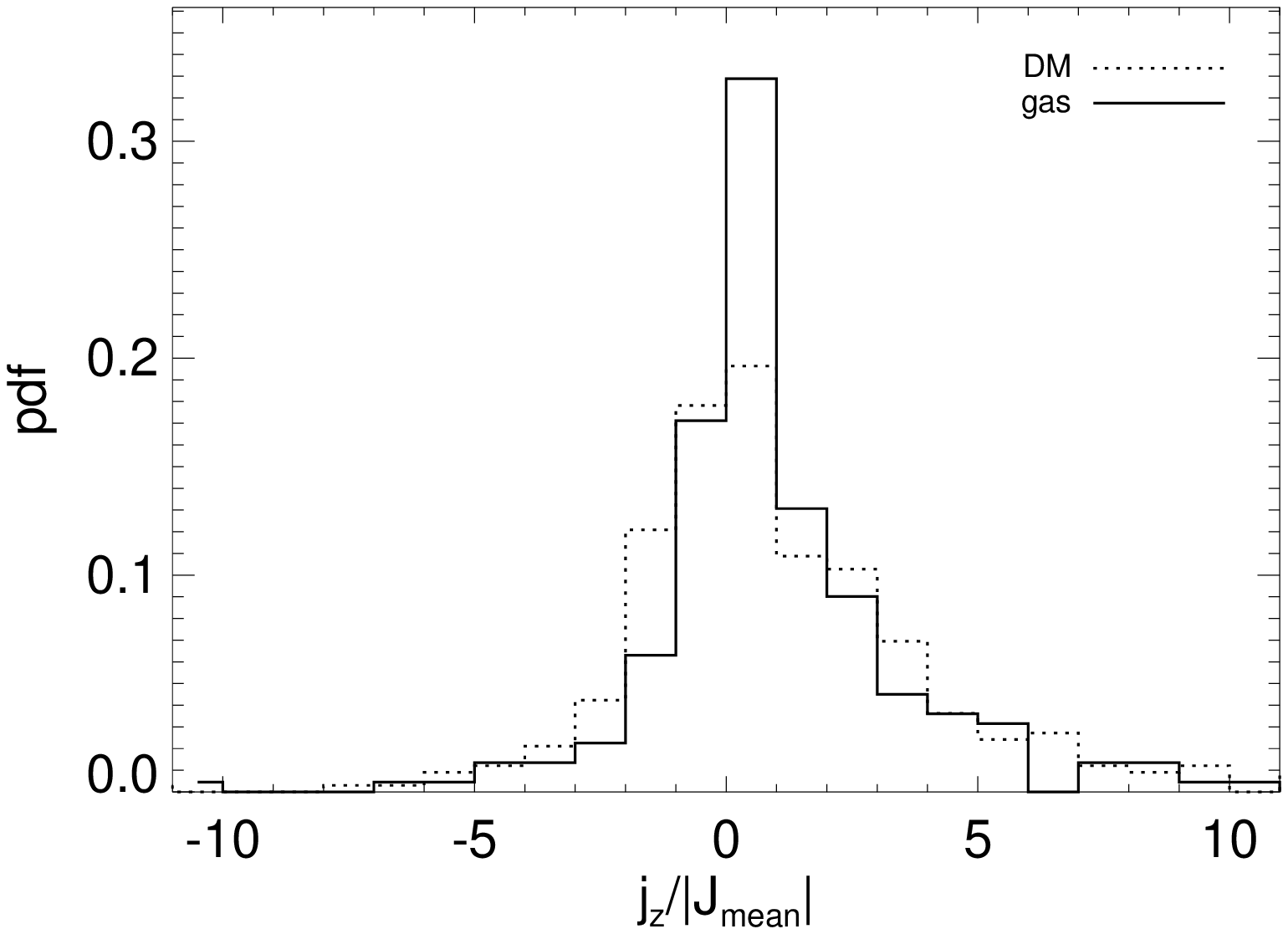}
\vspace{-0.4cm}\\
\includegraphics[width=0.35\textwidth]{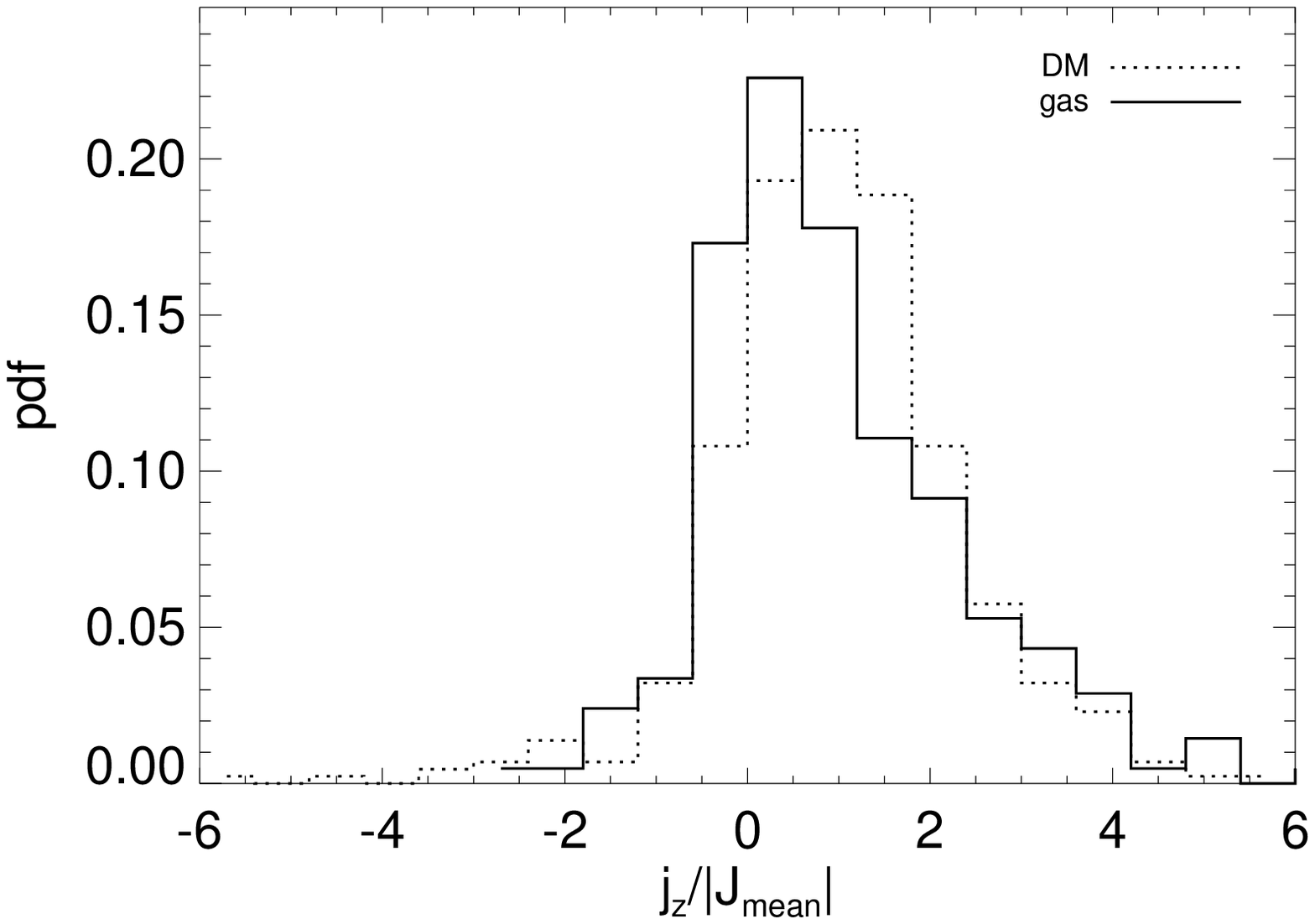}\\
\caption[]{\small
Distributions of the ratios between gas and dark-matter $j_{\rm z}$ values and the mean halo angular momentum for the three DCBH host candidates A (top), B (centre) and C (bottom) at $z=9$ for the run with a $T_{\rm eff}=10^5\, \rm K$ black body as popIII SED.
}
\label{fig:ratio}
\end{figure}

In Fig.~\ref{fig:eps} the distributions of the $\varepsilon$ values for the gas of the three DCBH candidates are plotted.
The trends reflect the structure of the haloes and the spread of the distributions is consistent with the variety of statistical properties of rotational patterns expected at these early times \cite[][]{Biffi2013, deSouza2013, Prieto2015}.
In particular, candidate A is an isolated object with a quite regular distribution, while candidate B shows a double peak that is linked to its internal substructure.
The candidate C shows a very irregular behaviour, as a consequence of its disturbed state due to the ongoing internal interactions.
Values of $\left| \varepsilon \right| >1$ refer to particles which are not dynamically bound to the structure.
Their amounts vary significantly in the three host candidates.
For candidate A, a fraction $\lesssim 20\%$ of gas parcels is going to escape, while the other two candidates feature larger amounts of escaping material, ranging from more than $25\%$ for candidate B up to about $ 75\%$ for candidate C.
The unbound material has little implications for candidates A and B, which are going to preserve most of their gas mass.
However, for candidate C the situation is more critical, as it is going to be left with a gas mass that is almost one order of magnitude smaller.
This complicates the formation of any intermediate-mass black hole from this halo.
A similar conclusion is reached in Fig.~\ref{fig:ratio}, where the ratio between $j_{\rm z}$ and the absolute value of the mean halo angular momentum, $ \left| j_{\rm mean} \right|$, for both gas and dark-matter is considered.
This ratio is linked to disk formation in the inner structure of the halo and provides a measure of the alignment of the angular momentum of every single particle with the mean angular momentum.

Candidate A shows a gas ratio distribution that is peaked around zero and declines symmetrically, as a consequence of its roughly spherical shape.
Candidates B and C, instead, have distributions that decline in a non-symmetric way, with a more prominent high-$j_{\rm z}$ tail.
This arises from the particles of the sub-halo that are located at distances larger than those of the main halo.
\\
The trends of gas and dark-matter are quite similar, although for $j_{\rm z} / \left| j_{\rm mean} \right| \simeq 0$ there is a higher gas peak in the gas distribution.
This originates mostly from the central regions of the halo, where gas settles in a spherical shape, and also explains why the difference is more marked in candidate A, rather than in the more irregular candidates B and C.
\\
The tail of the distributions for gas and dark-matter angular momenta have similar behaviours for values different from zero, meaning that the dynamical state of those gas particles is significantly influenced by the underlying dark-matter distribution.


\subsection{DCBH host turbulent motions} \label{sect:Re}

Turbulence is another important limiting factor for gas direct collapse, since it enhances gas fragmentation and hinders DCBH formation.
Strictly speaking, turbulence arises from the non-linear advection term, $ ( {\bf u} \cdot \nabla )~{\bf u} $, and the dissipative term, $\nu ~\nabla^2 {\bf u}$, in the equations of motion for a fluid whose velocity field is ${\bf u}$ and kinematic viscosity is $\nu$.
The (dimensionless) Reynolds number quantifies the effects of inertial forces with respect to viscosity forces and is defined as:
$Re \equiv  \,\parallel ( {\bf u} \cdot \nabla )~{\bf u}  \parallel /  \parallel \nu ~\nabla^2 {\bf u} \parallel  \,\sim u l / \nu $,
where $l$ is the typical scale of the turbulent motion.
When $ Re $ tends to zero the system is viscosity dominated and turbulence decays.
The scale at which $ Re = 1$ is denominated dissipation scale.
When $ Re \gg 1$ the advection term dominates and the influence of viscous forces is negligible. In most practical situations this usually happens for fluids with $Re$ values above 4000, that are commonly considered turbulent.
As an example, in the cool ISM, it is well established since long time that the Reynolds number assumes values between $10^5$ and $10^7$ \cite[][]{ Cordes1985, Armstrong1995}.
\\
Quantitatively, the kinematic viscosity of the gas, $\nu$, can be estimated as $\nu \sim c / n\sigma$, with $c$ sound speed, $n$ mean number density and $\sigma \sim 10^{-15}\,\rm cm^2$ typical interaction cross section \cite[][]{ElmegreenScalo2004}.
Consequently, $Re$ depends on the Mach number of the gas as well as on its thermal state.
\\
Despite these idealizations, in reality turbulence is not uniformly distributed, but shows clear spatial and temporal intermittency effects: regions particularly active coexist with regions completely inactive. In general, it is possible to take into account these effects, but they will not be considered here, because the expected deviations are small in comparison to the order-of-magnitude estimates presented below.

Fig.~\ref{fig:Re017} shows results for $Re$ as expected at $z=9$.
To have an idea of the global behaviour of the gas in each halo, we estimate $Re$ by employing the (physical) quantities describing each object, as computed at post-processing time, both via friend-of-friend algorithm (velocities, radii and masses at over-densities of 200 and 500 with respect to the background) and substructure finder (namely, velocity dispersion, maximum radial velocities and corresponding radii, as well as substructure masses).
Given the different nature of all these quantities, $Re$ results are not expected to coincide exactly, but should give us a broad overview of the orders of magnitudes reached case by case.
\\
We show them as function of the substructure mass, $M_{\rm sub}$, the mass within an over-density of 500, $M_{\rm 500}$, and the virial mass $M_{\rm vir}$ to get the corresponding $Re_{\rm sub}$, $Re_{\rm max}$, $Re_{\rm 500}$ and $Re_{\rm vir}$ values.
In particular,
$Re_{\rm sub}$ is obtained by substructure velocity dispersion and maximum radial velocities (blue triangles);
$Re_{\rm max}$ is obtained by maximum radial velocities and the radii at which such velocities are reached (green crosses);
$Re_{\rm 500}$ is obtained by velocity dispersions and radii within an over density of 500 (red diamonds);
$Re_{\rm vir}$ is obtained by velocity dispersion and virial radii (cyan asterisks). 
\\
We warn that the smallest haloes, despite their virial values can be obtained, do not always have well defined quantities for $Re_{\rm sub}$ and $Re_{\rm max}$ calculations, due to the limited number of their constituting particles.
\\
Bullet points highlight the values for the DCBH host candidates identified in the previous sections.
Consistently with what mentioned before, there are 3 friend-of-friend objects candidate to host a DCBH and they are over-plotted both on the sequences of red diamonds and of cyan asterisks.
Among these 3 halo candidates, 2 are composed by a main halo and a satellite, for resulting 5 substructures that are over-plotted by bullets on the green crosses and the blue triangles (black bullets refer to main haloes and magenta bullets to satellites).
\\
Independently from the details about $Re$ in the figure, the turbulent nature of the primordial haloes is striking.
The values obtained for bigger objects vary between roughly $ 10^5$ and $10^8$, depending on the assumptions, however, they are in broad agreement with previous studies of early star forming haloes \cite[][]{Maio2011} as well as with the values expected in turbulent ISM environments.
Moreover, in most cases Mach numbers are close to unity, which makes early turbulent motions nearly supersonic.

\begin{figure}
\includegraphics[width=0.5\textwidth]{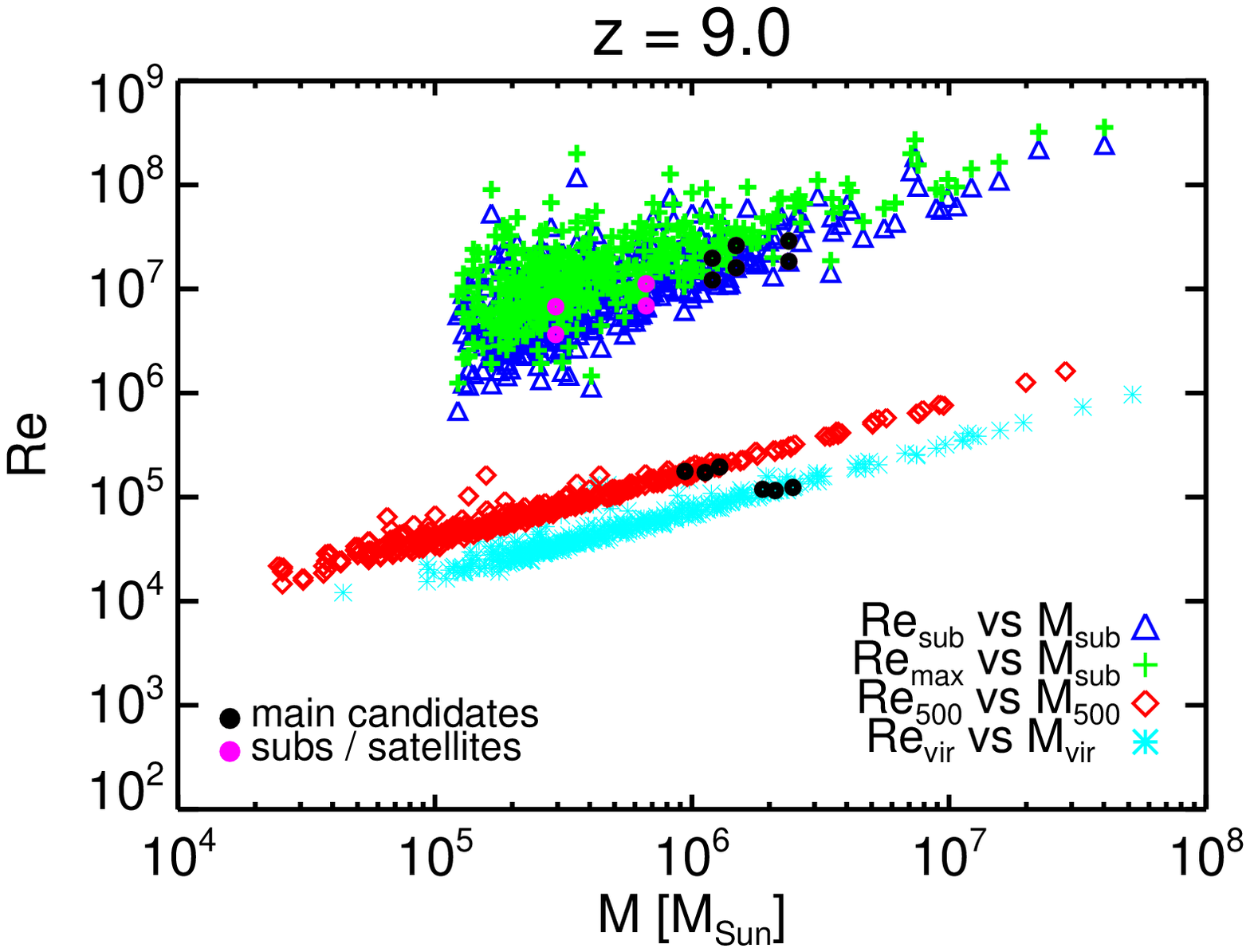}\\
\caption[]{\small
Reynolds number estimated with different approaches ($Re_{\rm sub}$, $Re_{\rm max}$, $Re_{\rm 500}$ and $Re_{\rm vir}$) as a function of substructure mass, $M_{\rm sub}$, mass within an over-density of 500, $M_{\rm 500}$, and virial mass, $M_{\rm vir}$, at $z=9$ for the run with a $T_{\rm eff}=10^5\, \rm K$ black body as popIII SED (see details in the text).
Black bullet points highlight the values for DCBH host candidates (3 friend-of-friend objects overplotted on the red diamonds and cyan asterisks).
Main haloes and substructures/satellite haloes are overplotted, respectively, by black and magenta bullets on the sequences of blue triangles and green crosses.
}
\label{fig:Re017}
\end{figure}
\begin{figure}
\includegraphics[width=0.5\textwidth]{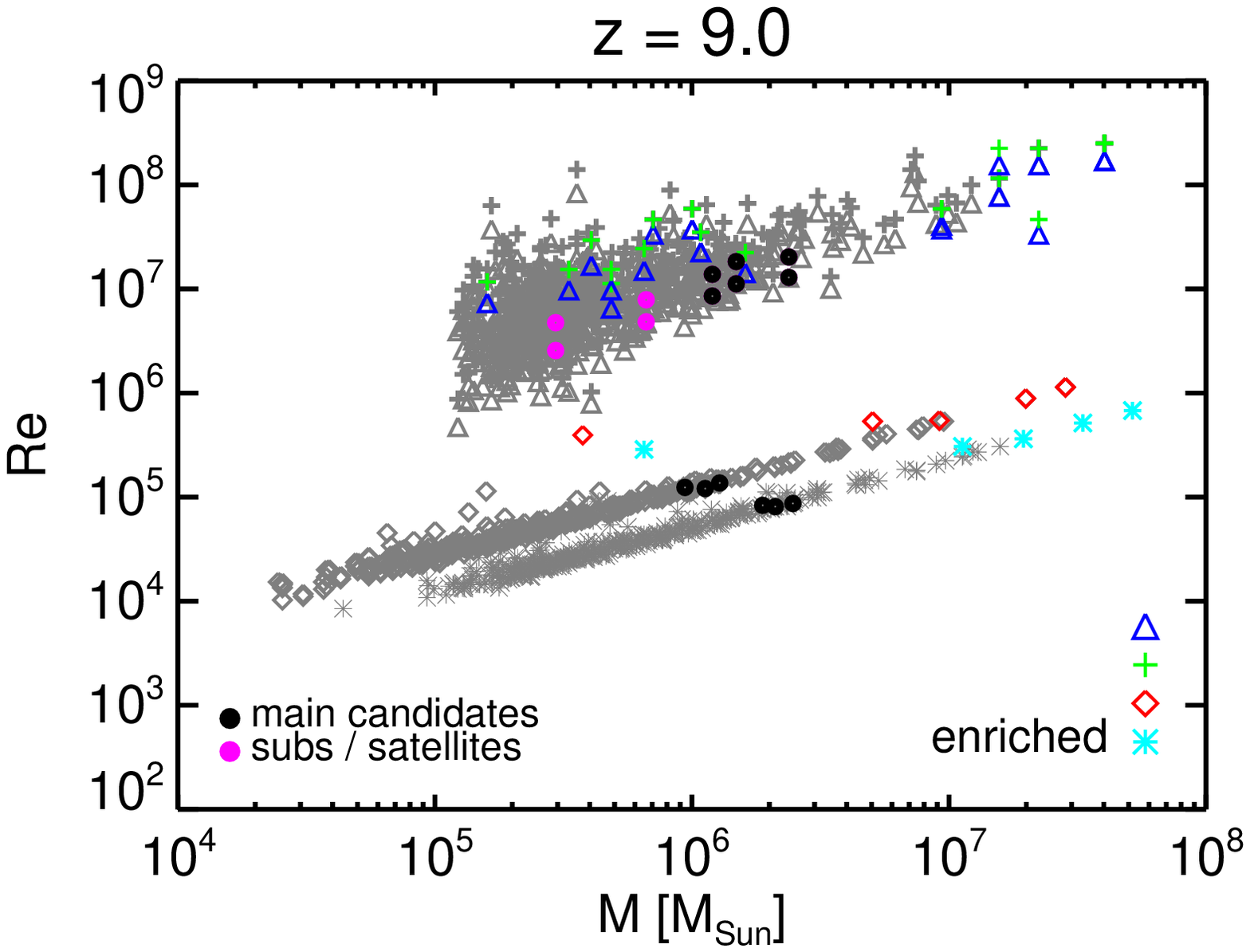}\\
\caption[]{\small
Reynolds numbers, estimated with the same approaches as Fig.~\ref{fig:Re017}, are shown in grey for pristine haloes and with coloured symbols for metal enriched haloes.
}
\label{fig:Re017compare}
\end{figure}

Fig.~\ref{fig:Re017} shows the effects of different ways of estimating $Re$, however it does not give us information on the role of metallicity for DCBH candidates.
For this reason, in Fig.~\ref{fig:Re017compare} the results for metal-free haloes (grey symbols) are directly compared to the results for haloes that are not metal-free (color symbols).\\
In this case, cyan asterisks refer to {\it metal enriched} haloes whose $Re$ numbers have been computed through virial quantities (i.e. $Re_{\rm vir}$ vs $M_{\rm vir}$).
They lie along the trend of the whole halo population, although their $Re > 10^5 $ values are higher than DCBH candidates (bullets) both at larger ($\sim 10^7\,\rm M_\odot$) and smaller ($\lesssim 10^6 \,\rm M_\odot$) masses.
Similarly, also red diamonds ($Re_{\rm 500}$ vs $M_{\rm 500}$ for metal enriched haloes) are above the bullet points, roughly following the general trend.
In the high-mass end, the larger $Re$ values are mainly due to the larger masses of haloes above $ \sim 10^7 \,\rm M_\odot$.
In the low-mass end (see red diamond and cyan asterisk at masses $ \lesssim 10^6 \,\rm M_\odot $), $Re$ values are affected by the metal enrichment process.\\
The values denoted by green crosses ($Re_{\rm max}$ vs $M_{\rm sub}$) and blue triangles ($Re_{\rm sub}$ vs $M_{\rm sub}$) are computed by using substructure information and refer to both enriched haloes and satellites.
Their distribution is quite sparse, however most of the points lie above the trend inferred from the bullets by up to 1~dex.
In this respect, DCBH candidates host lower chaotic motions than metal enriched haloes, although they are in line with the expectations for the most quiescent pristine haloes (grey symbols).\\
The physical reason for such behaviour relies on the additional entropy injected during metal pollution from nearby star forming regions.
Enriched material spreading into the surrounding objects affects the velocity and thermodynamical structure of hosting haloes, possibly causing an increase in the resulting $Re$ with respect to DCBH host candidates and pristine haloes.
This is particularly clear from the deviations from the general trend of the smaller enriched structures (asterisks and diamonds).
$Re$ values of larger polluted objects, instead, seem to be less sensitive to metal spreading, due to the higher degree of stochasticity implied by their bigger masses.\\
The wide scatter in the relations denoted by crosses and triangles indicates a strong influence of cosmic environment and substructure formation, too, that will influence velocity and thermodynamical features depending on whether haloes are in clustered or isolated regions (in fact, the spread is wider at the low-mass end, that is dominated by small structures and satellites).
The position of DCBH host candidates on the bottom edge of these two samples in the figure reveals their formation preferentially in weakly clustered regimes or in isolation, where shocks from structure formation \cite[that can trigger gas turbulent motions;][]{WiseAbel2007, Wise2012prad} are less common.\\
From these considerations it emerges that DCBH host candidates are on average less turbulent structures than metal enriched haloes (diamonds and asterisks) and are among the lowest-turbulence structures with pristine composition in the corresponding mass range (crosses and triangles).\\

\begin{figure}
\includegraphics[width=0.5\textwidth]{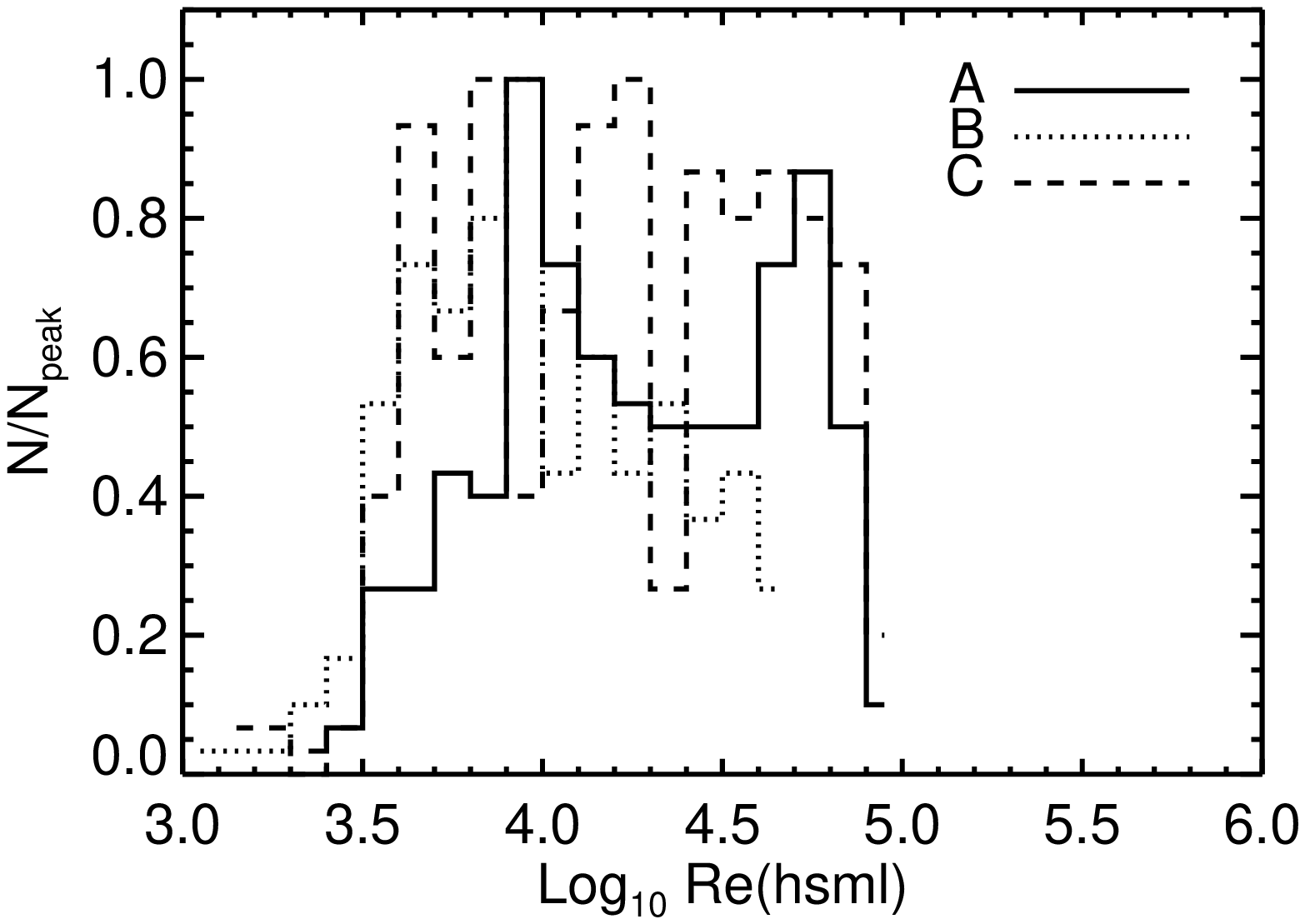}\\
\caption[]{\small
Distributions of the Reynolds number estimated at the hydro smoothing lenght scale for the three DCBH candidates A (solid line), B (dotted line) and C (dashed line). The distributions are normalised to their peak value.
}
\label{fig:Rehsml017}
\end{figure}

To judge better the level of turbulence within the DCBH hosting structures we compute the $Re$ values for their constituting gas particles, by employing the gas SPH smoothing lenght (hsml) and the gas velocity.
This allows us to assess turbulent motion locally rather than globally.
In Fig.~\ref{fig:Rehsml017} we show the distribution of $Re$ numbers at the SPH smoothing scales (hsml).
We find a broad trend within $Re \gtrsim 10^3$ and $\sim 10^5$ for all the three candidates, with peak values at
$Re\sim 10^4$-$10^{4.7}$ for candidate A, 
$Re\sim 10^4$ for candidate B and
$Re\sim 10^{3.6}$-$10^{4.2}$ for candidate C.
The distributions are very disperse, but all of them lie at $Re \gtrsim 10^{3.6} \sim 4000$, meaning that even at particle level, turbulent motions are relevant.
We do not find strong differences between the isolated candidate A and the candidates B and C, that are composed by 2 substructures.
The orders of magnitude of these local estimates are consistent with the orders of magnitude of the global estimates shown above, once taken into account the difference between the typical scales (radius vs. hsml) involved in the two cases. In practice, they show that turbulence plays an important role at all resolved scales.

We have checked that by adopting the mean square velocity, instead of the particle velocity, as typical velocity of the fluid, the results about $Re$ remain unchanged.

\subsection{DCBH host fate}

The presence of turbulence might affect direct collapse, providing additional non-thermal pressure support to the gas.
Since the turbulent dissipation timescale can be written as  $l / \sigma \approx \rm  4\,Myr~( {\it l} / 40\,pc) (\sigma / 10\,km/s )^{-1}$, with $l$ turbulent scale and $\sigma$ velocity dispersion \cite[][]{Semenov2016}, the timescales during which direct collapse events are possible lie between $\sim$ 4 and 40~Myr.
The former estimate is led by turbulence decay, while the latter by molecular-chemistry arguments.
In practice, the conditions for DCBH formation hold less than 40~Myr and hence the direct collapse should be as rapid as that.
As an example, the innermost isothermal core of the candidate A at $z=9$ has a typical dimension of $ \sim 100\,\rm pc$. Reasonably, this value is comparable or larger than the turbulent decay scale and, considering the typical $\sigma \sim 10\,\rm km/s$, the resulting dissipation timescale is $\lesssim 10 \,\rm Myr $.
\\
Hydro- and magneto-hydrodynamical simulations generally show turbulence dissipation on times that are much shorter than crossing times, both in subsonic and supersonic regimes \cite[][]{Stone1998, MacLow1998, Burkert2006, KimBasu2013, Semenov2016}, albeit still of the order of Myr.
\\
So, the occurrence of turbulent motions should delay DCBH formation until turbulence dissipates.
\\
At the same time, 2/3 of the candidates show substructures, which means that a pure direct collapse is rather unlikely and the process must be eventually accompanied by a merger \cite[][]{Latif2015, Becerra2015}.
\\
Merger events and substructure evolution (as e.g. in candidates B and C) are likely to inhibit gas direct collapse, enhancing gas fragmentation and halting accretion via tidal forces \cite[in agreement with][]{Chon2018}.
\\
Local rotational patterns and photo-evaporation effects (see previous sections) have maybe less severe implications, as they could cause the loss of only a fraction of the gas hosted in the halo potential wells.
\\
Thus, it turns out that the basic requirements necessary for DCBH formation, whenever met, do not guarantee the actual gas direct collapse.
In practice, the required necessary conditions are not sufficient for pure DCBH formation in 2/3 of the cases, while DCBH is probably expected for the remaining candidate A.
\\
What happens next will strongly depend on the surroundings of the host halos.
If gas photo-evaporation is stronger than the host potential wells (no matter whether they are generated by the whole halo or the newly formed DCBH) accretion will be inhibited.
Otherwise, accretion is expected to be episodic.
When averaged over several duty cycles, it should typically proceed at sub-Eddington rates \cite[][]{JohnsonBromm2007, Milosavljevic2009, Maio2013, Ricci2017}.
In cases when the surrounding environment is very dense, it is possible to reach super-Eddington rates \cite[][]{WyitheLoeb2012, AlexanderNatarajan2014, Madau2014, Inayoshi2016, Pezzulli2016, Pezzulli2017}.
The existence of massive black holes at $z \sim 7$ is challenging to explain, because it requires a growth process taking place in less than a billion years.
More likely, the growth mechanisms could be active for only a few 100s Myr, since massive black-hole seeds are usually expected at redshifts of $z\sim 6$--$20$, after the first structures form.
Currently, no observational detections of DCBHs are available.
Studies of possible identifications of DCBH candidates in the CANDELS/GOODS-S survey have claimed only two candidates with predicted mass greater than $10^5\,\rm M_\odot$, with robust X-ray detection and with photometric redshift $z>6$ \cite[][]{Pacucci2016}.
In theory, {\it ad hoc} seeding with $10^5\,\rm M_\odot$ black holes at $z\gtrsim 10$ of {\it all} the haloes with mass larger than $10^9\,\rm M_\odot$ (as commonly done in large numerical simulations) could lead to Eddington accretion rates at $z=9$--$6$ and to black hole masses higher than $10^9\,\rm M_\odot$ at $z \lesssim 7$.
Nevertheless, early attainment of a minimum mass of at least $10^6\,\rm M_\odot$ remains crucial for a fast growth.
These considerations suggest that DCBHs could be interesting massive seeds, since the $10^5\,\rm M_\odot$ black-hole seeds commonly adopted in large numerical simulations are easily justified with this channel, although it is quite unlikely that {\it all} the early haloes blindly seeded in simulations are actual hosts of DCBHs.
\\

\begin{figure}
\includegraphics[width=0.45\textwidth]{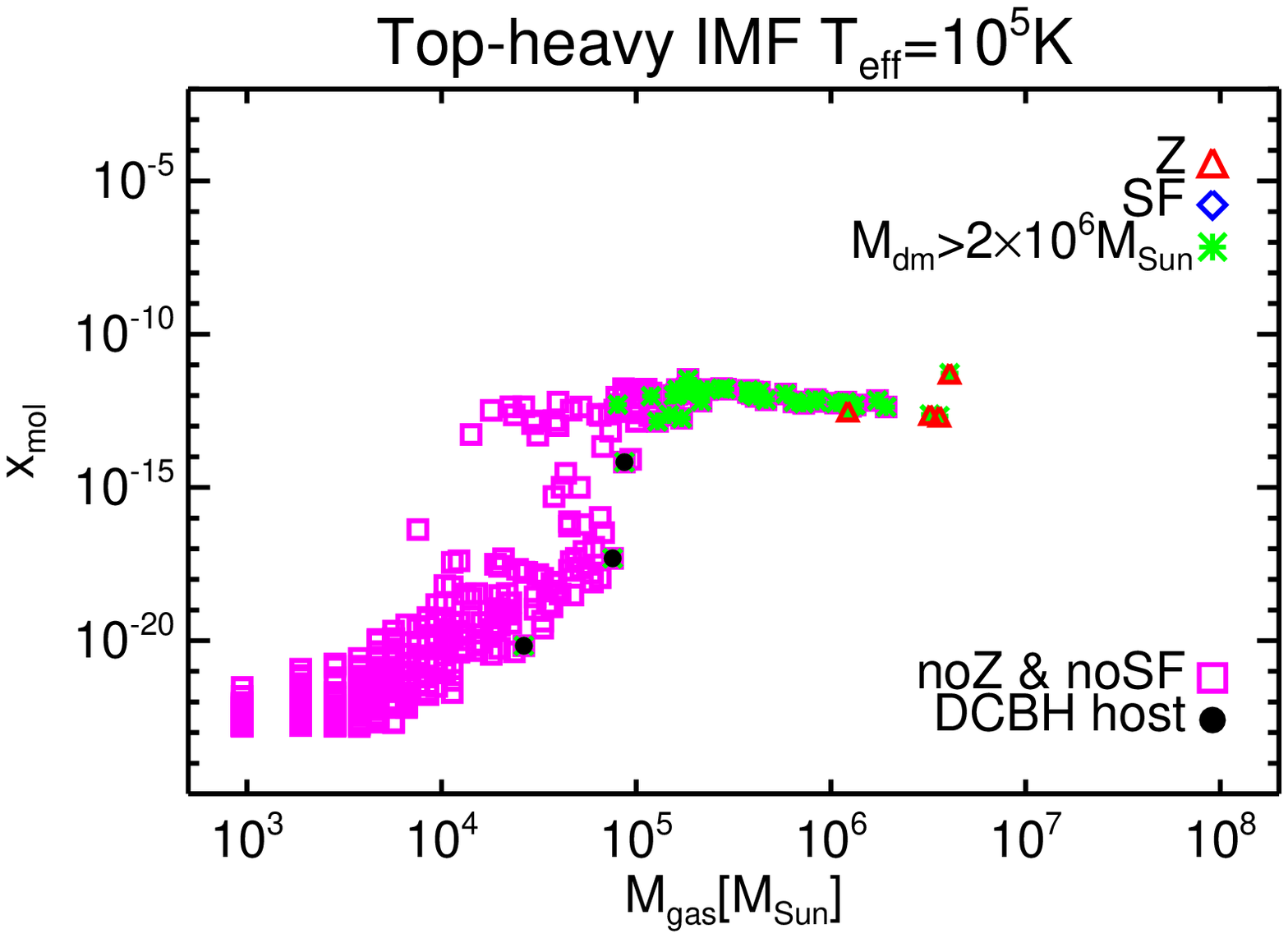}\\
\caption[]{\small
DCBH host candidates at $z=8.5$ for the run with a $T_{\rm eff}=10^5\,\rm K$ black body as popIII SED.
}
\label{fig:DCBHall}
\end{figure}

To conclude, we briefly investigate thermal and chemical properties of the halo populations identified at a time later than $z=9$, specifically at $z=8.5$ (when the Universe is about $0.58 \, \rm Gyr$ old).
In Fig.~\ref{fig:DCBHall}, DCBH host candidates (bullet points) for the run with top-heavy IMF and a $T_{\rm eff}=10^5$\,K black body as popIII SED (TH.1e5) are shown for redshift $z=8.5$.
Once compared to the corresponding results of Fig.~\ref{fig:DCBH011} and Fig.~\ref{fig:DCBH017} at $z=11.5$ and $z=9$, the changes in the molecular content are evident.
They are a consequence of the strong dissociating photon field established by primordial massive stars.
As mentioned before, at $z=9.5$ no DCBH host candidate is found, while a few Myrs later, at $z=9$, we have identified 3 haloes satisfying the basic requirements (previous Fig.~\ref{fig:DCBH017}).
Later on, at $z=8.5$ we find three additional systems.
Of the three $z=8.5$ DCBH candidates (mapped in Fig.~\ref{fig:groups018} in Appendix), two are composed by two substructures.
The third candidate is, instead, an isolated object (see host properties in Fig.~\ref{fig:DCBH3x3} in Appendix).
Notwithstanding the similarities, the three DCBH host candidates identified at $z=8.5$ do not coincide with the ones in Fig.~\ref{fig:DCBH017} at $z=9$.
Their comoving (physical) distances from the central star forming region are about
348 (37)~kpc/{\it h} 
179 (19)~kpc/{\it h} and 
48 (5)~kpc/{\it h},  respectively.
In comparison to the situation at $z=9$, these candidates are slightly farther from the radiation sources, but they are still found in proximity of denser filamentary regions.
Given the presence of substructures, rotational patterns, photo-heating and turbulent $Re$ numbers (Fig.~\ref{fig:Re018}, Fig.~\ref{fig:Rehsml018}), the formation of DCBHs in the $z=8.5$ candidates is as difficult as for the $z=9$ candidates.

\section{Discussion}\label{Sect:discussion}


We have employed cosmological N-body hydrodynamical chemistry calculations to trace the origin of primordial Direct Collapse Black Holes (DCBHs).
Our simulations follow photon propagation, gas atomic and molecular chemistry and heavy-element production from stars with different masses and metallicities during the first Gyr of the Universe.
Our detailed implementation takes into account multi-frequency radiative transfer from 150 frequency bins in the energy range [0.7, 100]~eV --  including 
H, He, D, H$_2$, HD, HeH$^+$ transitions as well as 
LW band ([11.2, 13.6]~eV), 
near-IR (at energies $\lesssim\rm 1.7~eV$) and 
UV ($\sim\rm [3, 100]~eV$) radiation.
This is coupled to non-equilibrium chemistry integration of e$^-$, H, H$^+$, H$^-$, He, He$^+$, He$^{++}$, D, D$^+$, H$_2$, H$_2^+$, HD, HeH$^+$ abundances, stellar evolution and metal enrichment from different species (He, C, N, O, Ne, Mg, Si, S, Ca, Fe, etc.) and stellar populations.
\\
Such implementation allows us to investigate the physical properties of early hosts of DCBHs and to compare the predictions of our calculations for different assumptions about the adopted IMF and SED of popIII stars.
\\
We find that DCBHs are very rare events favoured by the existence of powerful primordial sources.
Standard stellar sources are unlikely to establish a radiative cosmological background in the first Gyr, while massive sources are able to emit larger amounts of photons and to form a background field that can both dissociate molecules and photo-evaporate primordial haloes.
\\
Albeit exposed to significative LW radiation, ranging from a few up to $\sim 50 J_{21}$ at $z=9$, the basic environmental requirements necessary for their formation, whenever met, do not guarantee the actual gas direct collapse. In fact, a number of highly non-linear processes (merger events, substructure evolution, local rotational patterns, gas turbulence, photo-evaporation) play a significant role to halt or delay the birth of a DCBH.
In practice, the required {\it necessary} conditions turn out to be {\it sufficient} for pure DCBH formation only in 1/3 of the cases.
\\
It is noteworthy that alternative modelling accounting for increased physical viscosity would lower $Re$ (Sect.~\ref{sect:Re}), favouring DCBH formation, while, on the contrary, implementations reducing artificial viscosity would lead to opposite results \cite[][]{MM1997, Dolag2005, CullenDehnen2010, Price2012, Latif2015, BiffiValdarnini2015, Beck2016}.
Overall, viscous heating provides an important pathway to obtain an atomic gas phase within the centre of the halo and helps the formation of very massive objects.
\\
Our findings are subject to some {\it caveats}.
The exact numerical parameters adopted for the initial mass functions, wind prescriptions, initial-condition gas velocities and related issues are likely to induce some small changes \cite[][]{Campisi2011, Maio2011vb, Tescari2014}.
\\
Uncertainties might derive from the lack of a definitive treatment of metal diffusion mechanisms, which, despite several attempts, still remain an unsolved problem in astrophysics and further studies are still required for an accurate assessment.
\\
Theoretical stellar yields are affected by a plethora of uncertain physical processes in stars (such as explosion mechanisms, differential rotation, initial composition, magnetic fields, nuclear reaction rates, etc.) all of which can influence the final ratios.
Detailed values of popIII or popII-I metal yields for individual elements \cite[see e.g.][]{Francois2004, MaioTescari2015, Ma2015, Ma2017} are not expected to change gas hydrodynamics significantly, because metal content and cooling in the regimes investigated here are usually dominated by oxygen, that is fairly well explored.
\\
Besides that, molecular self-shielding is still under debate, due to its dependence on gas density and numerical resolution \cite[][]{Hartwig2015}.
High-resolution studies are needed to clarify its effects in promoting molecule formation and inhibiting DCBHs.
\\
Lately, studies by \cite{WG2017} have suggested that the effective temperature of popII stars might be larger than $10^4\,\rm K$, hence using a $10^4\,\rm K$ black-body spectrum can lead to underestimating the level of LW radiation required to photo-dissociate H$_2$.
That work shows little or no difference in the H$_2$ photo-dissociation rate for black-body effective temperatures of
$2\times 10^4$-$10^5\,\rm K$, while some variations appear between $10^4\,\rm K$ and $2\times 10^4\,\rm K$.
However, \cite{WG2017} developed a  one-zone model in which overestimates in the photodissociation rate are relatively small, just by a factor of $\sim 2$. Once embedded in three-dimensional calculations, the resulting effects are not expected to have a strong impact and, in fact, we did not find significant differences between the SL.1e4 and SL.4e4 cases analysed here.
Similarly, recently improved calculations including additional resonance contributions to the H$^-$ photodetachment rate give only very small corrections (less than 20 per cent) up to black-body temperatures of $10^5\,\rm K$ 
\cite[][]{Miyake2010, WG2017}.\\
Different stellar feedback modelling might have some implications on the DCBH scenario.
In practice, different velocities acquired by gas parcels through feedback mechanisms could have an effect on the optimal distance from the central source where DCBHs can be located, although we do not expect huge variations with respect to the results found here, as long as velocity values are realistically of the order of $\sim 10^2 \, \rm  km/s$.
\\

Despite the small-number statistics, it is instructive to make some comparisons to similar available works.\\
There is a vaste literature of three-dimensional radiative simulations studying pristine-gas collapse and the role of UV radiation 
\cite[][]{M2001, Yoshida2003, Oshea2004, Oshea2008}, as well as the birth of stars, galaxies and their feedback effects in a cosmological context
\cite[][]{Maio2010, Maio2011, Wise2012, Wise2012prad, Wise2014, Lopez2014, Mancini2015, Mancini2016, Ma2015, Ma2017, Ma2017dla}.
These studies have been performed with different codes (such as SPH Gadget code and AMR Enzo code) and by adopting different resolutions.
Their findings highlighted that feedback prescriptions are among the main causes of differences in the final results and have a crucial role for photon production and black-hole seeding.
\\
Recent pristine-gas simulations \cite[e.g.][and references therein]{Regan2017} have confirmed that LW radiation inevitably moves the gas in the host protogalaxy onto the isothermal atomic cooling track, without the deleterious effects of either photo-evaporating the gas or polluting it with heavy elements. Star formation and feedback modelling in these studies were absent, though.
We do find that LW radiation is responsible for dissociating H$_2$ and HD and for keeping the gas in the host protogalaxy almost isothermal \cite[][]{Maio2016}, however we do not rule out photo-evaporation, since this issue is tightly linked to the type of the original radiative sources considered.
A similar conclusion on gas photo-evaporation in early halos was reached by \cite{Wise2014}, who stressed its role in addition to heating and ionising the surrounding medium out to a (physical) radius\footnote{
We note here that the quoted upper limits might depend on to the details of the radiative implementation adopted.
}
of $10$-$15$~kpc at $z = 9$. These respectively reduce the in-situ and external cold gas supply that could feed future star formation.
\\
Photo-ionisation feedback is definitely responsible for evacuating gas from mini-haloes and for generating large HII regions around the central emitting source, but the implications for DCBH formation must be evaluated carefully.
\cite{Regan2016} have noted that, due to photo-ionisation, there could exist an optimal metal-free zone for DCBH formation between 1~kpc and 4~kpc from the emitting source.
Nevertheless, this is in tension with metal spreading, not included in that study. Previous calculations by \cite{Wise2012} found that one primordial SN is sufficient to enrich the entire star forming halo and surrounding $\sim5$~kpc to a metallicity of $10^{-3}Z_\odot$.
Metal enrichment would, then, inhibit DCBH formation at distances shorter than $5$~kpc.
These arguments suggest a distance $\gtrsim5$~kpc required for DCBH formation.
Consistently with \cite{Wise2012}, we find that DCBH conditions can be fulfilled only at radii above $\sim 5\,\rm kpc$, since closer regions are polluted by metal enrichment and subject to suppression of gas collapse through photo-ionisation feedback \cite[][]{Kannan2014}.
Metal pollution is not included in the work by \cite{Regan2016}.
This is also why we do not find eligible DCBH candidates in the SL.1e4 and SL.4e4 models: near the emitting source, DCBH formation is inhibited by metal spreading, while far away the radiation flux from standard solar-like or OB-type stars drops dramatically and the molecular gas fraction remains too high.
Furthermore, photo-heating is also responsible for increasing the thermal energy of the gas and causing its escape from mostly low-mass haloes. The impact can be dramatic for the case of powerful sources, while, for solar-like and OB-type scenarios they are much more modest \cite[for a more extended discussion see e.g.][and references therein]{Maio2016}.
\\
Other studies have investigated the possibility to have DCBH formation in pristine regions within halos where star formation has recently occurred \cite[][]{Dunn2018}.
They have relied on Gasoline and implemented a common stochastic model with star formation probability scaling with H$_2$ fractions, that are, in turn, tightly dependent on gas density.
They study the effects on massive black-hole seeds for different assumptions on the critical flux, although do not use any radiative transfer code to propagate photons nor implement stellar evolution calculations to spread heavy elements.
They confirm the strong dependence of DCBH formation on the adopted critical flux, finding results that are in line with previous ones \cite[][]{Habouzit2016}.\\
Lately, \cite{Barrow2018} have suggested observational features of DCBH, claiming that the upcoming JWST telescope might be able to detect and distinguish a young galaxy that hosts a DCBH at $z\sim 15$.
\\
The low number of DCBH candidates in comparison to the whole halo population at the redshifts of interest and the occurrence of DCBH candidates only in the TH.1e5 scenario suggest that DCBH formation is a rare event.
This is confirmed by recent calculations by \cite{Chon2016, Chon2018}, who explore the formation and collapse of pristine gas clouds in DCBH hosts. They follow a sink-particle scheme in high-density gas and focus on the first 0.1~Myr evolution of the accretion phase of supermassive stars finding that only
few gas clouds are able to collapse, while, in most cases, tidal forces of nearby galaxies act against gas collapse.
\\
A full statistical analysis of the DCBH number density would require much larger high-resolution boxes that are currently beyond computational capabilities.
We caution the reader that extrapolations from smaller to larger volumes are dangerous and lead to overestimates of the precise DCBH statistics, because small boxes, like the ones used here, are likely to collapse into one single object at $z\lesssim 6$.\\
In realistic large-scale cosmological environments, DCBH candidates could be present only around powerful popIII sources, that contribute $\sim 10^{-5}$ to the total star formation at redshift $z\sim6$ \cite[][]{Tornatore2007}, or lower \cite[][]{Wise2012}.
Thus, if one DCBH formed in our box and grew up to masses of $ 10^9\,\rm M_\odot$ by $z\sim 6$, we should infer an upper limit to the DCBH abundance of
$ < 2.7 \times 10^{-5} \, \rm Mpc^{-3}$.
The lower limit for the observed number density of supermassive black holes powered by masses of $\sim 1$-$2 \times 10^9 \,\rm M_\odot$ at $z \sim 6$ is
$ \gtrsim 1.1 \times 10^{-9} \, \rm Mpc^{-3}$ \cite[][]{Venemans2013}. 
This means that possible values for DCBH number densities should be bracketed within the range $\sim 10^{-9}$-$10^{-5} \,\rm Mpc^{-3}$.
The present-day number density of active galactic nuclei is about $10^{-4} \,\rm Mpc^{-3}$ \cite[][]{Shankar2009, Johnson2013} which is higher than our estimate and makes unlikely that all present-day supermassive black holes are originated from DCBHs \cite[consistently with e.g.][]{Sugimura2014}, but they could explain at least part of the population.
Previous semi-analytic works in the literature had highlighted the rarity of DCBH events based on the necessary assumptions mentioned at the beginning of this paper.
In light of our results, their estimates must be interpreted as upper limits for the actual occurrence of DCBHs, since we find that only 1/3 of the haloes where the necessary conditions are met could lead to pure DCBHs.
This is consistent with the numerical results by e.g. \citet{Latif2015}, who found that DCBH formation could be accompanied by mergers.
More quantitatively, we find that this should happen in 2/3 of the cases we identify.
\\
Furthermore, \citet{LatifVolonteri2015} have found in numerical simulations of isolated structures that rotational motions should not halt the formation of DCBHs, although we caution that dynamical effects might be responsible for lowering the resulting collapsed mass of about $20$ per cent or more, as a consequence of unbound gas escaping from the potential wells.
Independently from that, turbulence will always be present, possibly slowing down DCBH appearance.
From these considerations it emerges that most of the gas mass in the host candidates might collapse into a DCBH under a very turbulent regime with mean square velocities close to the local sound speed.
\\
At variance with the conventional gas direct collapse proposal, \cite{Mayer2010} suggested that the formation of super-massive black holes of $10^8$-$10^9\,\rm M_\odot$ could occur simply via major mergers of gas-rich galaxies at $z>6$ and no need to suppress cooling and star formation.
Crucial ingredients of such conjecture are a temperature floor of $2 \times 10^4\,\rm K$, to mimic turbulence pressure support, and a rapidly accreting nuclear disks due to efficient angular momentum loss during mergers.
In their actual calculations, performed with the Gasoline code, radiative cooling is shut off at $2 \times 10^4\,\rm K$ to avoid strong gravitational instabilities and widespread fragmentation of gas-rich disks \cite[][]{K2005}.
Despite this forcing, it has not been possible to show how two super-massive black holes bind during a galaxy merger with gas, because of the difficulty of modelling a wide range of spatial scales \cite[][]{MB2019}.
\\
\cite{Valiante2016} also investigated semi-analytically the relative role of light and heavy seeds as progenitors of the first  supermassive black holes at $z>6$. Although light seeds are unlikely to produce large black-hole masses, the authors found a strong dependence on the interplay between chemical, radiative and mechanical feedback effects, which could easily enhance metal and/or dust cooling.
Furthermore, in agreement with our results, they noted the importance of the adopted stellar mass range (IMF) for primordial stars that dramatically affects the history of cold gas. \\
We expect that primordial baryonic streaming velocities originated at decoupling \cite[][]{TH2010, Maio2011vb, Greif2011} might delay initial gas evolution, but they have quite modest effects on DCBHs and their hosts \cite[][]{LNS2014, Hirano2017}.
\\
The additional presence of early X-ray photons \cite[][]{Inayoshi2015, Latif2015} or cosmic rays \cite[][]{Jasche2007} might enhance free electrons and H$_2$ formation around $10^4\,\rm K$, limiting the role of DCBHs in primordial epochs and lowering the (still vague) expectations for their occurrence \cite[][]{Habouzit2016, Smith2017}.
This would make the formation of supermassive black hole seeds much more problematic and would complicate the explanation of the observed population of supermassive black holes at $z\simeq 7$ \cite[][]{Fan2006}.
\\
Throughout this work we have assumed a $\Lambda$CDM scenario.
We warn the reader, though, that in some particular cases \cite[such as for warm dark matter,][]{MaioViel2015} the background cosmological model can influence the baryonic-structure evolution, mostly at higher redshift.
This can play a role during the very beginning of the onset of star formation, but in the epochs of interest here baryon evolution tends to dominate and alleviate the discrepancies due either to alternative cosmologies \cite[][]{Maio2006} or to possible non-Gaussianities \cite[][]{MaioIannuzzi2011, Maio2011cqg, Maio2012, MaioKhochfar2012}.

\section{Conclusions} \label{Sect:conclusions}
In this study, we have tried to quantify DCBH appearance and their host features under three scenarios for primordial radiative emissions.
We have considered a scenario of powerful very massive primordial stars (TH.1e5), a scenario with OB-type primordial stars (SL.4e4) and a conservative scenario with solar-like primordial stars (SL.1e4).
While in the latter two scenarios newly formed stars are not able to provide sufficient amounts of LW radiation to dissociate molecules in close pristine haloes and to cause DCBH events, in the former one star formation provides enough photons to dissociate H$_2$ in many low-mass haloes.
As a result, three DCBH host candidates are identified and studied in detail.
They are found in the filamentary structures of the cosmic web and have dark-matter masses around a few times $10^6\,\rm M_\odot$.
This means that they are smaller than typical star forming haloes (which have masses of $10^7$--$10^8\,\rm M_\odot$), but large enough not to be disrupted by nearby feedback effects (as it happens to $\lesssim 10^6\,\rm M_\odot$ haloes).
Their evolution is rather complex, because DCBH host candidates present neat evidences of substructure formation, local gas motions and turbulent patterns.
In particular, 2/3 of the candidates are composed by a main halo and a smaller (bound) satellite. Thus, pure DCBH formation {\it should not be possible}, since the satellite should merge with the main halo before or during the direct collapse, inducing gas compression, cooling and star formation.
In all the DCBH host candidates, gas angular momentum, albeit small in average, displays non-null $z$-component at $\varepsilon \sim 1$ or higher, meaning that some rotational motions are present and some material might escape from the halo.
Therefore, even if rotational motions did not alter dramatically the pathway to DCBH formation, they could affect the final mass of more than $20$ per cent.
An important point is related to local turbulent motions in the primordial halo population, with DCBH host candidates featuring large Reynolds numbers.
This is {\it in contrast} with a rapid direct collapse of gas into a black hole and implies time delays of the order of $\sim 4$--$40$~Myr.
\\
These findings are relevant for the seeding and growth of accreting supermassive black holes at high redshift and demonstrate that DCBHs could be born in the early Universe only under very particular conditions.
Our results rule out the possibility that primordial solar-like or OB-type stars might have contributed significantly to the establishment of a LW background and to any direct-collapse event.
Despite early DCBH host candidates could have existed, it seems difficult to sustain the necessary conditions for their formation for a very long period of time.
Furthermore, in most cases such necessary conditions are {\it not sufficient} for pure DCBH formation.
\\
Joint efforts of upcoming international facilities devoted to study the infant and the high-energy Universe, such as 
SKA \cite[][]{Koopmans2015}, 
ATHENA \cite[][]{Nandra2013},
JWST \cite[][]{Gardner2009},
E-ELT \cite[][]{Puech2010, Maiolino2013}
and
WFIRST \cite[][]{Whalen2013}, 
will shed light on the still unanswered questions.
\\



\begin{acknowledgements}

We warmly acknowledge the anonymous referee for useful and constructive comments that significantly improved the quality of the manuscript.
We also acknowledge constructive interactions with S.~Chon, M.~Habouzit, L.~Hernquist, C.~Pfrommer, R. Schneider and M.~Volonteri.
U.~M. acknowledges kind hospitality at the Harvard-Smithsonian Center for Astrophysics and, particularly, the Institute for Theory and Computation of the Harvard University, Cambridge, MA (USA).
He has completed this work through a research grant awarded by the German Research Fundation (DFG) project n. 390015701 and through the HPC-Europa3 Transnational Access Programme project n. HPC17ERW30.
S.~B. acknowledges financial support from PRIN-MIUR 2015W7KAWC and the INFN INDARK grant.
The numerical simulations and analysis have been performed under the PRACE-2IP programme, grant agreement n. RI-283493, with resources allocated at the NOTUR computing centre, Norway, project number NN9903K, and supported by the facilities of the Italian Computing Center (CINECA), the Leibniz Institute for Astrophysics and the Max Planck Institute for Astrophysics, Germany.
We acknowledge the NASA Astrophysics Data System and the JSTOR archive for their bibliographic tools.

\end{acknowledgements}


\bibliographystyle{aa1}
\bibliography{bibl}


\appendix

\begin{figure}
\includegraphics[width=0.45\textwidth]{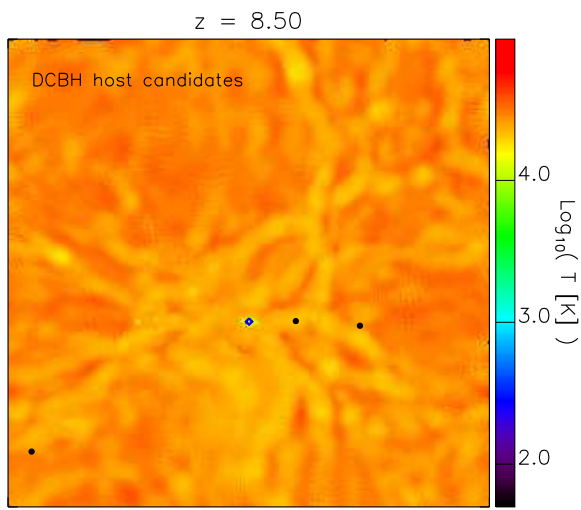}\\
\caption[]{\small
Mass-weighted temperature map and DCBH host candidates (bullet points) identified at $z=8.5$ for the run with top-heavy popIII SED.
}
\label{fig:groups018}
\end{figure}

\begin{figure*} 
\includegraphics[width=0.3\textwidth]{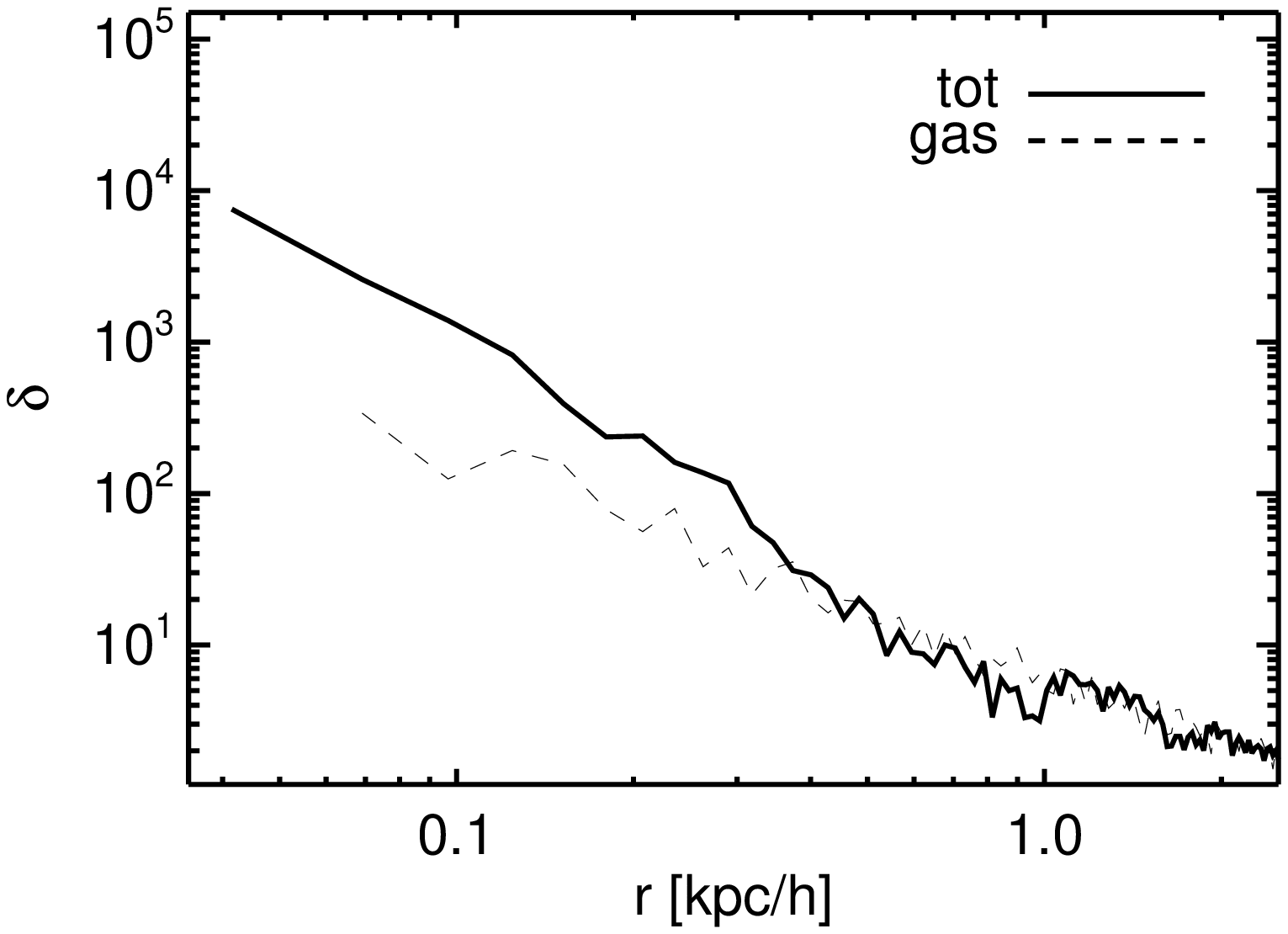}  	
\includegraphics[width=0.3\textwidth]{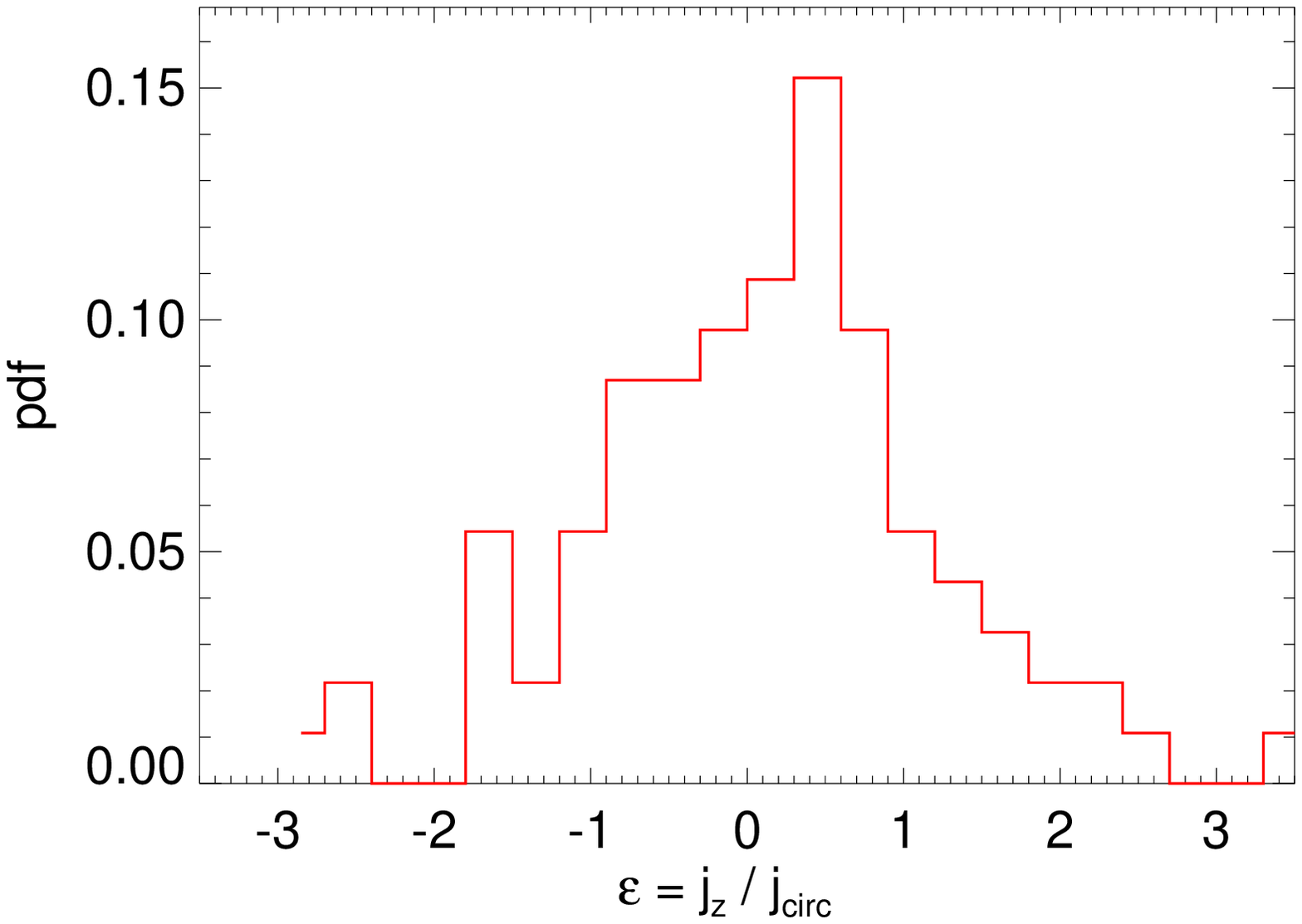}
\includegraphics[width=0.3\textwidth]{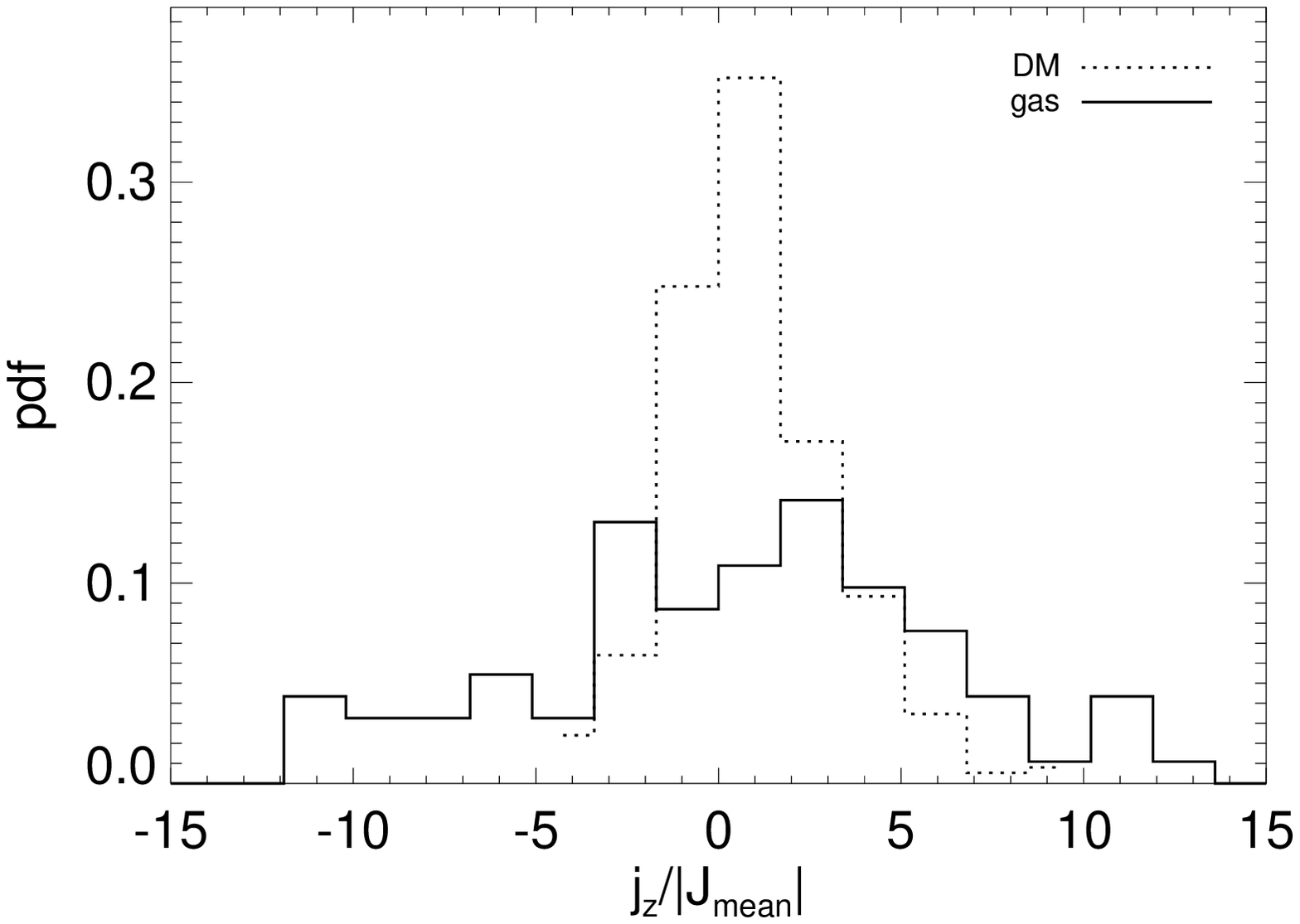}\\
\includegraphics[width=0.3\textwidth]{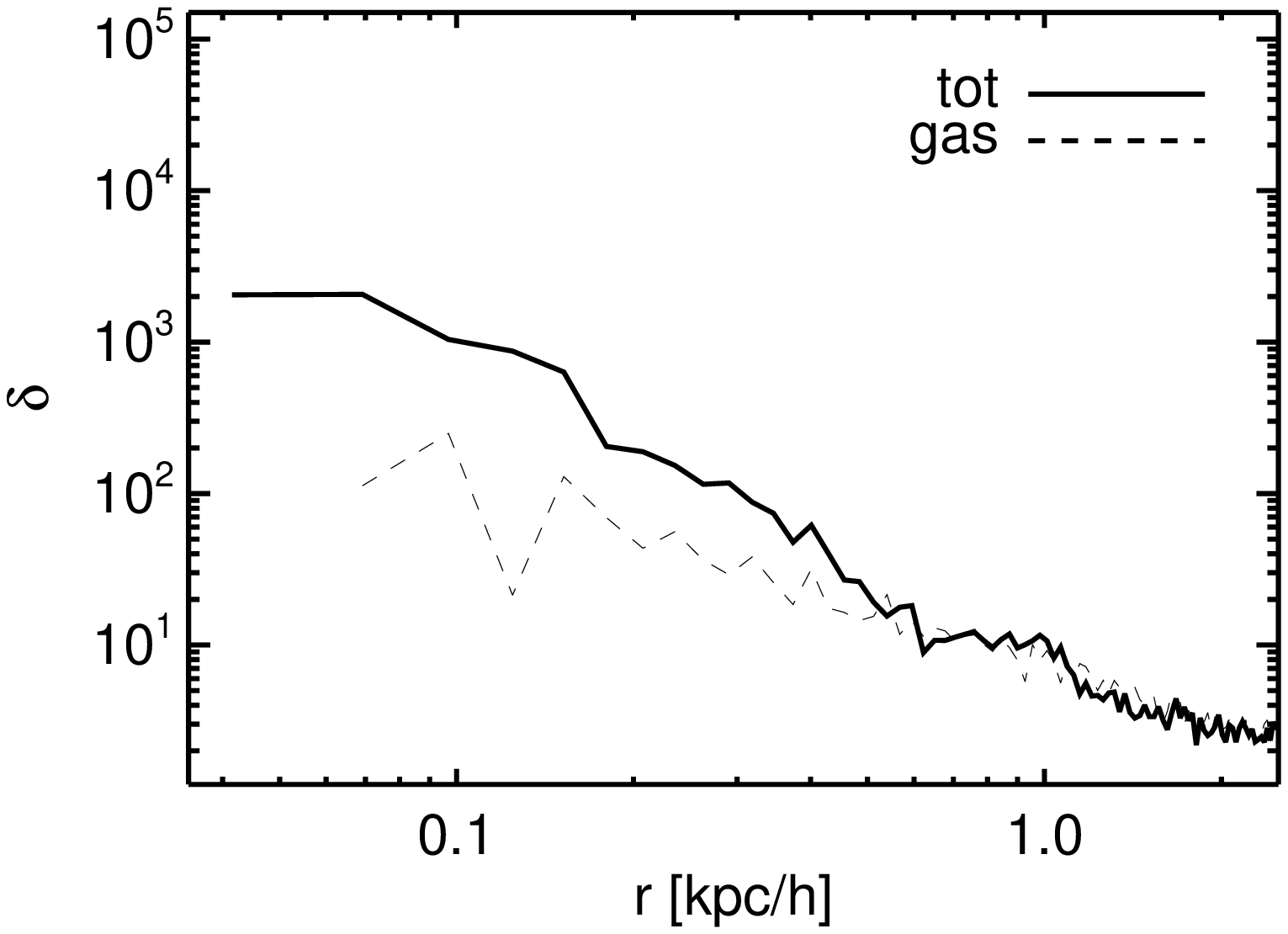}  	
\includegraphics[width=0.3\textwidth]{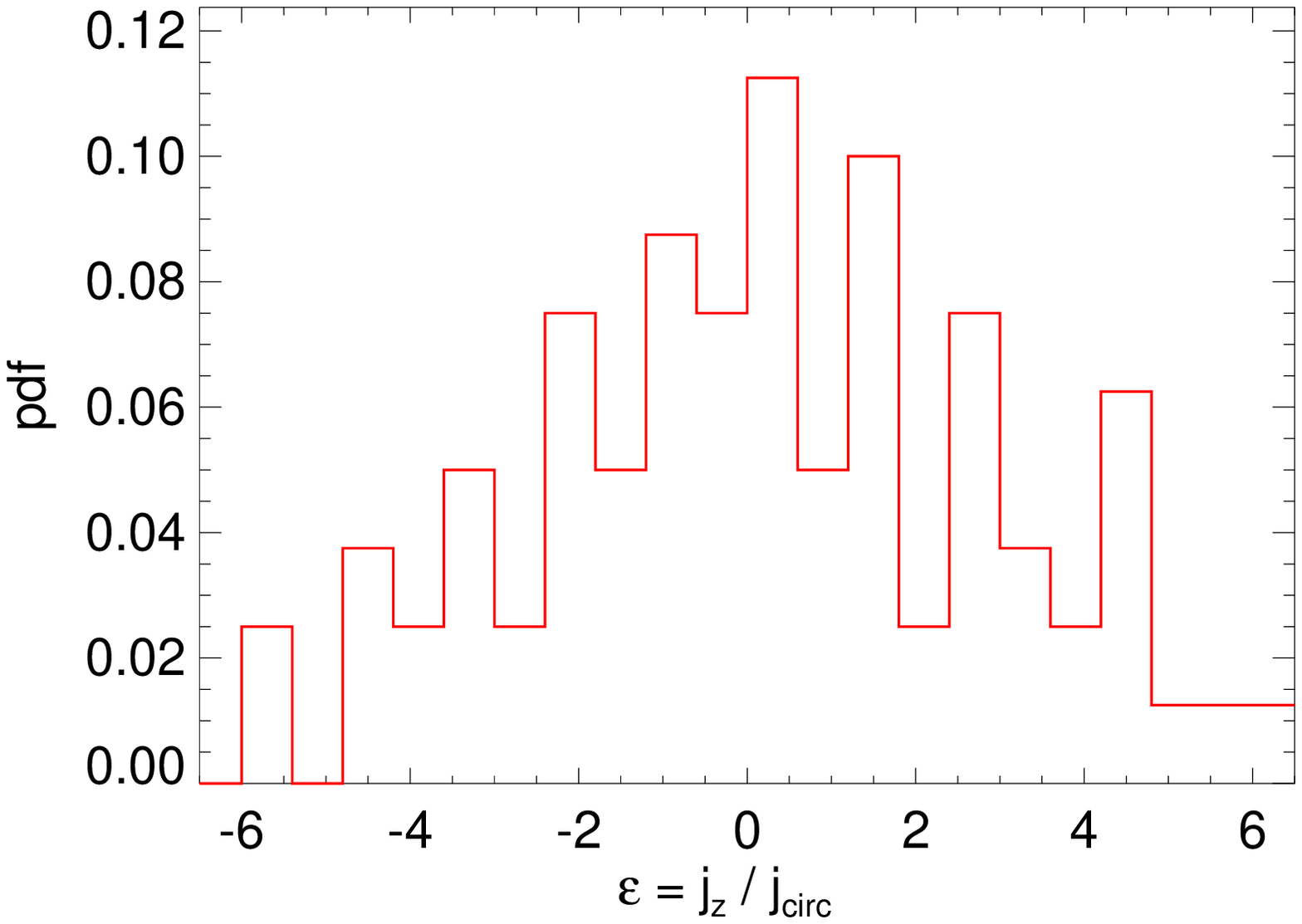}
\includegraphics[width=0.3\textwidth]{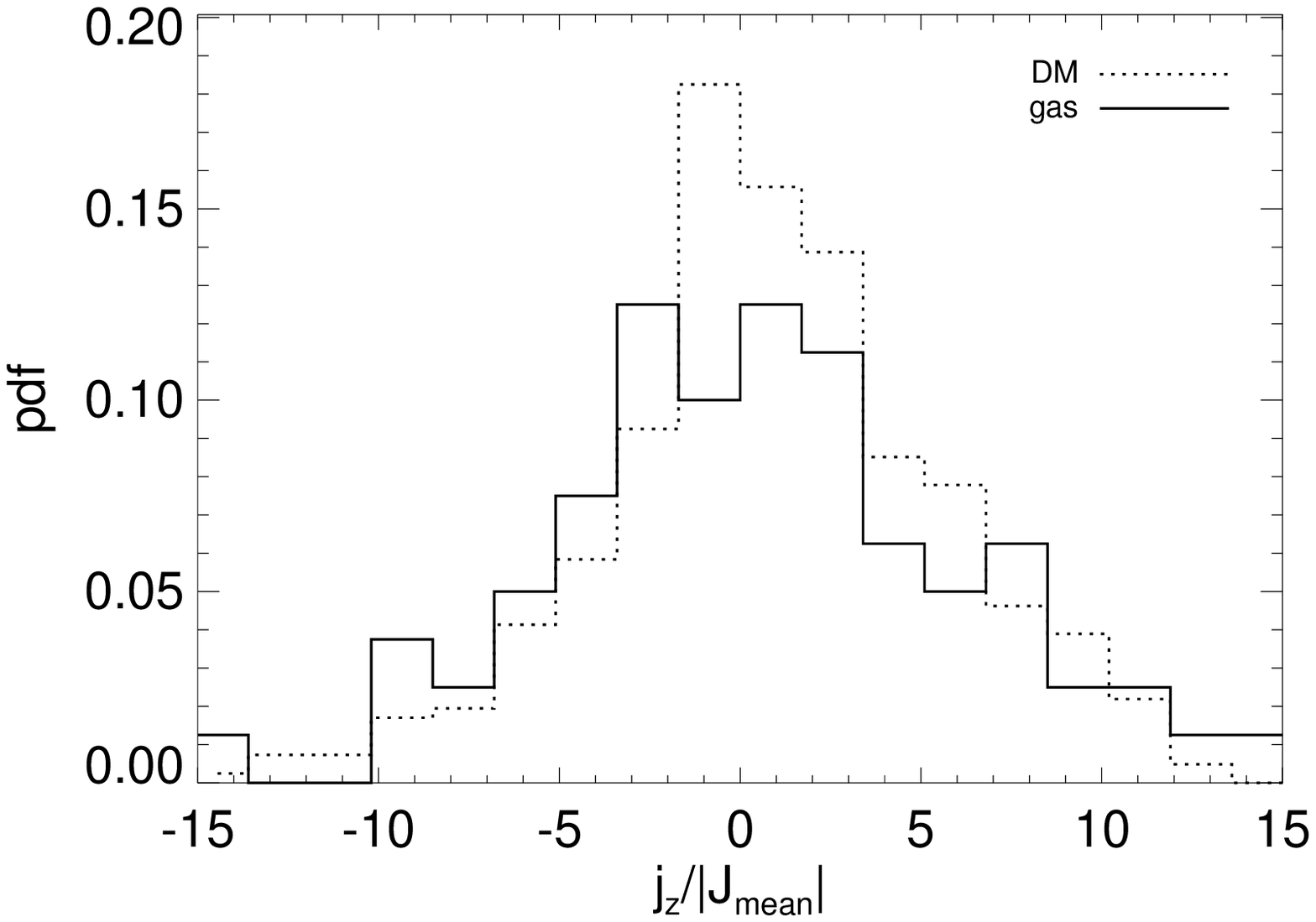}\\
\includegraphics[width=0.3\textwidth]{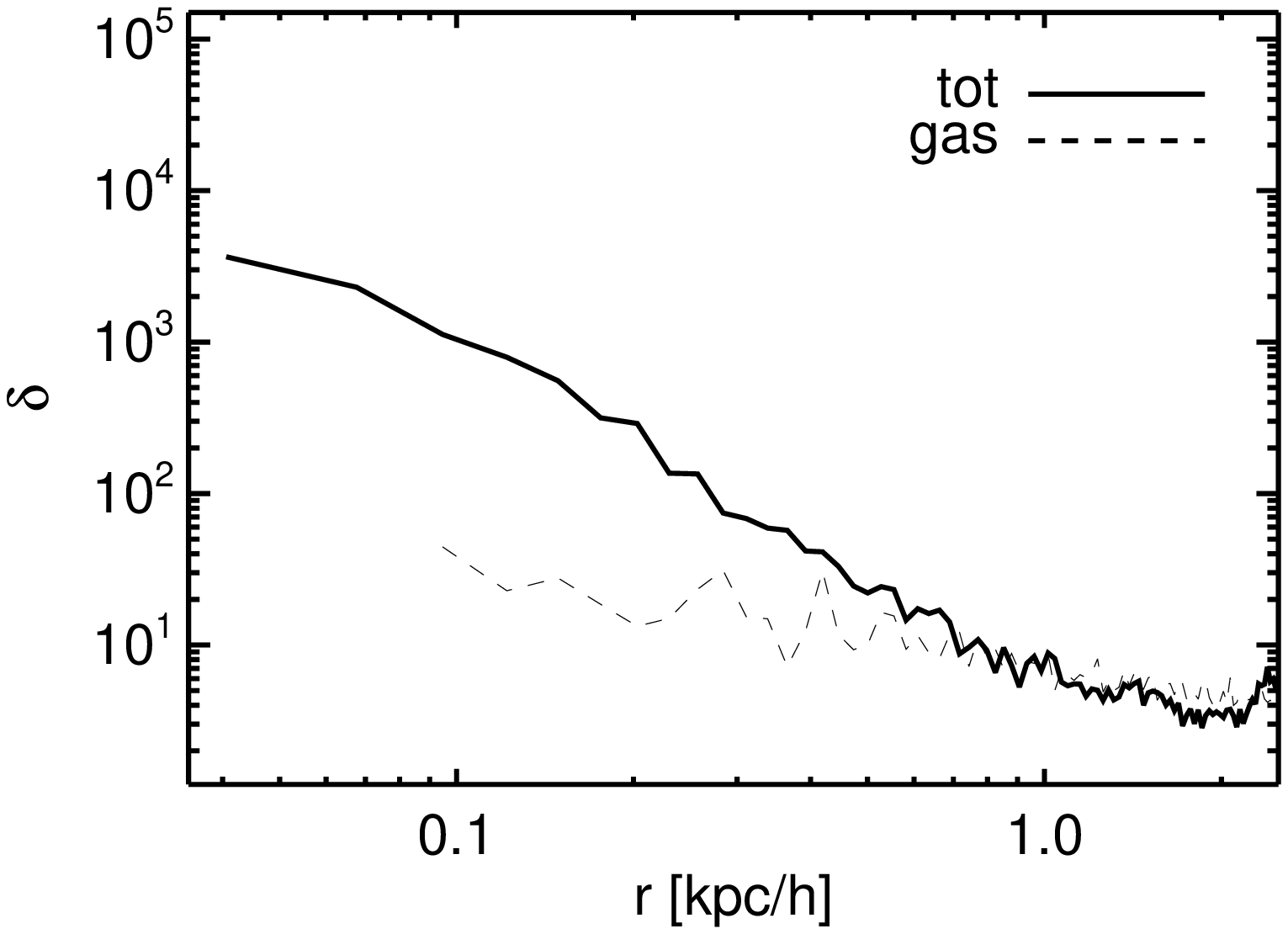}  	
\includegraphics[width=0.3\textwidth]{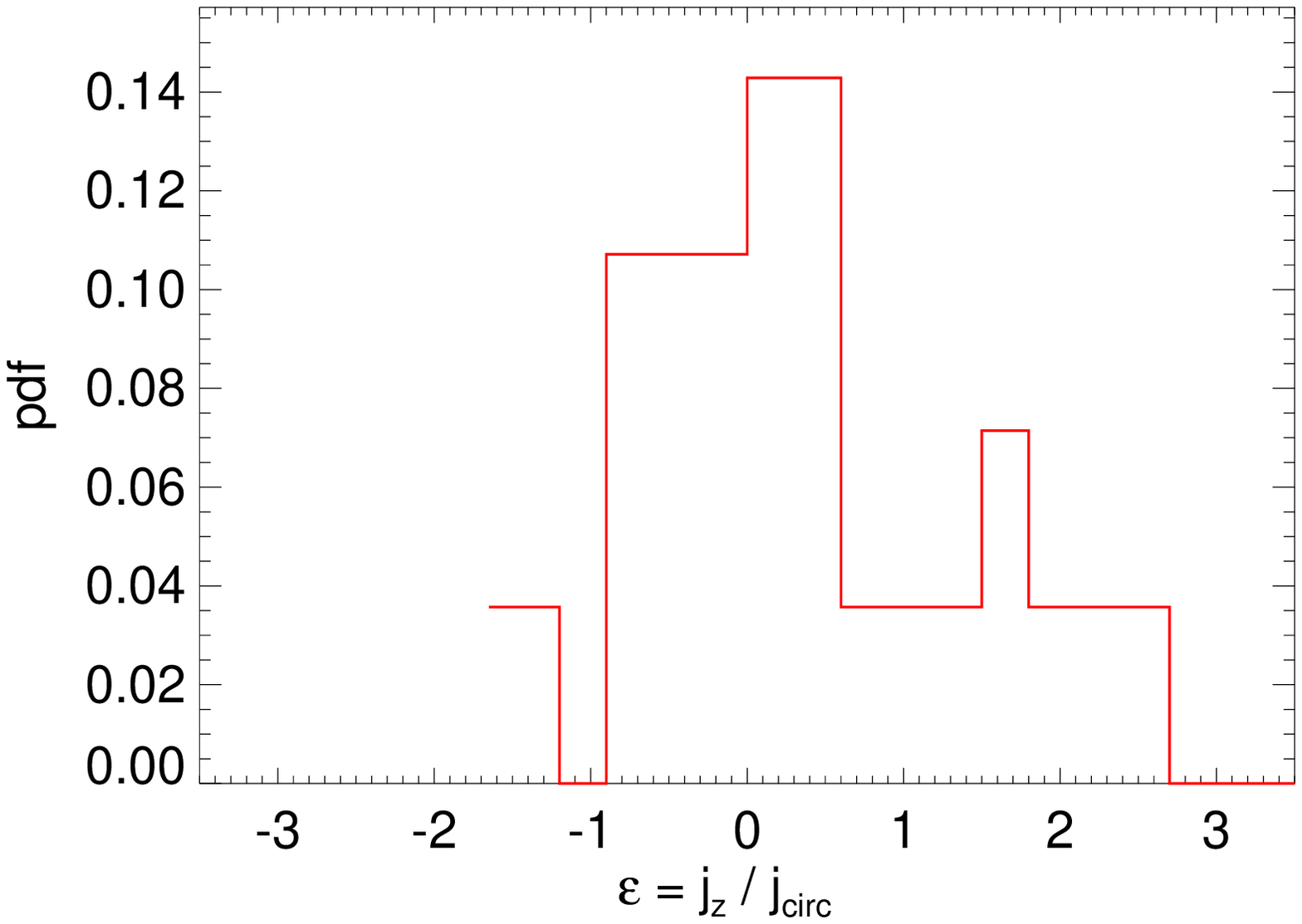}
\includegraphics[width=0.3\textwidth]{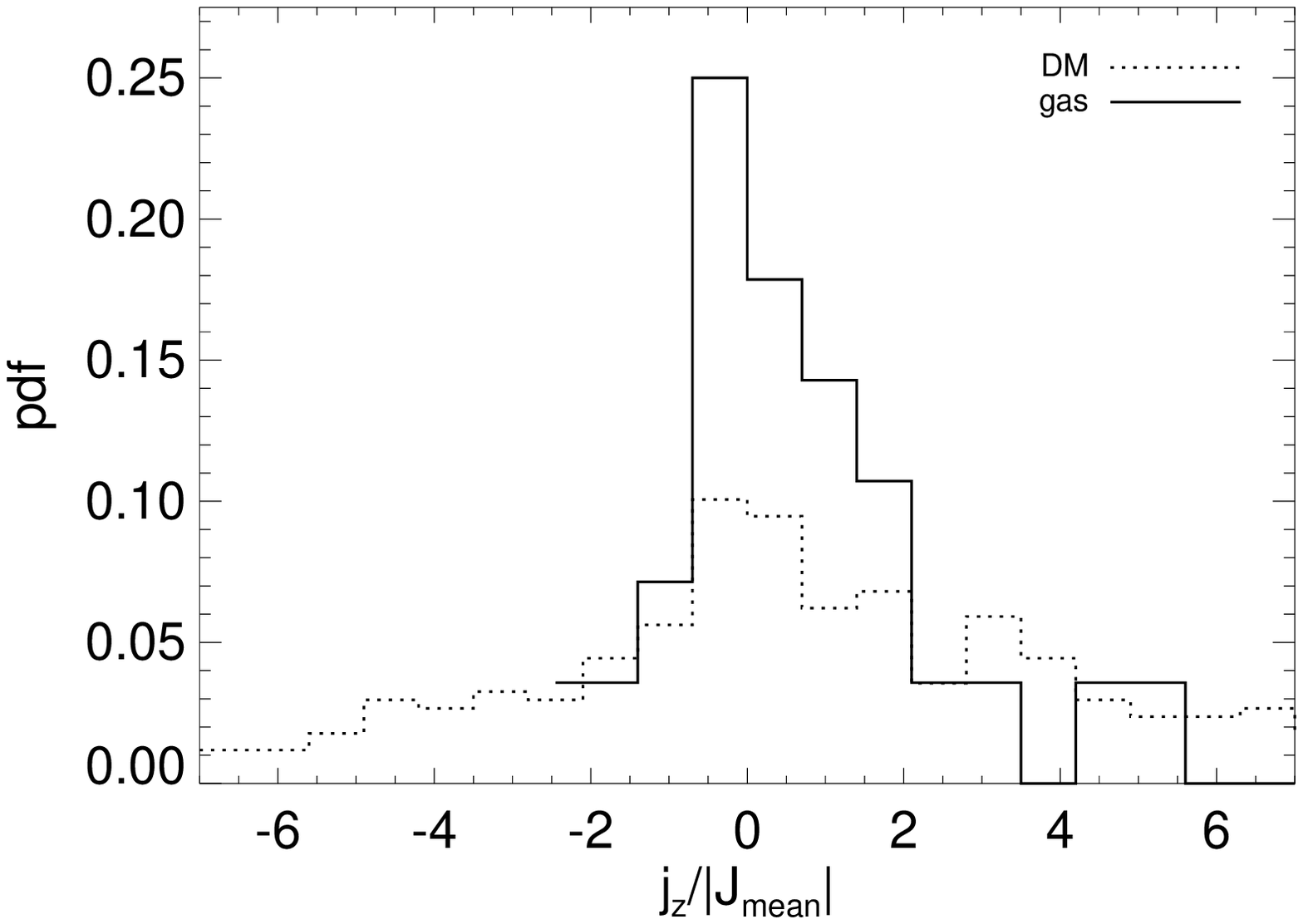}\\
\caption[]{\small 
From top to bottom, the distributions of gas and total-matter profiles (left), circularity (centre) and angular momentum ratios (right) of the three DCBH host candidates at $z=8.5$ in the run with top-heavy popIII sources (TH.1e5) are shown.
}
\label{fig:DCBH3x3}
\end{figure*}

\begin{figure}
\includegraphics[width=0.5\textwidth]{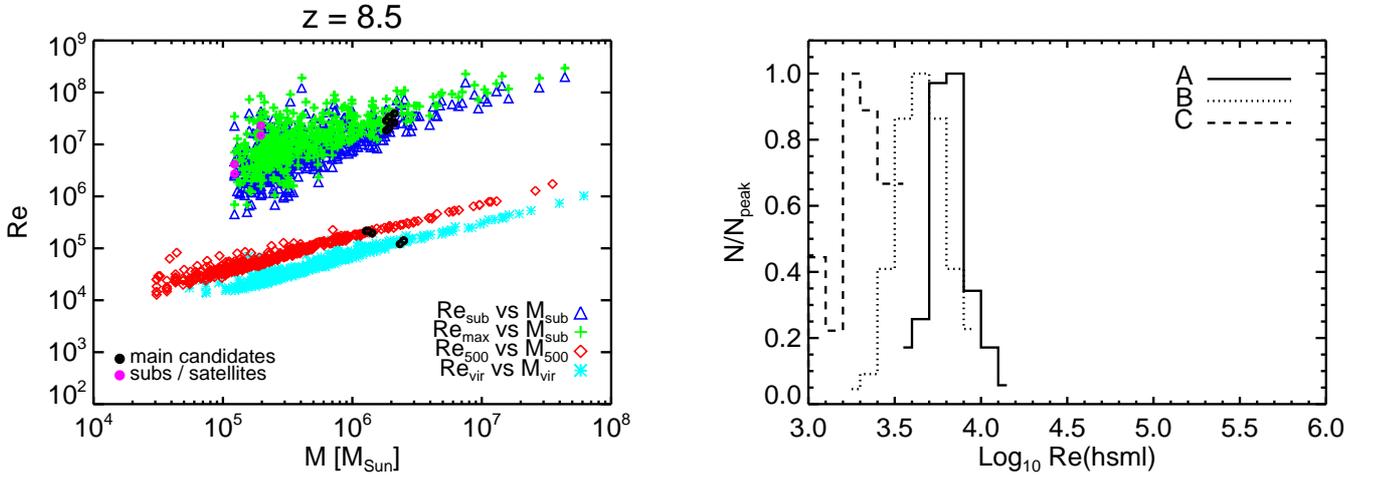}\\
\caption[]{\small
Halo population (symbols) and DCBH host candidates (bullets) Reynolds numbers at redshift $z=8.5$ estimated with different approaches (see legends and text in Sect.~\ref{sect:Re}).
}
\label{fig:Re018}
\end{figure}

\begin{figure}
\includegraphics[width=0.5\textwidth]{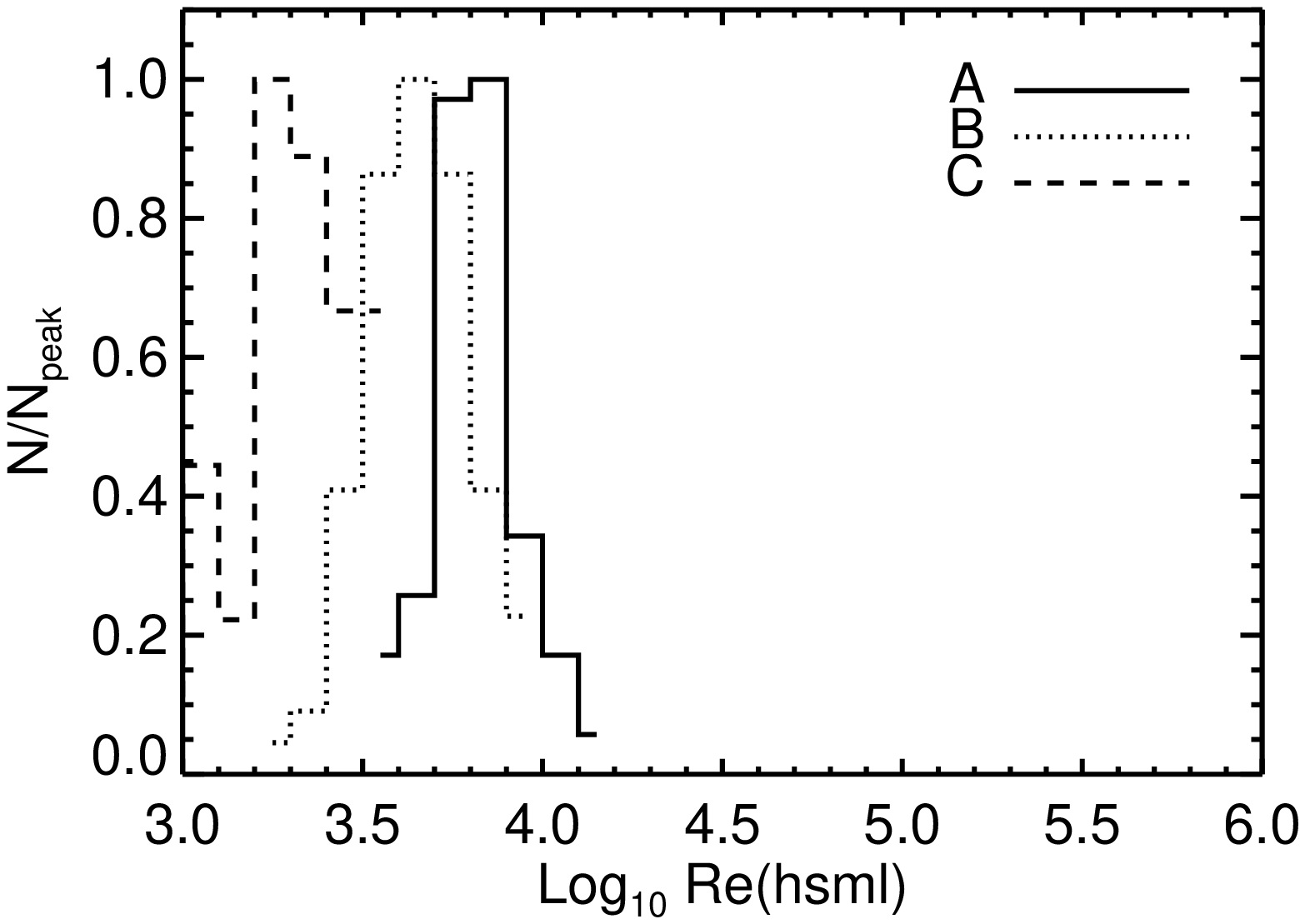}\\
\caption[]{\small 
Distributions of the Reynolds number estimated at the hsml scale for the three DCBH candidates, namely A (solid line), B (dotted line) and C (dashed
line). The distributions are normalised to their peak value.
Although identified with the same name, the three DCBH host candidates at $z=8.5$ do not correspond to the ones at $z=9$ (see discussion in Sect.~\ref{sect:Re}).
}
\label{fig:Rehsml018}
\end{figure}

For sake of completeness, here we briefly report results for $z=8.5$ DCBH host candidates, as already done and extensively discussed in the main body of the text for $z=9$.
The mass-weighted temperature map with highlighted DCBH host candidates at $z=8.5$ for the run with top-heavy popIII SED is depicted in Fig.~\ref{fig:groups018}, while, in Fig.~\ref{fig:DCBH3x3}, corresponding density profiles, circularity and angular momentum ratio distributions of the three DCBH host candidates are displayed.
As mentioned, these are host haloes different from the ones identified at $z=9$.\\
In Fig.~\ref{fig:Re018}, expected Reynolds numbers, $Re$, for both the entire halo population and the DCBH host candidates at $z=8.5$ are displayed. $Re$ values have been estimated by following the different approaches discussed in Sect.~\ref{sect:Re} about Fig.~\ref{fig:Re017}.
The distributions of the Reynolds number estimated at the hsml scale for the three DCBH candidates are shown in Fig.~\ref{fig:Rehsml018}.
The moderately high $Re$ number values point towards a mildly turbulent medium (see discussion in Sect.~\ref{sect:Re}).


\end{document}